\documentclass[aps,prd,preprint,superscriptaddress,floatfix,showpacs]{revtex4}
\usepackage{graphicx}

\begin{document}

\newcommand{\figwidth}{70mm}

\preprint{KEK-CP-139}

\title{Heavy quark expansion parameters from lattice NRQCD}

\newcommand{\Tsukuba}%
{Institute of Physics, University of Tsukuba,
 Tsukuba, Ibaraki 305-8571, Japan}

\newcommand{\RCCP}%
{Center for Computational Physics, University of Tsukuba,
 Tsukuba, Ibaraki 305-8577, Japan}

\newcommand{\ICRR}%
{Institute for Cosmic Ray Research, University of Tokyo,
 Kashiwa, Chiba 277-8582, Japan}

\newcommand{\KEK}%
{High Energy Accelerator Research Organization (KEK),
 Tsukuba, Ibaraki 305-0801, Japan}

\newcommand{\Hiroshima}%
{Department of Physics, Hiroshima University,
Higashi-Hiroshima, Hiroshima 739-8526, Japan}

\newcommand{\YITP}%
{Yukawa Institute for Theoretical Physics, Kyoto University,
 Kyoto 606-8502, Japan}

\newcommand{\BNL}%
{RIKEN BNL Research Center, Brookhaven National Laboratory,
 Upton, NY, 11973}

\author{S.~Aoki}
\affiliation{\Tsukuba}


\author{M.~Fukugita}
\affiliation{\ICRR}

\author{S.~Hashimoto}
\affiliation{\KEK}

\author{K-I.~Ishikawa}
\affiliation{\Tsukuba}
\affiliation{\RCCP}

\author{N.~Ishizuka}
\affiliation{\Tsukuba}
\affiliation{\RCCP}

\author{Y.~Iwasaki}
\affiliation{\Tsukuba}

\author{K.~Kanaya}
\affiliation{\Tsukuba}

\author{T.~Kaneko}
\affiliation{\KEK}

\author{Y.~Kuramashi}
\affiliation{\KEK}

\author{M.~Okawa}
\affiliation{\Hiroshima}



\author{N.~Tsutsui}
\affiliation{\KEK}

\author{A.~Ukawa}
\affiliation{\Tsukuba}
\affiliation{\RCCP}

\author{N.~Yamada}
\altaffiliation[Present address: ]{\BNL}
\affiliation{\KEK}

\author{T.~Yoshi\'{e}}
\affiliation{\Tsukuba}
\affiliation{\RCCP}

\collaboration{JLQCD Collaboration}
\noaffiliation

\date{\today}


\begin{abstract}
  We present a lattice QCD calculation of the heavy quark
  expansion parameters $\mu_{\pi}^2$ and $\mu_G^2$
  for heavy-light mesons and heavy-light-light baryons.
  The calculation is carried out on a 20$^3\times$48 lattice
  at $\beta$ = 6.0 in the quenched approximation, using the
  lattice NRQCD action for heavy quarks.
  We obtain the parameters $\mu_{\pi}^2$ and $\mu_G^2$ in
  two different methods: a direct calculation of the matrix
  elements and an indirect calculation through the mass
  spectrum, and confirm that the both methods give
  consistent results.
  We also discuss an application to the lifetime ratios.
\end{abstract}


\maketitle

\section{Introduction}

The heavy quark expansion (HQE)
\cite{Neubert:1997gu,Bigi:1997fj} is a fundamental tool in
the study of heavy quark physics. 
The inclusive decay rate of heavy hadrons containing a
single heavy quark may be expanded in terms of inverse heavy
quark mass $1/m_Q$ using the Operator Product Expansion
(OPE) technique, which enables us to calculate the inclusive
rates in a model independent manner
\cite{Chay:1990da,Bigi:1993fe,Manohar:1993qn,Blok:1993va}.
In particular, the determination of the
Cabibbo-Kobayashi-Maskawa (CKM) matrix elements $|V_{cb}|$
and $|V_{ub}|$ through the corresponding semi-leptonic
branching fractions relies on HQE.

It requires, however, several nonperturbative parameters as
coefficients in HQE.
At the order $1/m_Q^2$ the nonperturbative parameters
\begin{eqnarray}
  \mu_{\pi}^2(H_Q) 
  & \equiv &
  \frac{1}{2M_{H_Q}}
  \left\langle H_Q \left| \bar{Q}(i\vec{D})^2 Q \right| H_Q \right\rangle,
  \label{eq:mu_pi}
  \\
  \mu_G^2(H_Q) 
  & \equiv &
  \frac{1}{2M_{H_Q}}
  \left\langle H_Q \left| \bar{Q}\vec{\sigma}\cdot\vec{B} Q \right| H_Q
  \right\rangle,
  \label{eq:mu_G}
\end{eqnarray}
appear in general.
Here, $Q$ denotes a heavy quark field defined in the Heavy
Quark Effective Theory (HQET), and $|H_Q\rangle$ represents
a heavy-light meson or a heavy-light-light baryon state
(for $b$ hadrons, $H_b$ = $B$, $B^*$, $\Lambda_b$,
$\Sigma_b$, $\Sigma_b^*$, \textit{etc.}). 
Both parameters have mass dimension two, since they include
a (spatial) covariant derivative squared $\vec{D}^2$ or a
chromomagnetic operator $\vec{B}$. 
The inclusive decay rate of $H_Q$ is written in terms of
$\mu_{\pi}^2(H_Q)$ and $\mu_G^2(H_Q)$ as 
\begin{equation}
  \label{eq:inclusive_decay_rate}
  \Gamma(H_Q\rightarrow X_f) =
  \frac{G_F^2 m_Q^5}{192\pi^3}
  \left[
    c_3^f
    \left(
      1 - \frac{\mu_{\pi}^2(H_Q)-\mu_G^2(H_Q)}{2m_Q^2}
    \right)
    + 2 c_5^f\,
    \frac{\mu_G^2(H_Q)}{m_Q^2}
    + \cdots
  \right],
\end{equation}
where the coefficients $c_3^f$ and $c_5^f$ are
perturbatively calculable.
On the other hand, the parameters
$\mu_{\pi}^2(H_Q)$ and $\mu_G^2(H_Q)$ have to be extracted
from some experimental data or to be calculated
nonperturbatively. 
Several methods to determine $\mu_\pi^2$ and $\mu_G^2$ have
been studied, and some of them are summarized in
Section~\ref{sec:Heavy_Quark_Expansion_Parameters}.

In this work we calculate $\mu_\pi^2$ and $\mu_G^2$ in
quenched lattice QCD using the NRQCD action including
$O(1/m_Q)$ terms for heavy quark. 
Since the matrix element of power divergent operator
$\bar{Q}(i\vec{D})^2Q$ suffers from large perturbative
uncertainty in the matching calculation with the continuum
operator \cite{Martinelli:1995vj}, we consider their
difference between different hadron states,
like $\mu_\pi^2(\Lambda_b)-\mu_\pi^2(B)$,
in which the power divergence cancels.
This kind of difference is also interesting in its own
right, as it appears in the evaluation of lifetime
difference of $b$ hadrons \cite{Neubert:1997we}.

One of the advantages of this calculation is that we can
choose several quark masses in the calculation so that the
heavy quark mass dependence of the hadron masses and matrix 
elements may be studied.
We calculate both matrix elements $\mu_\pi^2$ and $\mu_G^2$
and compare them with the corresponding mass spectrum and
its heavy quark mass dependence.
Another advantage in the use of the NRQCD lattice action
is that the statistical signal in the Monte Carlo
calculation is much better than in the static limit
\cite{Hashimoto:1994nd}.

This paper is organized as follows.
In Section~\ref{sec:Heavy_Quark_Expansion_Parameters}, the
implications for the heavy quark expansion parameters from
heavy hadron spectrum and the results of the previous
nonperturbative calculations are discussed.
In Section~\ref{sec:Lattice_calculation} we describe our
lattice calculation in detail.
The results for hadron masses and heavy quark expansion parameters
are shown in Section~\ref{sec:Results}.
The consistency check between the calculation of matrix
elements and spectrum is also presented.
Our results are applied to the lifetime ratio of
different $b$ hadrons in Section~\ref{sec:Phenomenological_application}.
The conclusions are given in
Section~\ref{sec:Conclusions}.

\section{Heavy Quark Expansion Parameters}
\label{sec:Heavy_Quark_Expansion_Parameters}
In this section we briefly review the determination of the
HQE parameters from mass spectrum and from some
nonperturbative techniques.
The determination through the measurements of several
mass and energy moments in the inclusive 
$B\rightarrow X_cl\nu$ and $B\rightarrow X_s\gamma$ 
decays is another possibility 
\cite{Dikeman:1995ad,Kapustin:1995nr,Falk:1995me,Falk:1995kn,Gremm:1996yn},
which is not covered in the following. 

\subsection{Implications from spectroscopy}
\label{subsec:Implications_from_spectroscopy}
The HQE parameters $\mu_{\pi}^2$ and $\mu_G^2$ defined in 
(\ref{eq:mu_pi}) and (\ref{eq:mu_G}) can be indirectly
obtained through heavy hadron masses, using the HQE of
hadron masses 
\begin{equation}
  M_{H_Q} = m_Q 
  + \overline{\Lambda}
  + \frac{\mu_{\pi}^2(H_Q)-\mu_G^2(H_Q)}{2m_Q}
  + O\left(\frac{1}{m_Q^2}\right),
  \label{eq:mass_formula}
\end{equation}
where $\overline{\Lambda}$ is the residual energy difference 
between $M_{H_Q}$ and $m_Q$ surviving in the infinite heavy
quark mass limit. 
The parameters $\mu_{\pi}^2$ and $\mu_G^2$ appear in the
correction term of $O(1/m_Q)$.
Considering proper mass differences, certain
combinations of $\overline{\Lambda}$, $\mu_{\pi}^2$ and $\mu_G^2$
can be extracted as shown below.

The notation $\lambda_1$ and $\lambda_2$ is often used 
instead of $\mu_\pi^2$ and $\mu_G^2$
for $B$ and $B^*$ mesons in the literature.
The relation between $\lambda_{1,2}$ and $\mu_{\pi,G}^2$ is
given by 
\begin{eqnarray}
\label{eq:symmetry_relation_pi}
  \lambda_1 & \equiv & -\mu_\pi^2(B) = -\mu_\pi^2(B^*),\\
\label{eq:symmetry_relation_G}
  \lambda_2 & \equiv & \frac{1}{3}\mu_G^2(B) = - \mu_G^2(B^*),
\end{eqnarray}
and the HQE of meson masses in (\ref{eq:mass_formula})
becomes 
\begin{eqnarray}
  M_B &=& m_b + \overline{\Lambda}
  -\frac{\lambda_1+3\lambda_2}{2m_b} +
  O\left(\frac{1}{m_b^2}\right), 
  \label{eq:M_B}
  \\
  M_{B^*} &=& m_b + \overline{\Lambda}
  -\frac{\lambda_1- \lambda_2}{2m_b} +
  O\left(\frac{1}{m_b^2}\right). 
  \label{eq:M_Bstar}
\end{eqnarray}
The parameter $\lambda_2$ may be evaluated through the
hyperfine splitting of ground state $B$ mesons as
\begin{equation}
  M_{B^*}-M_B 
  \left(\simeq \frac{4\lambda_2}{2m_b} \right)
  = 46 \;\mathrm{MeV},
\end{equation}
or, equivalently
\begin{equation}
  \lambda_2 \simeq
  \frac{1}{4} \left(M_{B^*}^2-M_B^2\right) 
  = 0.12 \; \textrm{GeV}^2,
\end{equation}
at the leading order.

For $\Lambda_b$ baryon, the parameter $\mu_G^2(\Lambda_b)$
vanishes, since the light degrees of freedom is spin
singlet inside $\Lambda_b$.
The relations
\begin{eqnarray}
  \mu_\pi^2(\Sigma_b) = \mu_\pi^2(\Sigma_b^*),
  \label{eq:symmetry_relation_pi_for_baryon}
  \\
  \frac{1}{2}\mu_G^2(\Sigma_b) = -\mu_G^2(\Sigma_b^*),
  \label{eq:symmetry_relation_G_for_baryon}
\end{eqnarray}
hold for $\Sigma_b$ and $\Sigma_b^*$ baryons, as they are
related by spin rotations, 
analogous to (\ref{eq:symmetry_relation_pi}) and
(\ref{eq:symmetry_relation_G}) for $B^{(*)}$ mesons.

The spin-averaged meson mass becomes independent of
$\lambda_2$ 
\begin{equation}
  M_{\bar{B}} \equiv \frac{M_B+3M_{B^*}}{4} =
  m_b + \overline{\Lambda} - \frac{\lambda_1}{2m_b} + 
  O\left(\frac{1}{m_b^2}\right),
\end{equation}
but $\lambda_1$ cannot be extracted solely from this
expression, as it appears together with the lowest order
parameter $\overline{\Lambda}$.
In order to proceed further, we have to consider a similar
relation for the $D$ meson and take a mass difference to
obtain 
\begin{equation}
  M_{\bar{B}} - M_{\bar{D}} = m_b - m_c
  - \lambda_1 
  \left(\frac{1}{2m_b}-\frac{1}{2m_c}\right)
  + O\left(\frac{1}{m_{b,c}^2}\right).
\end{equation}
The leading dependence on the heavy quark masses $m_b$ and 
$m_c$ can be subtracted out if we take a double mass
difference 
\begin{equation}
  \mu_\pi^2(\Lambda_b)-\mu_\pi^2(B)
  =
  2 \frac{(M_{\Lambda_b}-M_{\Lambda_c}) - (M_{\bar{B}}-M_{\bar{D}})}
  {\frac{1}{M_{\bar{B}}}-\frac{1}{M_{\bar{D}}}}
  + O\left(\frac{1}{m_{b,c}}\right),
\end{equation}
from which we obtain
\begin{equation}
  \mu_{\pi}^2(\Lambda_b)-\mu_{\pi}^2(B) =
  -0.01 \pm 0.03 \; \mathrm{GeV}^2.
  \label{eq:diff_mupi}
\end{equation}
This argument relies on HQE truncated at order $1/m_Q$,
which is questionable for charmed mesons and baryons.
Therefore, for the use of the HQE parameter $\mu_\pi^2$ in
other phenomenological analysis, some independent
theoretical calculations are desirable.

\subsection{Nonperturbative calculations}
\label{sec:Nonperturbative_calculations}
The determination of $\lambda_1$ using the QCD sum rule has
been attempted by two groups and reached conflicting results
$\lambda_1$ = $-$0.5$\pm$0.2~GeV$^2$ \cite{Ball:1993xv} and
$-$0.1$\pm$0.05~GeV$^2$ \cite{Neubert:1996wm}.
Their difference is explained to come from non-diagonal
matrix elements like 
$\langle B|\bar{Q}(i\vec{D})^2 Q|B'\rangle$,
where $B'$ is an excited state of $B$ meson
\cite{Bigi:1997fj}. 
Since there is no definite way to evaluate these matrix
elements at present, it is not straightforward to improve
the determination of $\lambda_1$ within the QCD sum rule
technique. 

The lattice QCD can also be used to determine the HQE
parameters.
In the lattice calculation of the matrix element 
$\langle B|\bar{Q}(i\vec{D})^2 Q|B\rangle$
the subtraction of quadratic divergence is essential, since
otherwise the perturbative expansion to relate lattice and
continuum operators poorly converges
\cite{Martinelli:1995vj}.
First lattice calculation with such nonperturbative
subtraction was done by Crisafulli \textit{et al.}
\cite{Crisafulli:1995} using the HQET on the lattice, 
which was updated in Gimenez \textit{et al.}
\cite{Gimenez:1997av},
and the result is 
$\lambda_1$ = 0.09 $\pm$ 0.14~GeV$^2$.

Another possible approach on the lattice is to fit the
measured mass spectrum for various heavy quark masses with
the mass relation (\ref{eq:mass_formula}).
Ali~Khan \textit{et al.} \cite{AliKhan:1999yb} performed
such analysis for $b$ flavored mesons and baryons using
the lattice NRQCD for heavy quark.
Their result is
$\lambda_1$ = $-$0.1 $\pm$ 0.4~GeV$^2$ for $B$ meson.
Kronfeld and Simone \cite{Kronfeld:2000gk} performed similar
analysis with a larger set of lattice data of heavy-light
mesons, and quoted 
$\lambda_1$ = $-$0.45 $\pm$ 0.12~GeV$^2$.
The calculation of $\mu_\pi^2$ for $b$ baryon is available
only from Ali~Khan \textit{et al.} \cite{AliKhan:1999yb}.
They quoted $\mu_\pi^2(\Lambda_b)$ = $-$1.7 $\pm$ 3.4 GeV$^2$.

For the parameter $\lambda_2$, 
Gimenez \textit{et al.} \cite{Gimenez:1997av}
found $\lambda_2$ = 0.07 $\pm$ 0.01 GeV$^2$ from the direct
calculation of the matrix element.
Ali~Khan \textit{et al.} \cite{AliKhan:1999yb} estimated
$\lambda_2(B_d)$ = 0.069 $\pm$ 0.019 GeV$^2$ and
$\lambda_2(B_s)$ = 0.078 $\pm$ 0.012 GeV$^2$
from the hyperfine splitting measured on the lattice.

The difference of $\overline{\Lambda}$ between several heavy
hadrons is only estimated from the mass difference.
Ali~Khan \textit{et al.} \cite{AliKhan:1999yb} estimated 
$\overline{\Lambda}(\Lambda_b)-\overline{\Lambda}(B)$ 
= 415 $\pm$ 156 MeV,
$\overline{\Lambda}(\Sigma_b)-\overline{\Lambda}(\Lambda_b)$ 
= 176 $\pm$ 152 MeV and
$\overline{\Lambda}(B_s)-\overline{\Lambda}(B_d)$ 
= 81 $\pm$ 31 MeV.

In this work we calculate $\mu_\pi^2$ and $\mu_G^2$ on the
lattice for ground state mesons and baryons.
We use the both methods, namely the direct measurement of
the matrix elements and the extraction from the heavy hadron
spectrum.
The difference of $\overline{\Lambda}$ is also evaluated
from the mass difference.

\section{Lattice calculation}
\label{sec:Lattice_calculation}
In this section we present the details of our lattice
calculation, which include the definition of the NRQCD
action, simulation parameters, and the method to
extract the matrix elements.
The matching of lattice operators onto their continuum
counterpart is also discussed.

\subsection{Lattice NRQCD}
We use the lattice NRQCD action
\cite{Thacker:1990bm,Lepage:1992tx} for heavy quark. 
The particular form of the action used in this work is the
same as in \cite{Ishikawa:1999xu,Aoki:2002bh}.
\begin{equation}
  S_{\mathrm{NRQCD}}= 
  \sum_{x,y} Q^{\dagger}(x)( \delta_{x,y} - K_{Q}(x,y) ) Q(y).
  \label{eq:NRQCD_lattice}
\end{equation}
The kernel to describe the time evolution of heavy quark is
given by
\begin{equation}
  K_{Q}(x,y) 
  \equiv
  \left( 1-\frac{a H_{0}}{2 n} \right)_{t+1}^n
  \left( 1-\frac{a \delta H}{2} \right)_{t+1}
  \delta^{(-)}_{4}
  U^{\dagger}_{4}(t)
  \left( 1-\frac{a \delta H}{2} \right)_t
  \left( 1-\frac{a H_{0}}{2 n} \right)_t^n,
  \label{eq:evolution_kernel_Q}
\end{equation}
where the index to label the spatial coordinate is suppressed.
The operator $\delta^{(-)}_4$ is defined as 
$\delta^{(-)}_4(x,y) \equiv \delta_{x_4-1,y_4}
 \delta_{\vec{x},\vec{y}}$, and
\begin{eqnarray}
  H_0 
  & \equiv & 
  -\frac{\Delta^{(2)}}{2m_Q}, 
  \label{eq:H_0}
  \\
  \delta H 
  & \equiv &
  - c_B \frac{g}{2m_Q}\vec{\sigma}\cdot\vec{B}.
  \label{eq:delta_H}
\end{eqnarray}
$\Delta^{(2)}$ is a lattice covariant Laplacian
\begin{eqnarray}
  \Delta^{(2)} Q(x)
  &=& 
  \sum_{i=1}^3\Delta^{(2)}_iQ(x) 
  \nonumber\\
  &=& 
  \sum_{i=1}^3
  \left[
    U_i(x)Q(x+\hat{i}) + U_i^{\dagger}(x-\hat{i})Q(x-\hat{i})
    - 2 Q(x)
  \right],
\end{eqnarray}
and the chromo-magnetic field $\vec{B}$ is defined as the
clover-leaf type on the lattice \cite{Lepage:1992tx}.
The parameter $n$ in the evolution kernel
(\ref{eq:evolution_kernel_Q}) is a positive integer
introduced to stabilize unphysical momentum modes
\cite{Thacker:1990bm,Lepage:1992tx}.
With these definitions the lattice NRQCD action
(\ref{eq:NRQCD_lattice}) deduces to the usual continuum
NRQCD action 
\begin{equation}
  \mathcal{L}_{\mathrm{NRQCD}}^{cont} =
  Q^{\dagger} \left[
    D_0 + \frac{\vec{D}^2}{2M}
    + g \frac{\vec{\sigma}\cdot\vec{B}}{2M}
  \right]
  Q
  \label{eq:NRQCD_continuum}
\end{equation}
in the limit of vanishing lattice spacing.

The parameters appearing in the NRQCD action
(\ref{eq:NRQCD_lattice}), $m_Q$ and $c_B$ at this order,
have to be matched onto their continuum counterparts using
perturbation theory.
The matching of heavy quark mass $m_Q$ is done through the
calculation of hadron masses as described later.
On the other hand, the one-loop calculation for $c_B$ is
unfortunately not yet available, so we use the tree level
value $c_B$ = 1.
However, we apply the mean field improvement of the gauge
link variable $U_\mu(x)\rightarrow U_\mu(x)/u_0$
\cite{Lepage:1992xa} everywhere it appears, with $u_0$ a
mean link value defined through the plaquette expectation
value 
$u_0 \equiv \langle \frac{1}{3} \mathrm{Tr} U_P \rangle$.
With the mean field improvement we expect that the tree
level matching is reasonably good.
Furthermore the final predictions for the matrix elements
deduced from our analysis are given in the static limit,
which is irrelevant to the parameter $c_B$.

The four-component heavy quark field $h$ used to construct
the hadron interpolating fields is related to
the two-component nonrelativistic field $Q$ through the
Foldy-Wouthuysen-Tani (FWT) transformation
\begin{equation}
  h = R
  \left(
    \begin{array}{c}
      Q\\
      0
    \end{array}
  \right),
\end{equation}
with the rotation matrix $R$ given by
\begin{equation}
  R = 1-\frac{\vec{\gamma}\cdot\vec{\Delta}}{2m_Q}
\end{equation}
at order $1/m_Q$.
Our convention for the gamma matrices is
\begin{equation}
  \gamma_4 =
  \left(
    \begin{array}{cc}
      \mathrm{I} & 0\\
      0 & -\mathrm{I}
    \end{array}
  \right),
  \;\;\;\;
  \vec{\gamma}=
  \left(
    \begin{array}{cc}
      0 & -i\vec{\sigma}\\
      i\vec{\sigma} & 0
    \end{array}
  \right),
\end{equation}
and the spatial covariant derivative is defined as
\begin{equation}
  \Delta_i Q(x)=
  \frac{1}{2}
  \left[
    U_i(x)Q(x+\hat{i})-U_i^{\dagger}(x-\hat{i})Q(x-\hat{i})
  \right].
\end{equation}

\subsection{Simulation details}
Our calculation is carried out in quenched lattice QCD at
$\beta$ = 6.0 on a $20^3\times48$ lattice. 
Gauge configurations are generated with the single plaquette
action, and 515 configurations are analyzed.

The NRQCD action including $O(1/m_Q)$ described in the
previous subsection is adapted for heavy quarks. 
Five heavy quark masses $am_Q$ = 1.3, 2.1, 3.0, 5.0, and 10.0
are simulated to study the $m_Q$ dependence of hadron masses
and matrix elements. 
The details on the parameters for heavy quark are shown
in Table~\ref{tab:NRQCD_parameters}.

For light quarks, the $O(a)$-improved Wilson action
\cite{Sheikholeslami:1985ij} with the non-perturbatively
tuned coefficient $c_{\mathrm{SW}}$ = 1.769
\cite{Luscher:1996ug} is used. 
Three hopping parameters $K$ = 0.13331, 0.13384, and 0.13432
are employed to extrapolate to the chiral limit
$K_c$ = 0.135284(8).
The inverse lattice spacing $a^{-1}$ = 1.85(5)~GeV is
determined through the $\rho$ meson mass $m_{\rho}$ = 770~MeV.

The strange quark mass $am_s$=0.0460(22)
is fixed using $m_K/m_{\rho}$=0.644 as an input.

\subsection{Hadron masses}
The hadron masses are measured through the asymptotic
behavior of two-point functions
\begin{equation}
  C(J;t) =
  \sum_{\vec{x}}
  \langle J(\vec{x},t) {J^{(S)}}^{\dagger}(\vec{0},0)\rangle
  \rightarrow e^{-E_{sim} t},
  \label{eq:C}
\end{equation}
for sufficiently large time separation $t$.
With the NRQCD action, for which the bare heavy quark mass
is subtracted from the formulation, we obtain the binding
energy $E_{sim}$ from the two-point function.
The interpolating operator $J$ is chosen such that it shares 
the same quantum number with the hadron of interest.
The hadrons and their interpolating operators we consider in
this work are the following. 
\begin{eqnarray}
  B &=& \bar{d}\gamma_4\gamma_5 h, \\
  B^* &=& \bar{d}\gamma_i h, \\
  \Lambda_b(s_z=+1/2) &=& 
  \epsilon_{abc}(u^a C\gamma_5 d^b)h^c_{\uparrow}, \\
  \Lambda_b(s_z=-1/2) &=& 
  \epsilon_{abc}(u^a C\gamma_5 d^b)h^c_{\downarrow}, \\
  \Sigma_b(s_z=+1/2) &=&
  -\frac{1}{\sqrt{3}} \epsilon_{abc}
  (u^a C\gamma_3 d^b)h^c_{\uparrow}
  +\sqrt{\frac{2}{3}} \epsilon_{abc}
  \left(u^a C\frac{\gamma_1-i\gamma_2}{2} d^b\right)h^c_{\downarrow}, \\
  \Sigma_b(s_z=-1/2) &=&
  -\sqrt{\frac{2}{3}} \epsilon_{abc}
  \left(u^a C\frac{\gamma_1+i\gamma_2}{2}d^b\right)h^c_{\uparrow}
  +\frac{1}{\sqrt{3}} \epsilon_{abc}(u^a C\gamma_3  d^b)h^c_{\downarrow}, \\
  \Sigma_b^*(s_z=+3/2) &=&
  \epsilon_{abc}
  \left(u^a C\frac{\gamma_1-i\gamma_2}{2} d^b\right)h^c_{\uparrow}, \\
  \Sigma_b^*(s_z=+1/2) &=&
  \sqrt{\frac{2}{3}} \epsilon_{abc}(u^a C\gamma_3 d^b)h^c_{\uparrow}
  +\frac{1}{\sqrt{3}} \epsilon_{abc}
  \left(u^a C\frac{\gamma_1-i\gamma_2}{2} d^b\right)h^c_{\downarrow}, \\
  \Sigma_b^*(s_z=-1/2) &=&
  \frac{1}{\sqrt{3}} \epsilon_{abc}
  \left(u^a C\frac{\gamma_1+i\gamma_2}{2}d^b\right)h^c_{\uparrow}
  +\sqrt{\frac{2}{3}} \epsilon_{abc}(u^a C\gamma_3  d^b)h^c_{\downarrow}, \\
  \Sigma_b^*(s_z=-3/2) &=&
  \epsilon_{abc}
  \left(u^a C\frac{\gamma_1+i\gamma_2}{2} d^b\right)h^c_{\uparrow}.
\end{eqnarray}
Although the notations motivated from the $b$ hadron spectrum
are used, we use them for general heavy quark mass we
consider. 
The light quark fields $u$ and $d$ denote the relativistic
up and down quark fields, respectively.
The heavy quark field $h$ has a subscript $\uparrow$ or
$\downarrow$, which represents its spin component in the $z$
direction. 
We assume the Dirac representation of gamma matrices, and
$s_z$ means the $z$ component of the spin of baryons.
The charge conjugation matrix $C$ has a representation
$C=\gamma_0 \gamma_2$.
The superscript $a$, $b$ or $c$ denotes a color index of
quarks. 

The smeared operator $J^{(S)}$ is used at the source in
(\ref{eq:C}) to enhance the overlap with the ground state.
It is defined such that the heavy quark field is smeared
according to an exponential form $e^{-a \cdot r^b}$ around
the light quark field fixed at the origin.
$r$ is a distance from the origin, and the parameters $a$
and $b$ are measured for the pion wave function.
Thus, they depend on the light quark mass, as listed in
Table~\ref{tab:NRQCD_parameters}. 
Although it is not an optimal choice for heavy hadrons, the
plateau is satisfactory as we demonstrate later.

The hadron mass is obtained through the relation
\begin{equation}
  M_{\mathrm{had}} = 
  (Z_m M_0-E_0) + E_{\mathrm{sim}},
  \label{eq:binding_energy_and_mass}
\end{equation}
where $Z_m$ is the mass renormalization factor which relates
the bare quark mass $M_0$ with the pole mass and $E_0$ is
the energy shift of the heavy quark. 
These factors are perturbatively calculated at the one-loop
level in \cite{Ishikawa:1999xu} for our choice of heavy
quark action.
We summarize these factors in Table~\ref{tab:mass_constant}.

\subsection{Matrix elements}
To calculate the expansion parameters $\mu_{\pi}^2$ and
$\mu_G^2$ from three-point functions, 
we construct a ratio
\begin{equation}
  R_i(J;t,t^{\prime})
  =
  \frac{\sum_{\vec{x},\vec{y}}\langle J(\vec{x},t) O_i(\vec{y},t^{\prime})
    {J^{(S)}}^{\dagger}(\vec{0},0) \rangle
  }{\sum_{\vec{x}}\langle J(\vec{x},t) {J^{(S)}}^{\dagger}(\vec{0},0)
    \rangle},
  \label{eq:R}
\end{equation}
with $O_i$ either the kinetic operator
\begin{equation}
  O_{\pi} = \bar{Q} (i\vec{D})^2 Q,
  \label{eq:O_pi}
\end{equation}
or the chromo-magnetic operator
\begin{equation}
  O_G = \bar{Q} (\vec{\sigma}\cdot\vec{B}) Q.
  \label{eq:O_G}
\end{equation}
The interpolating operator $J$ is one of the operators
listed in the previous subsection.
The asymptotic behavior of the ratio yields the
corresponding matrix element. 
We fix the position of the operator at $t^{\prime}$ = 9 
and move the sink $t$. 

\subsection{Operator renormalization}
\label{subsec:Operator_renormalization}

The matching of the operators $O_\pi$ and $O_G$ with their
continuum counterpart is known only at the tree level except
in the static limit, where one-loop coefficients are known
\cite{Maiani:az,Flynn:1991kw}.
The perturbative expansion is especially dangerous for the
kinetic operator $O_\pi$, since it mixes with lower
dimensional operators $\bar{Q}D_0Q$ and $\bar{Q}Q$ and thus
power divergences appear. 
We, therefore, consider the differences of matrix elements
with different hadron states, such as
$\mu_\pi^2(\Lambda_b)-\mu_\pi^2(B)$, in which the effect of
mixed operators cancel at the leading order in $1/m_Q$.
The effect remains at finite values of $1/m_Q$, and
hence we take the infinite heavy quark mass limit after
measuring the differences at several values of $m_Q$.

The other operator $O_G$ does not mix with lower dimensional
ones in the static limit.
However, once the $1/m_Q$ correction is introduced,
the mixing with $O_\pi$ and
the other lower dimensional operators appears since the NRQCD action
contains the $\vec{\sigma}\cdot\vec{B}$ term.
Hence, we again consider the difference among different
hadron states to cancel the mixing contribution
and take the infinite heavy quark mass limit.

One-loop matching of the lattice operators to the
corresponding continuum operators is known in the static
limit.
For the kinetic operator $O_\pi$ it is calculated in
\cite{Maiani:az} and the multiplicative part is given as
\begin{equation}
  Z_\pi = 1 + 0.0687\, g^2.
  \label{eq:Z_pi}
\end{equation}
Since we apply the tadpole improvement using the plaquette
expectation value and its effect is to multiply the link
variable by $1/u_0$, the corresponding one-loop contribution 
$\frac{1}{12} g^2$ has to be subtracted from the one-loop
coefficient, and thus we obtain 
\begin{equation}
  \tilde{Z}_\pi = 1 - 0.0146\, g^2,
  \label{eq:Z_pi_tilde}
\end{equation}
whose numerical value at $\beta$ = 6.0 is 0.975 if we use
the boosted coupling $\tilde{g}^2=g_0^2/u_0^4$ = 1.70.
For the spin-chromomagnetic operator $O_G$ the one-loop
calculation is found in \cite{Flynn:1991kw} as
\begin{equation}
  Z_G = 1 + g^2 
  \left(
    - \frac{3}{16\pi^2}\ln m_Q^2 a^2 + 0.437
  \right), 
  \label{eq:Z_G}
\end{equation}
where $m_Q$ denotes the heavy quark mass arising from the
continuum theory.
The tadpole improvement amounts to multiply $1/u_0^4$ and the
one-loop coefficient is modified as
\begin{equation}
  \tilde{Z}_G = 1 + g^2 
  \left(
    - \frac{3}{16\pi^2}\ln m_Q^2 a^2 + 0.104
  \right),
  \label{eq:Z_G_tilde}
\end{equation}
and its numerical value is 1.12 for the $b$ quark mass
$m_b$ = 4.6~GeV.
For both cases the tadpole improvement acts to greatly
reduce the perturbative coefficients.

\section{Results}
\label{sec:Results}
In this section, we present the results for hadron masses
and matrix elements.
The heavy quark mass dependence of the matrix elements from
the direct calculation is studied carefully by two methods.
We also make a comparison between the results from the direct calculation
and from the indirect calculation.
All errors of measured quantities are estimated by
the single elimination jackknife procedure.

\subsection{Hadron masses}
\label{subsec:Hadron_masses}
\begin{figure}
  \includegraphics[width=\figwidth,angle=-90]{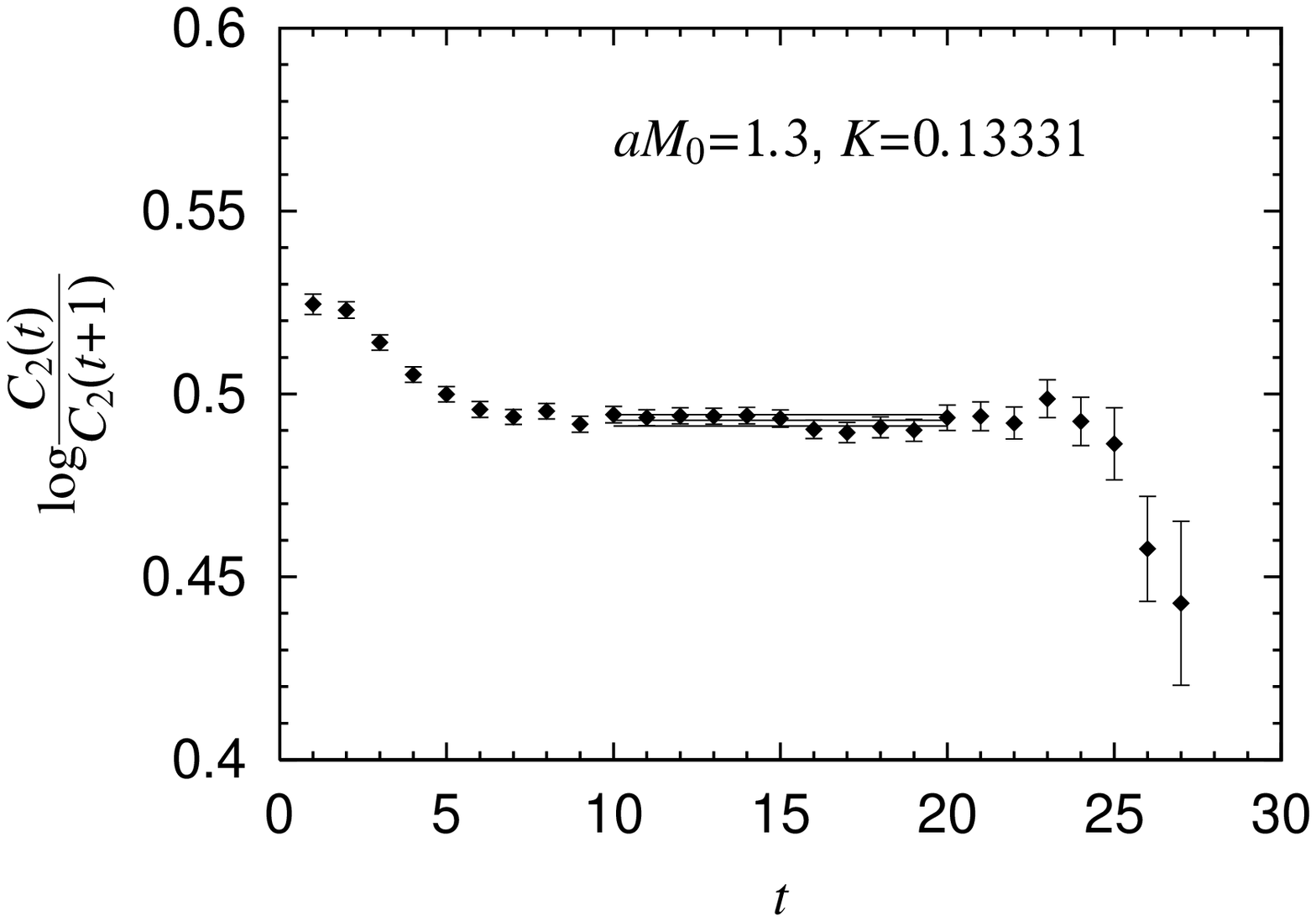}
  \\
  \includegraphics[width=\figwidth,angle=-90]{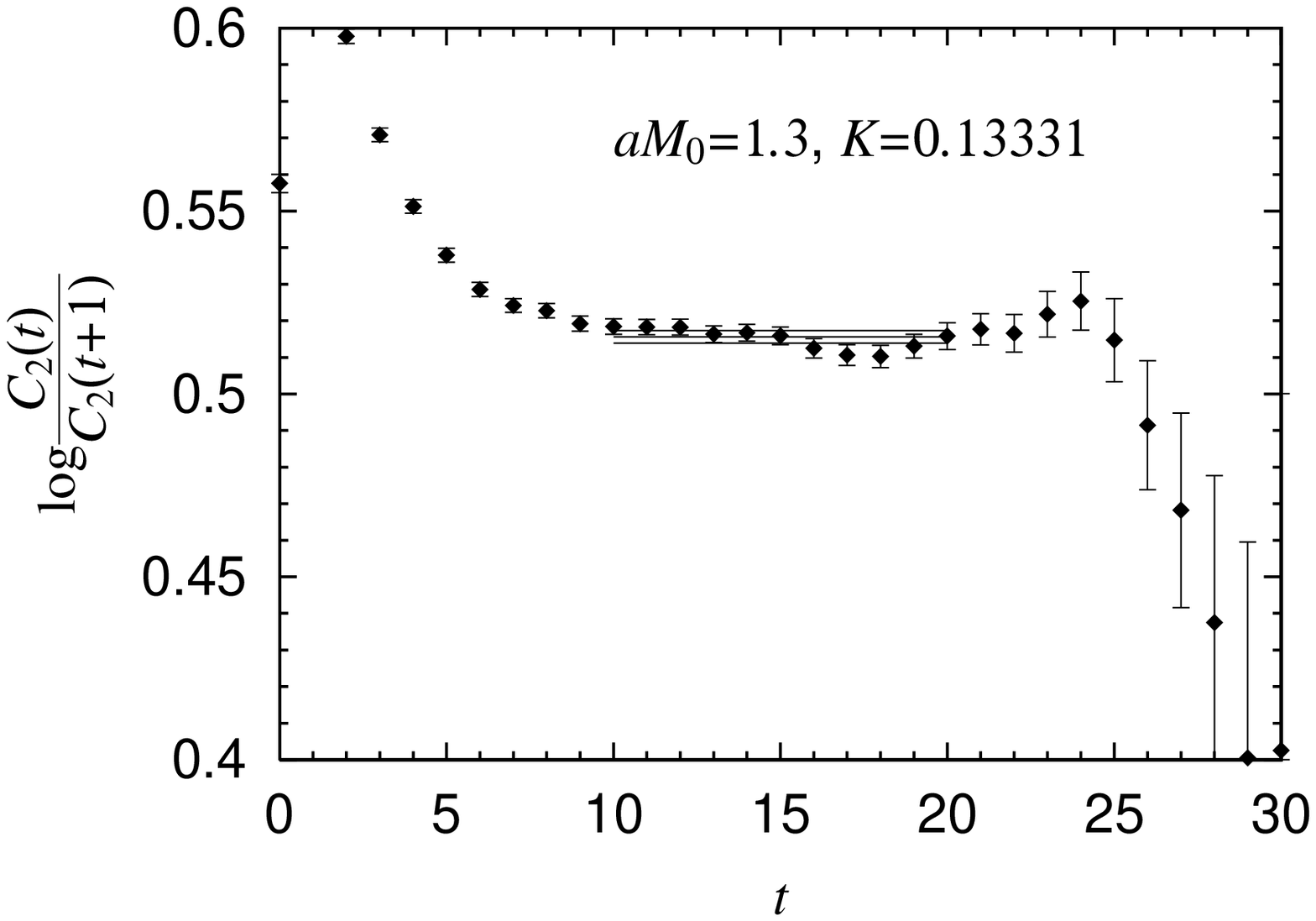}
  \caption{
    Effective mass plot for the $B$ (top panel) and $B^*$
    (bottom panel) mesons at $K$=0.13331 and
    $aM_0$=1.3. 
    Solid lines represent the fitting result with an error
    band of one standard deviation.
  }
  \label{fig:effective_mass_mesons}
\end{figure}

\begin{figure}
  \includegraphics[width=\figwidth,angle=-90]{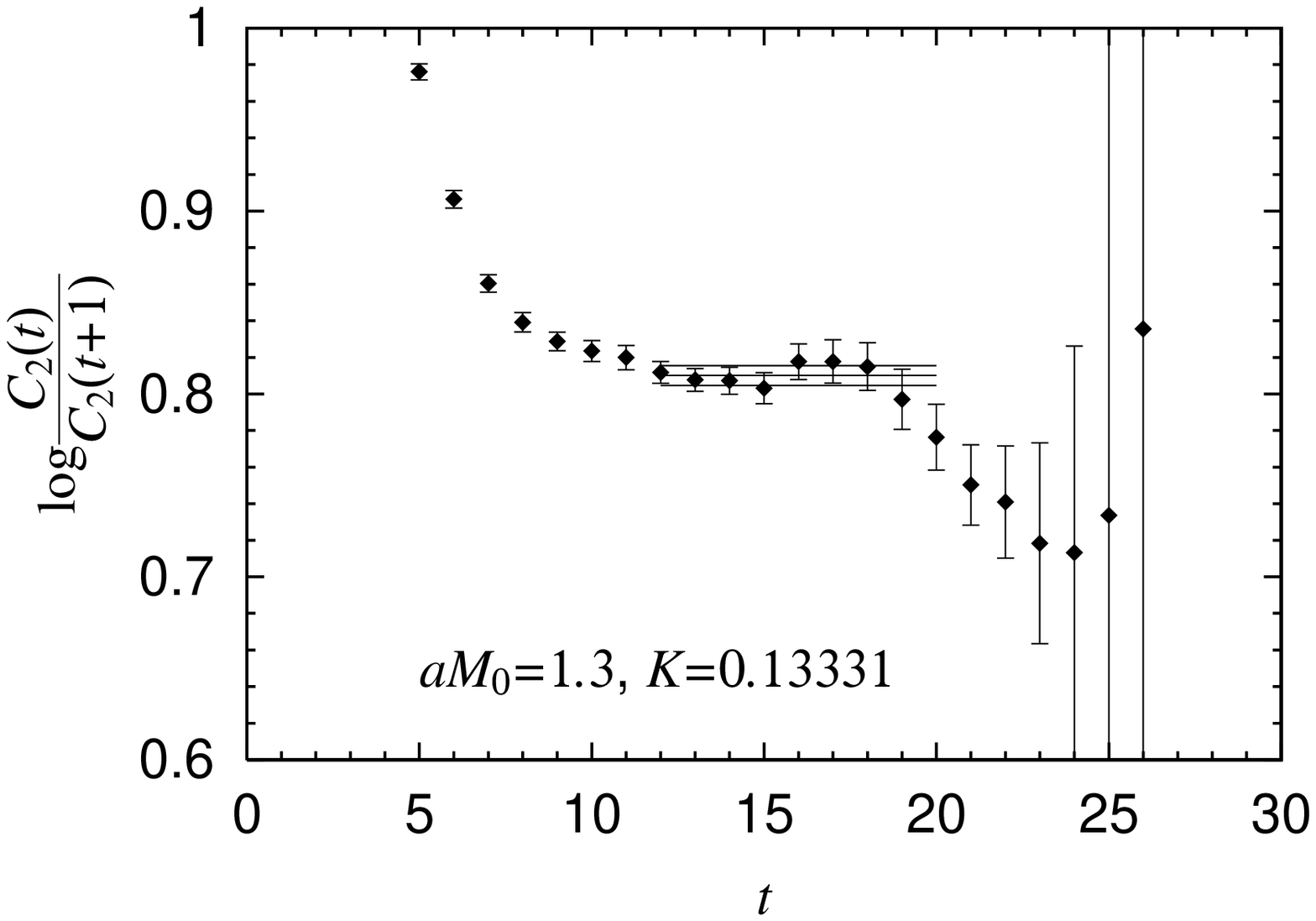}
  \includegraphics[width=\figwidth,angle=-90]{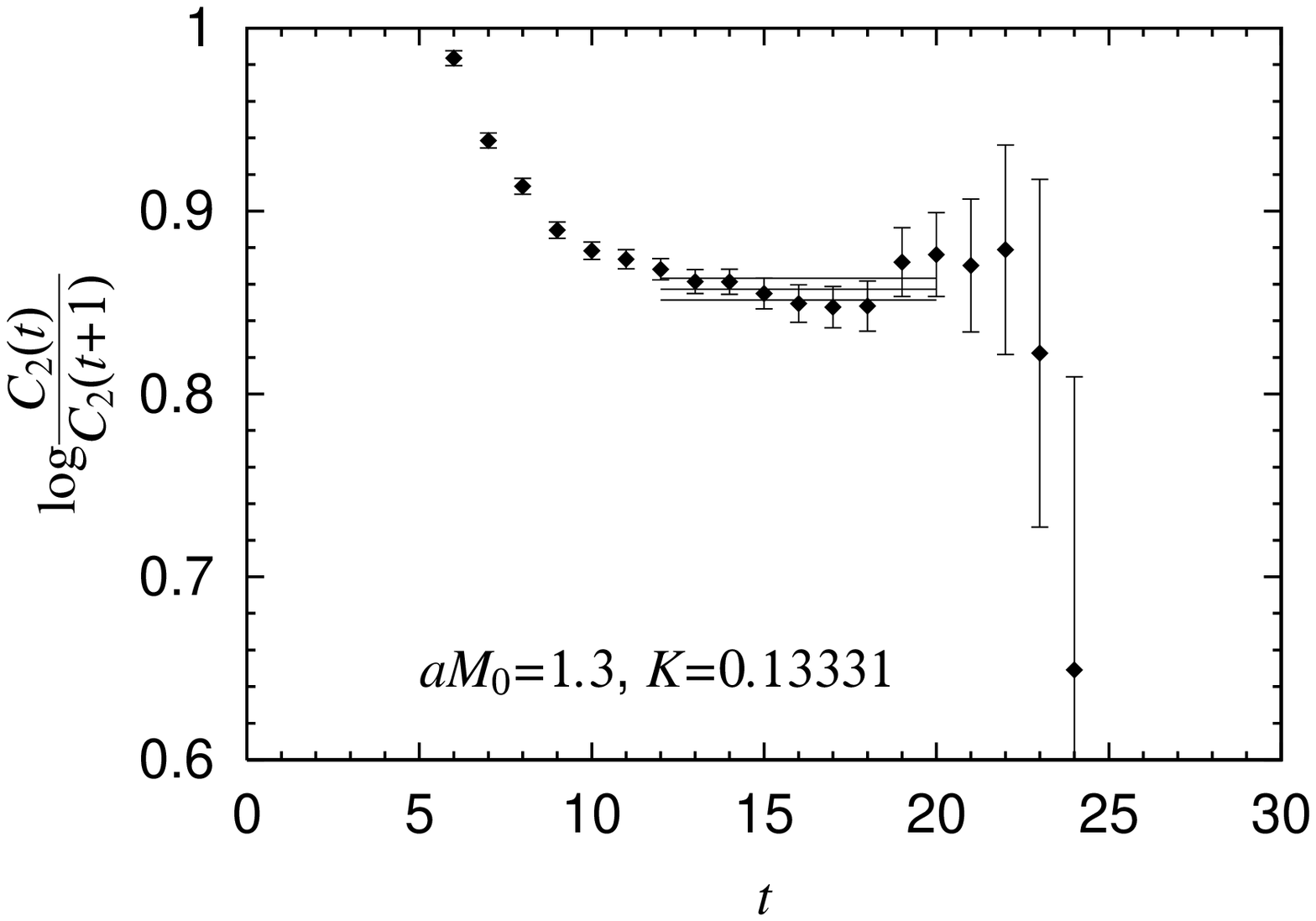}
  \includegraphics[width=\figwidth,angle=-90]{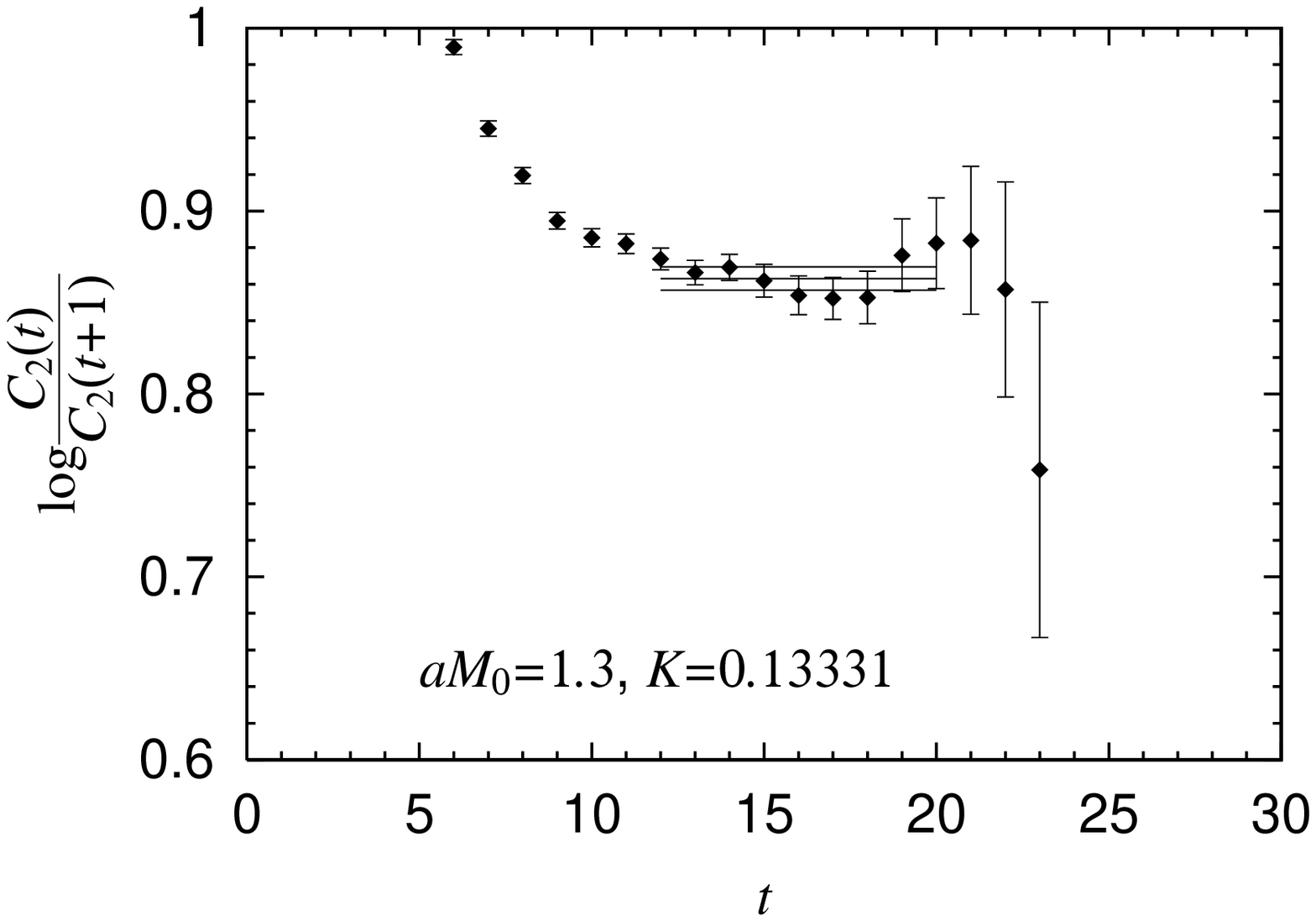}
  \caption{
    Effective mass plot for the $\Lambda_b$ (top panel) and
    $\Sigma_b$ (middle panel) and $\Sigma_b^*$ (bottom
    panel) baryons at $K$=0.13331 and $aM_0$=1.3. 
    Solid lines represent the fitting result with an error
    band of one standard deviation.
  }
  \label{fig:effective_mass_baryons}
\end{figure}

In Figures~\ref{fig:effective_mass_mesons} and
\ref{fig:effective_mass_baryons} 
we show the typical effective mass plots for relevant mesons
and baryons.
The plateau is convincing for the $B$ and $B^*$ mesons
(Figure~\ref{fig:effective_mass_mesons}) in the time region
starting around $t$ = 8, while it starts later in time for
baryons (Figure~\ref{fig:effective_mass_baryons}) and is
dominated by statistical fluctuations after $t$ = 20. 
We therefore fit the data in the time interval [10,20] for
mesons and in [12,20] for baryons.
The results for the binding energy are summarized in
Table~\ref{tab:binding_energy}.

\begin{figure}
  \includegraphics[width=\figwidth,angle=-90]{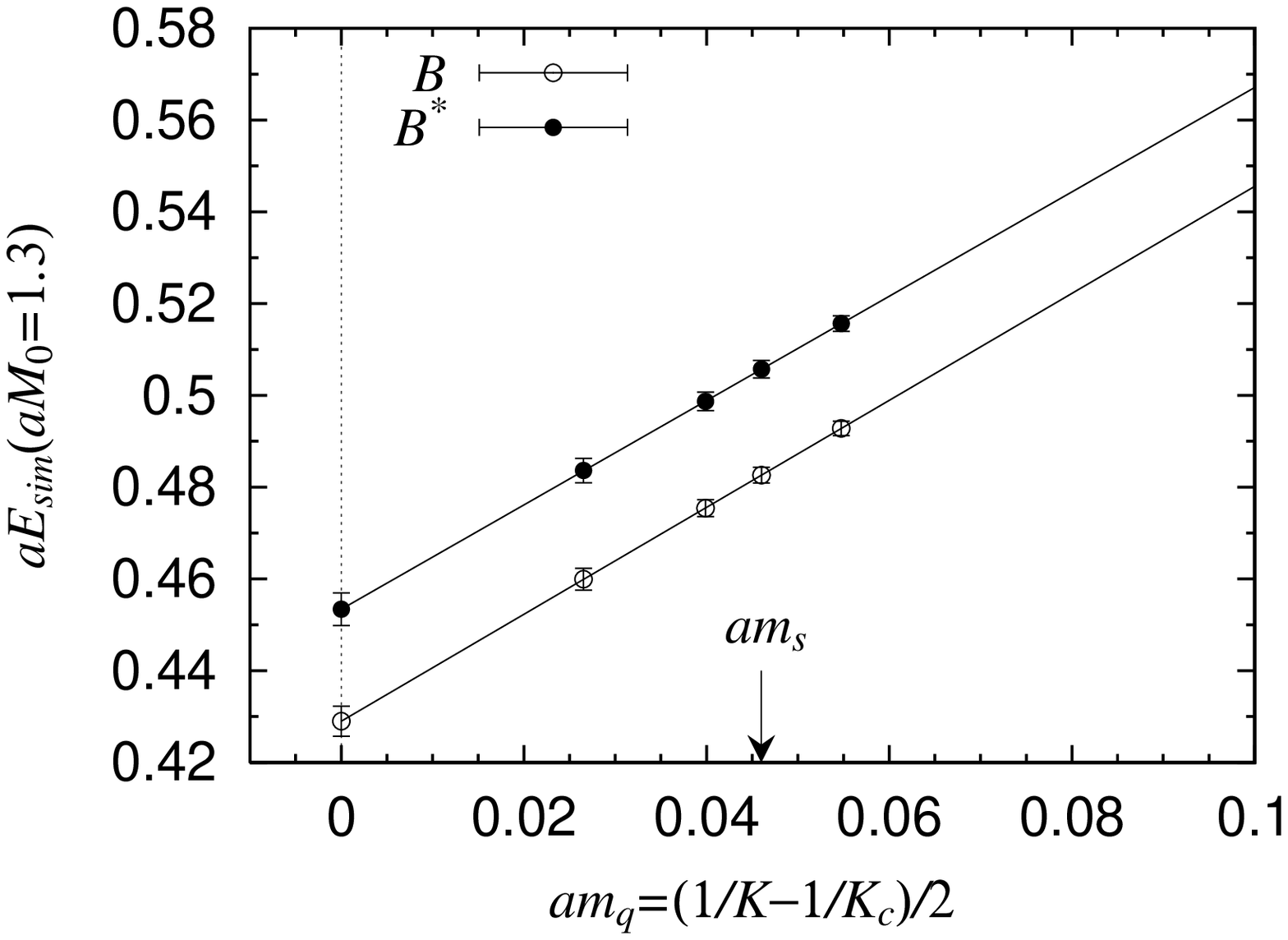}
  \includegraphics[width=\figwidth,angle=-90]{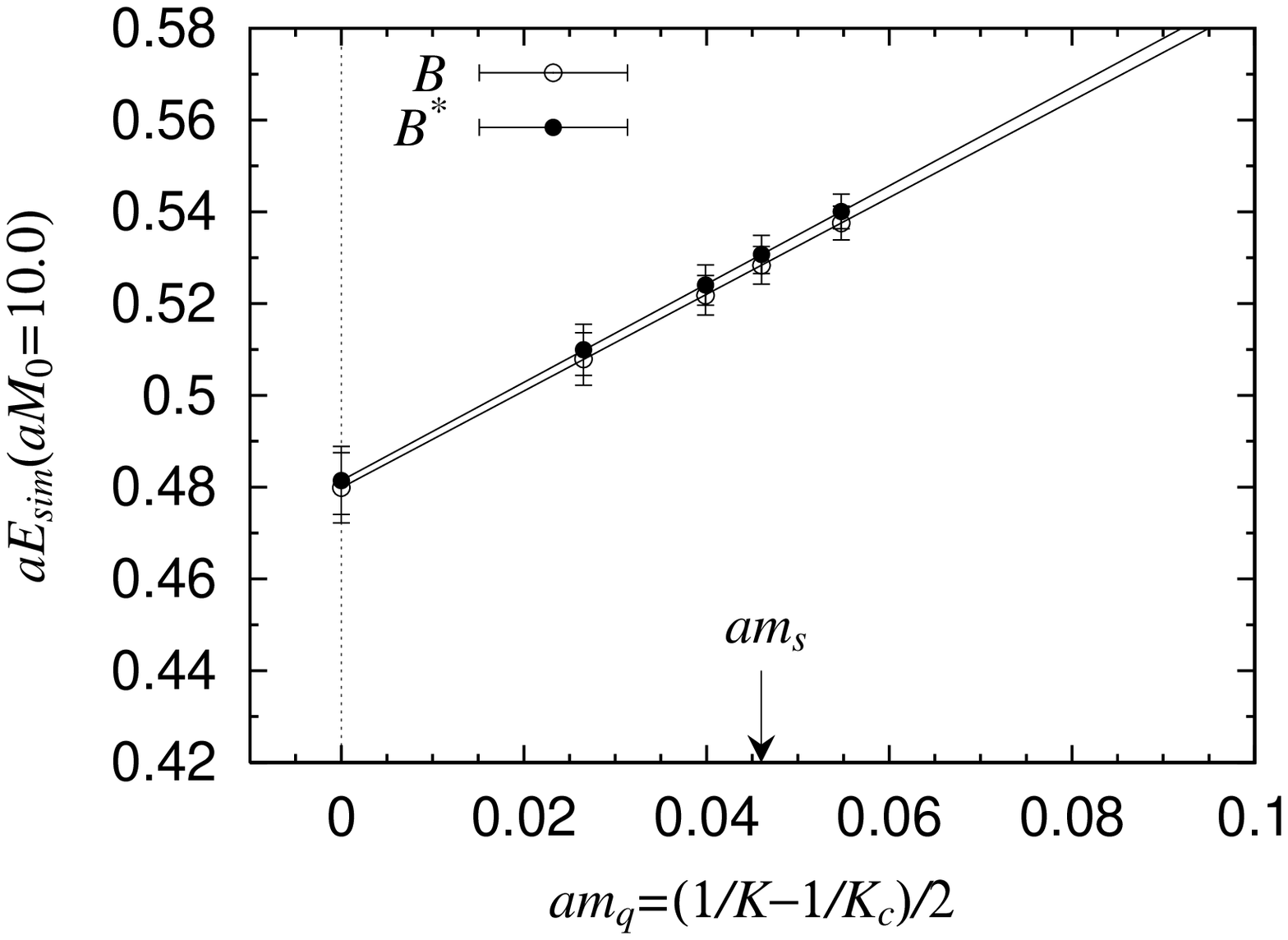}
  \caption{
    Binding energy of the $B$ and $B^*$ mesons
    as a function of light quark mass
    at $aM_0$=1.3 (top panel) and 10.0 (bottom panel).
  }
  \label{fig:lqdepm.all}
\end{figure}

\begin{figure}
  \includegraphics[width=\figwidth,angle=-90]{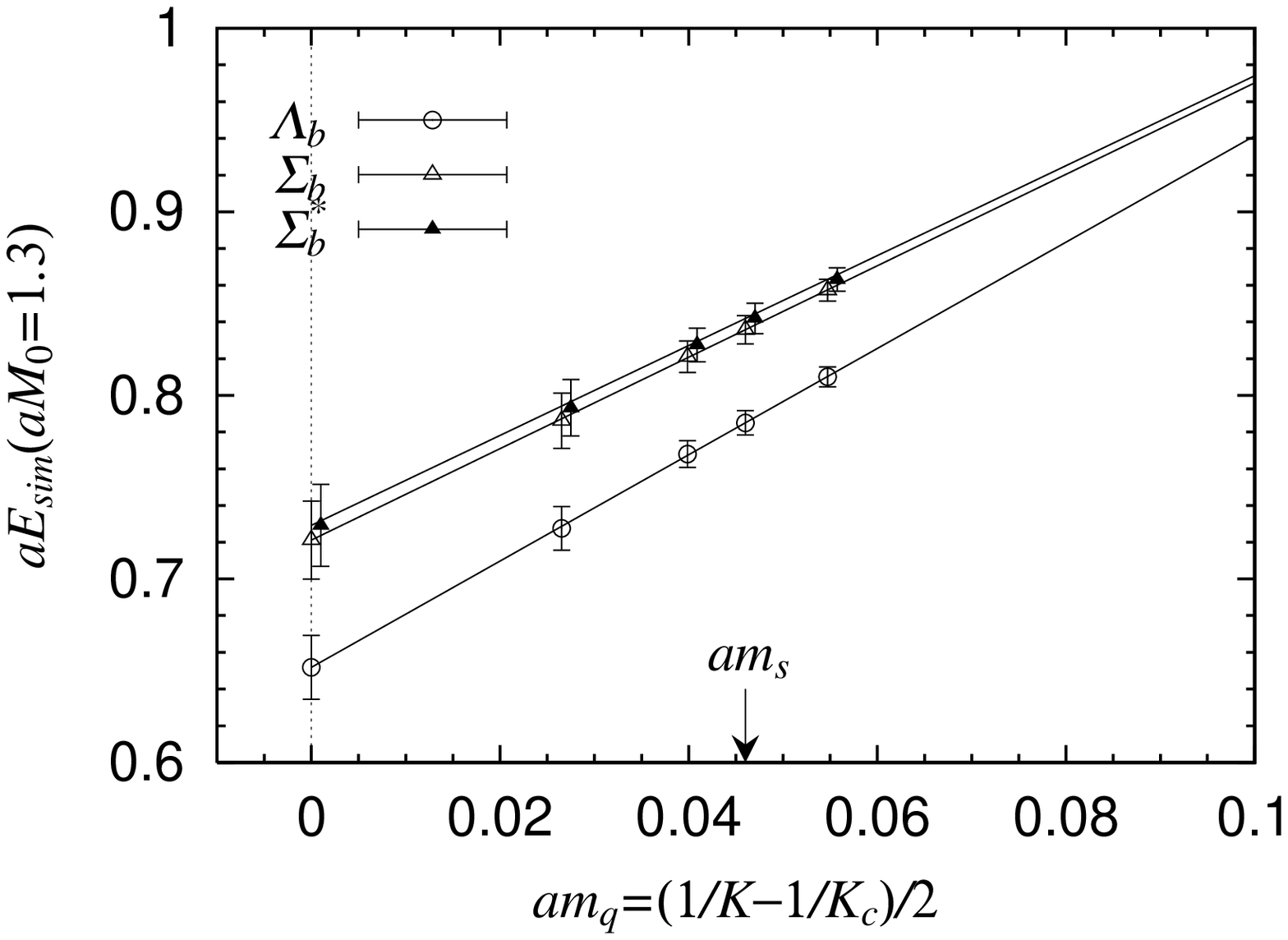}
  \includegraphics[width=\figwidth,angle=-90]{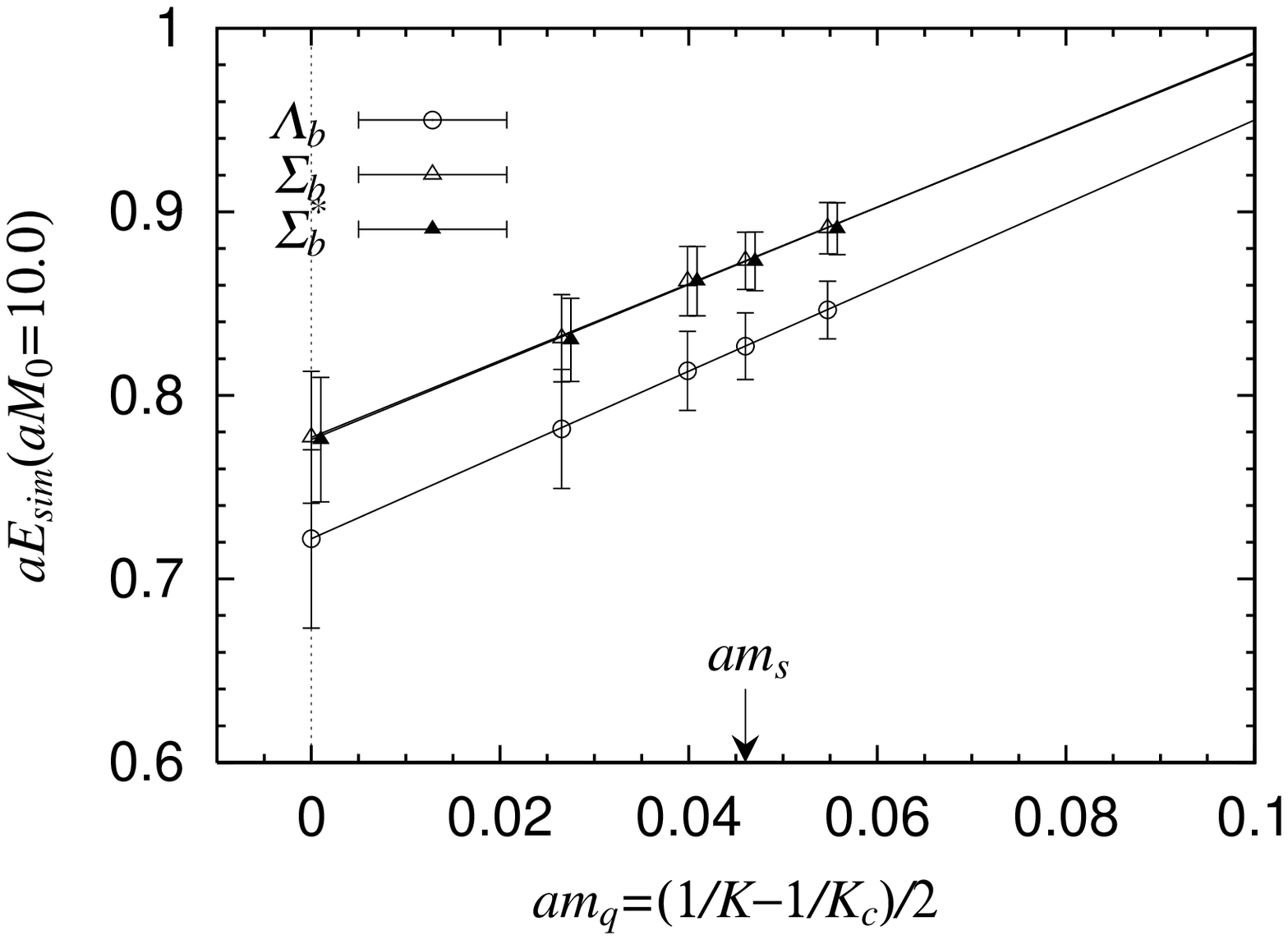}
  \caption{
    Binding energy for the $\Lambda_b$, $\Sigma_b$ and $\Sigma_b^*$ baryons
    as a function of light quark mass
    at $aM_0$=1.3 (top panel) and 10.0 (bottom panel).
  }
  \label{fig:lqdepb.all}
\end{figure}

Because the light quark mass dependence
of the binding energy is well described by
a linear function as shown in
Figure~\ref{fig:lqdepm.all} and \ref{fig:lqdepb.all},
we can extrapolate (interpolate) the binding energy to the
chiral limit (to the strange quark).
The binding energy at the chiral limit and the strange quark
is also presented in Table~\ref{tab:binding_energy}.

\subsection{Matrix elements}
\begin{figure}
  \includegraphics[width=\figwidth,angle=-90]{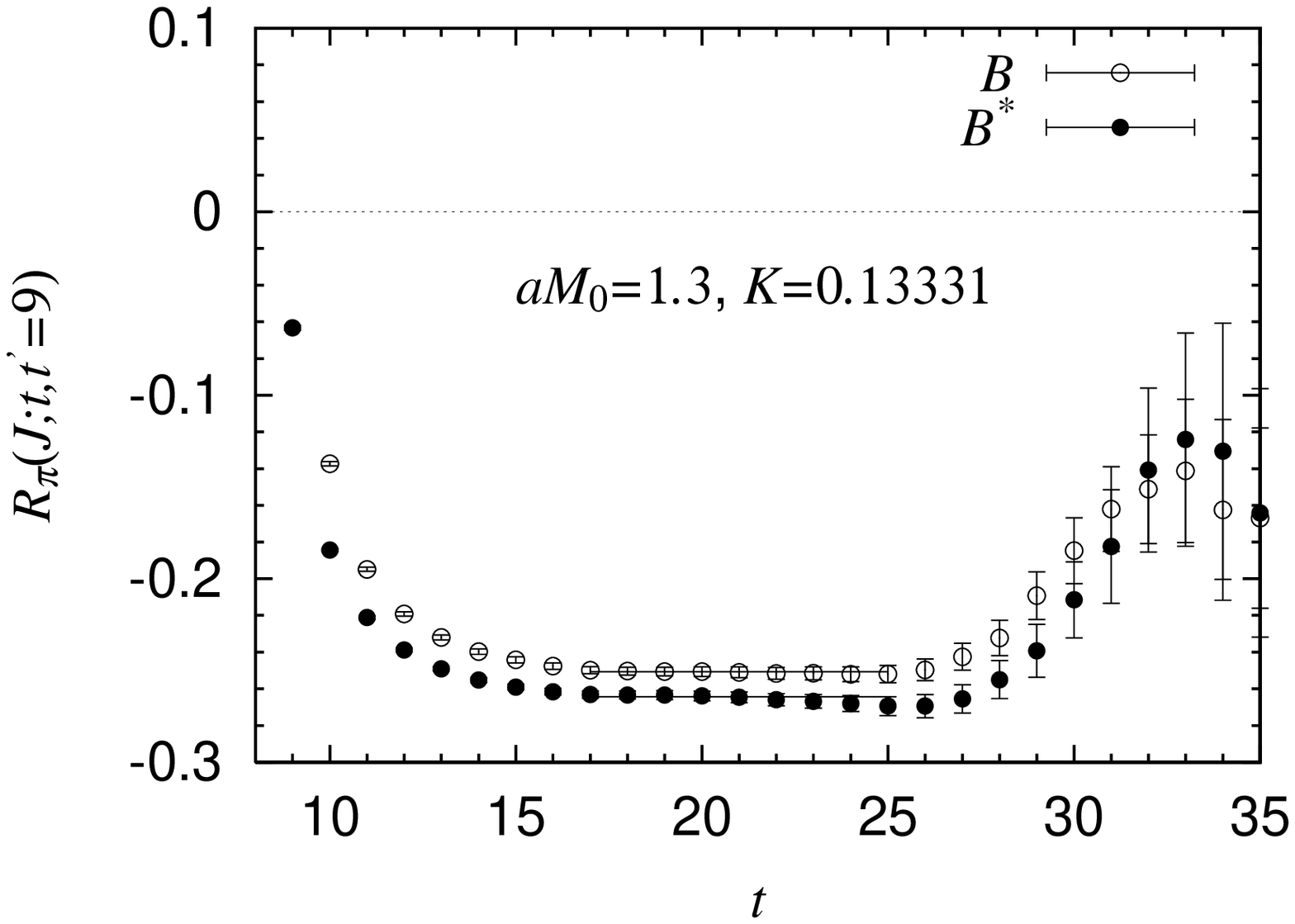}\\
  \includegraphics[width=\figwidth,angle=-90]{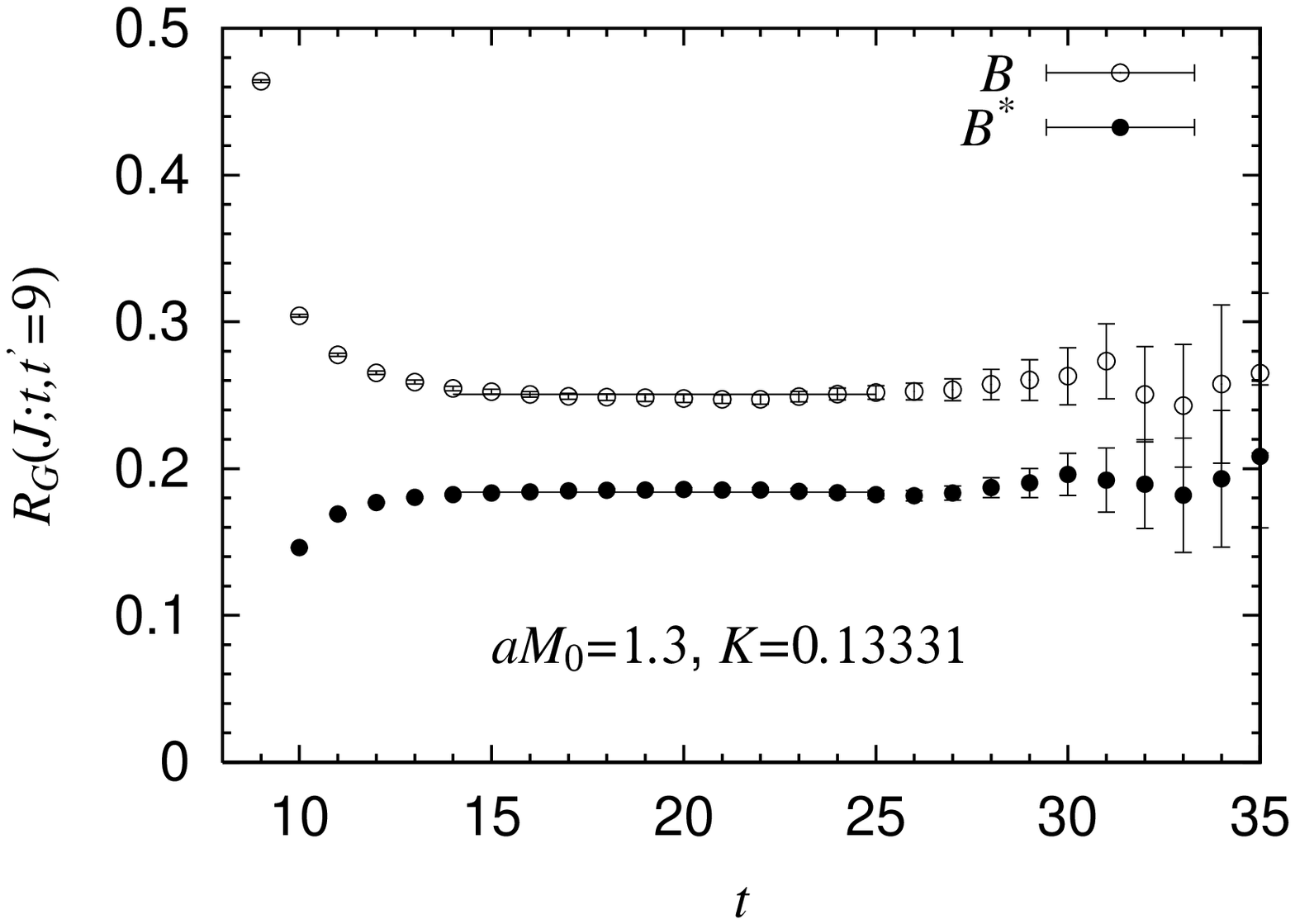}
  \caption{
    Ratio $R_i(J;t,t^{\prime}=9)$ for $\mu_{\pi}^2$ (top
    panel) and for $\mu_G^2$ (bottom panel)
    at $K$=0.13331 and $aM$=1.3.
    Open (filled) symbols are the data for the $B$ ($B^*$) meson.
    Solid lines represent a constant fit with an fit
    interval [17,25] for $\mu_\pi^2$ or [14,25] for
    $\mu_G^2$. 
  }
  \label{fig:ratiom}
\end{figure}

\begin{figure}
  \includegraphics[width=\figwidth,angle=-90]{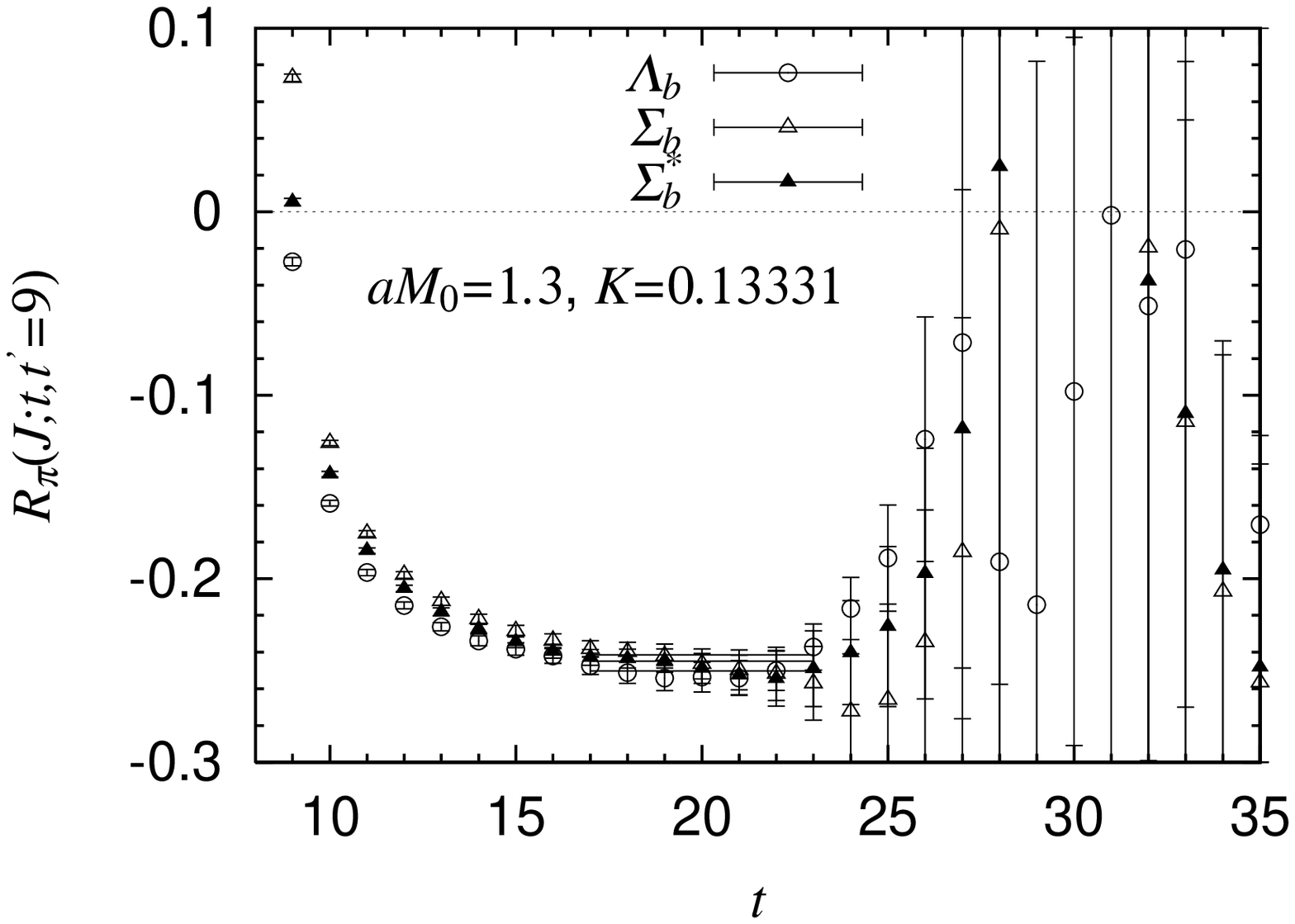}\\
  \includegraphics[width=\figwidth,angle=-90]{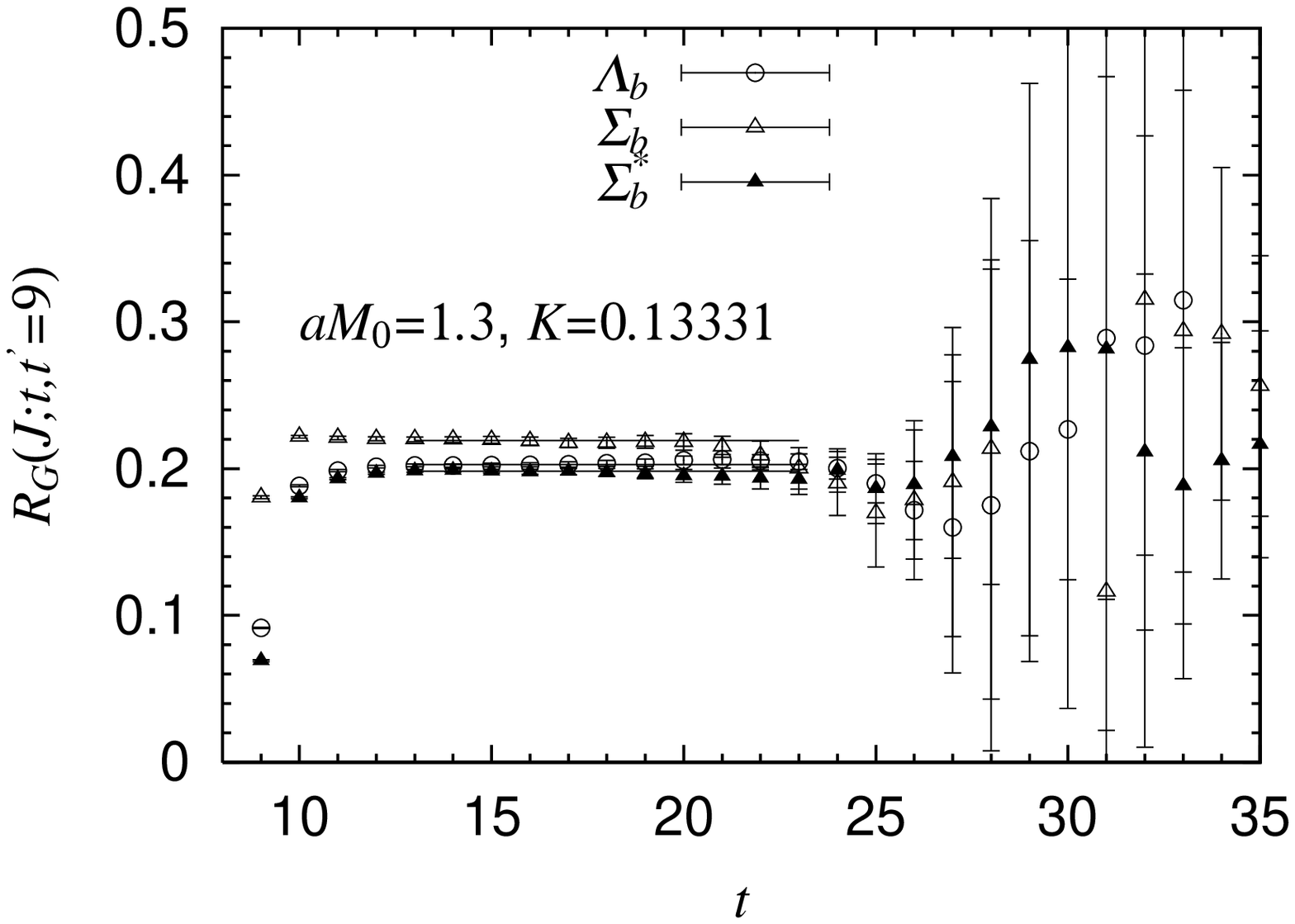}
  \caption{
    Ratio $R_i(J;t,t^{\prime}=9)$ for $\mu_{\pi}^2$ (top
    panel) and for $\mu_G^2$ (bottom panel)
    at $K$=0.13331 and $aM$=1.3.
    Open circles, open triangles and filled triangles are
    data for $\Lambda_b$, $\Sigma_b$ and $\Sigma_b^*$
    baryons, respectively.
    Solid lines represent a constant fit with an fit
    interval [17,23] for $\mu_\pi^2$ or [13,23] for
    $\mu_G^2$. 
  }
  \label{fig:ratiob}
\end{figure}

The ratio $R_i(J;t,t')$ defined in (\ref{eq:R}) is shown as
a function of $t$ in Figure~\ref{fig:ratiom} for $B$ and
$B^*$ mesons.
It shows a statistically cleanest data with heaviest light
($K$ = 0.13331) and lightest heavy ($aM$ = 1.3) quarks. 
The plateau is very convincing and appears earlier in $R_G$
than in $R_\pi$, and then we fit the data with a constant in
the time interval [17,25] for $R_\pi$ or [14,25] for $R_G$.
For other mass parameters the data are noisier, but we can
identify the plateau in the same time interval.
Similar plots for baryons ($\Lambda_b$, $\Sigma_b$, and
$\Sigma_b^*$) are shown in Figure~\ref{fig:ratiob}.
Since the statistical error dominates earlier in time we
truncate the fit range at $t$ = 23.
The results for the matrix elements $\mu_\pi^2$ and
$\mu_G^2$ are summarized in
Table~\ref{tab:matrix_elements_mu_pi} and
\ref{tab:matrix_elements_mu_G}, respectively.

\begin{figure}
  \includegraphics[width=\figwidth,angle=-90]{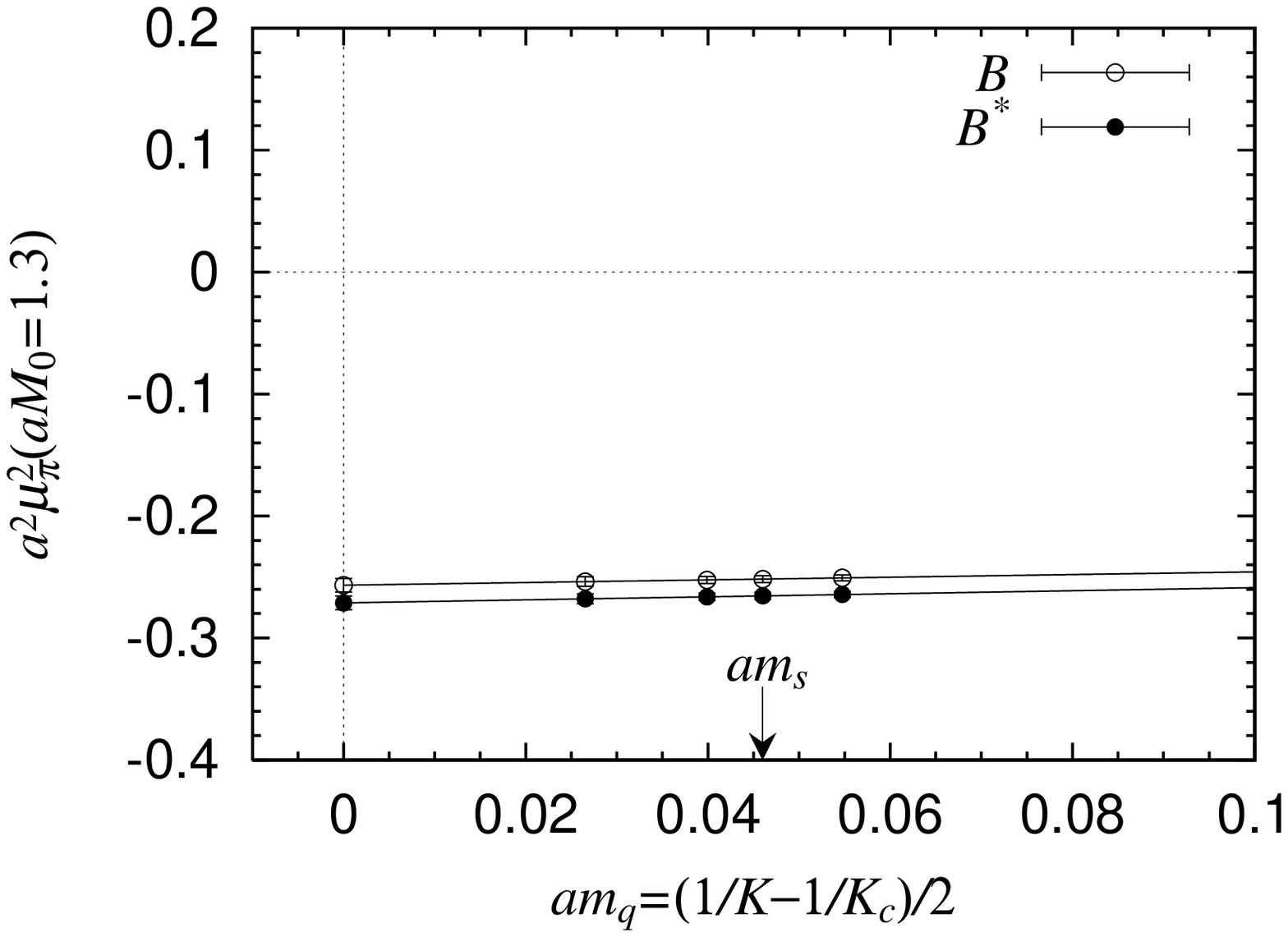}\\
  \includegraphics[width=\figwidth,angle=-90]{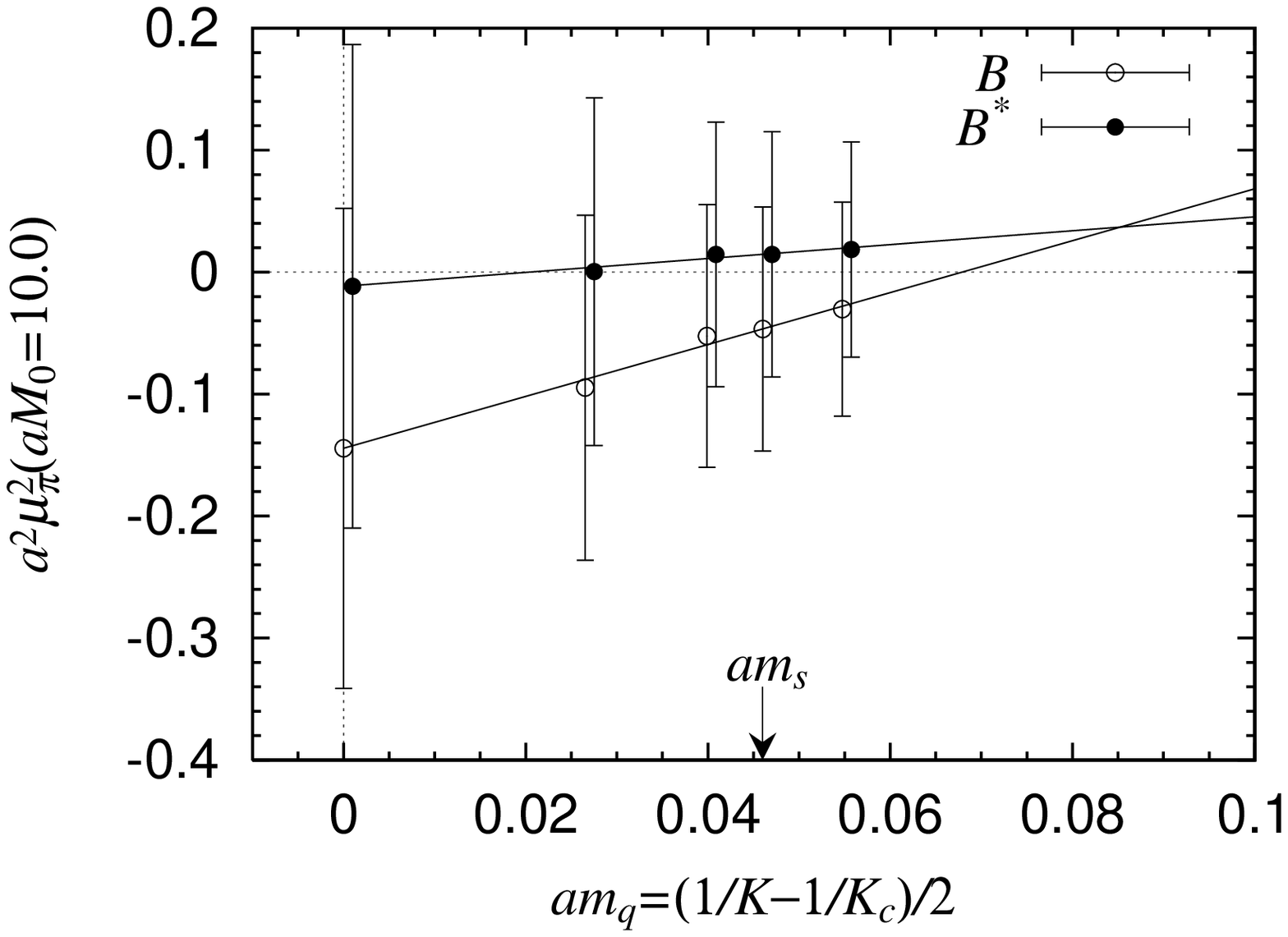}\\
  \caption{
    Matrix element $\mu_\pi^2$ for the $B$ and $B^*$ mesons
    as a function of light quark mass
    at $aM_0$=1.3 (top panel) and 10.0 (bottom panel).
  }
  \label{fig:lqdep3m.all.pi2}
\end{figure}

\begin{figure}
  \includegraphics[width=\figwidth,angle=-90]{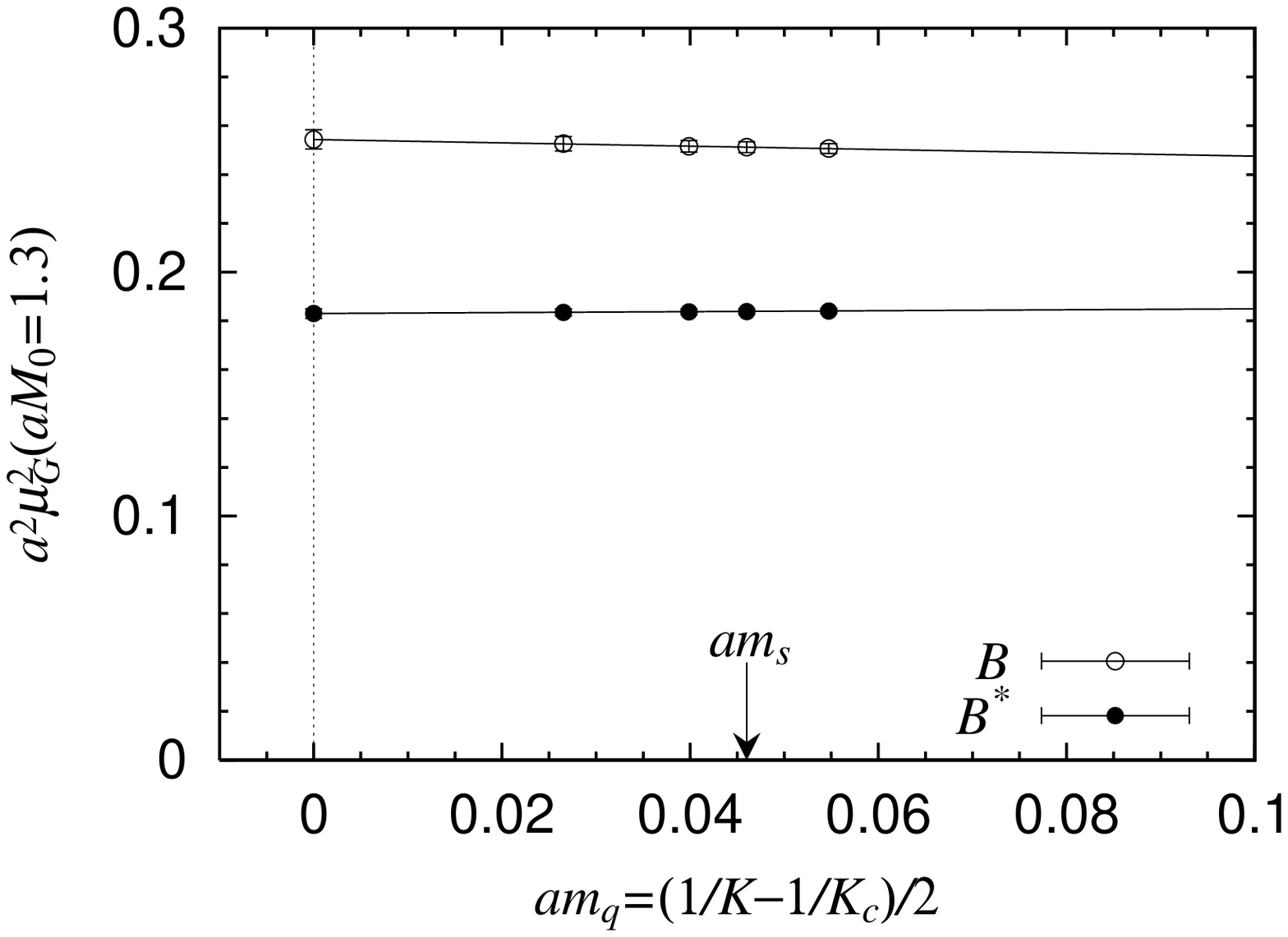}\\
  \includegraphics[width=\figwidth,angle=-90]{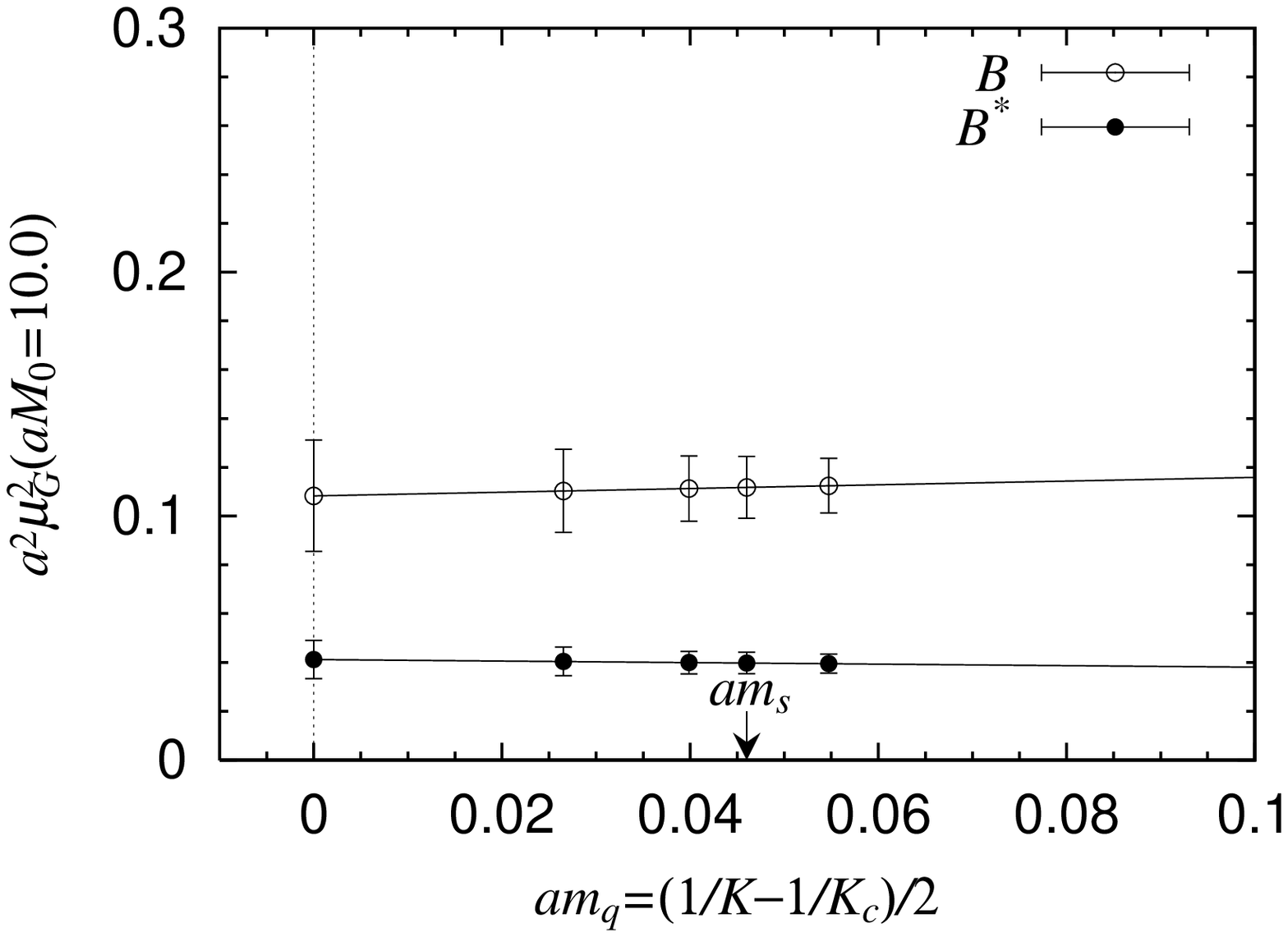}\\
  \caption{
    Matrix element $\mu_G^2$ for the $B$ and $B^*$ mesons
    as a function of light quark mass
    at $aM_0$=1.3 (top panel) and 10.0 (bottom panel).
  }
  \label{fig:lqdep3m.all.G2}
\end{figure}

\begin{figure}
  \includegraphics[width=\figwidth,angle=-90]{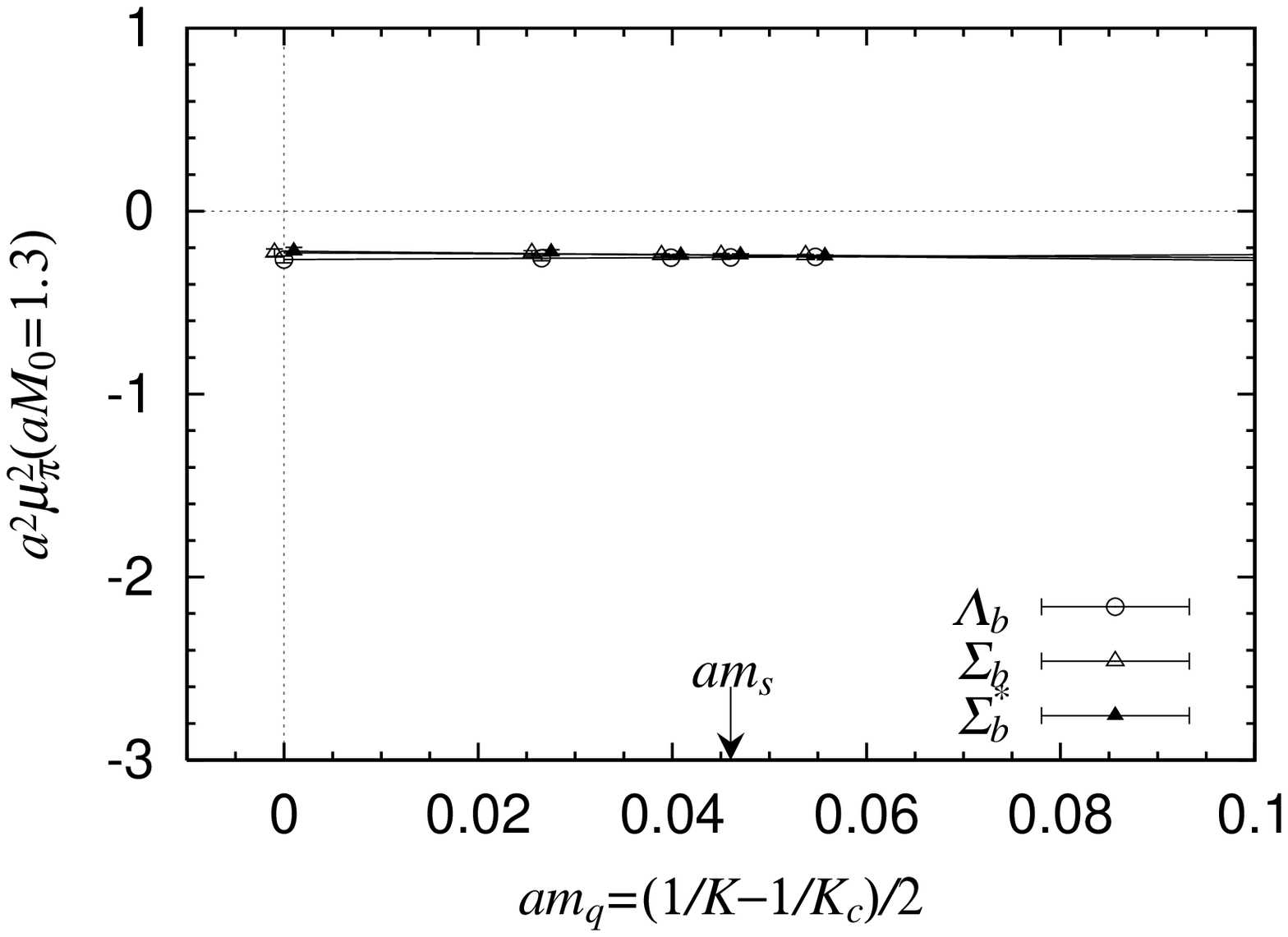}\\
  \includegraphics[width=\figwidth,angle=-90]{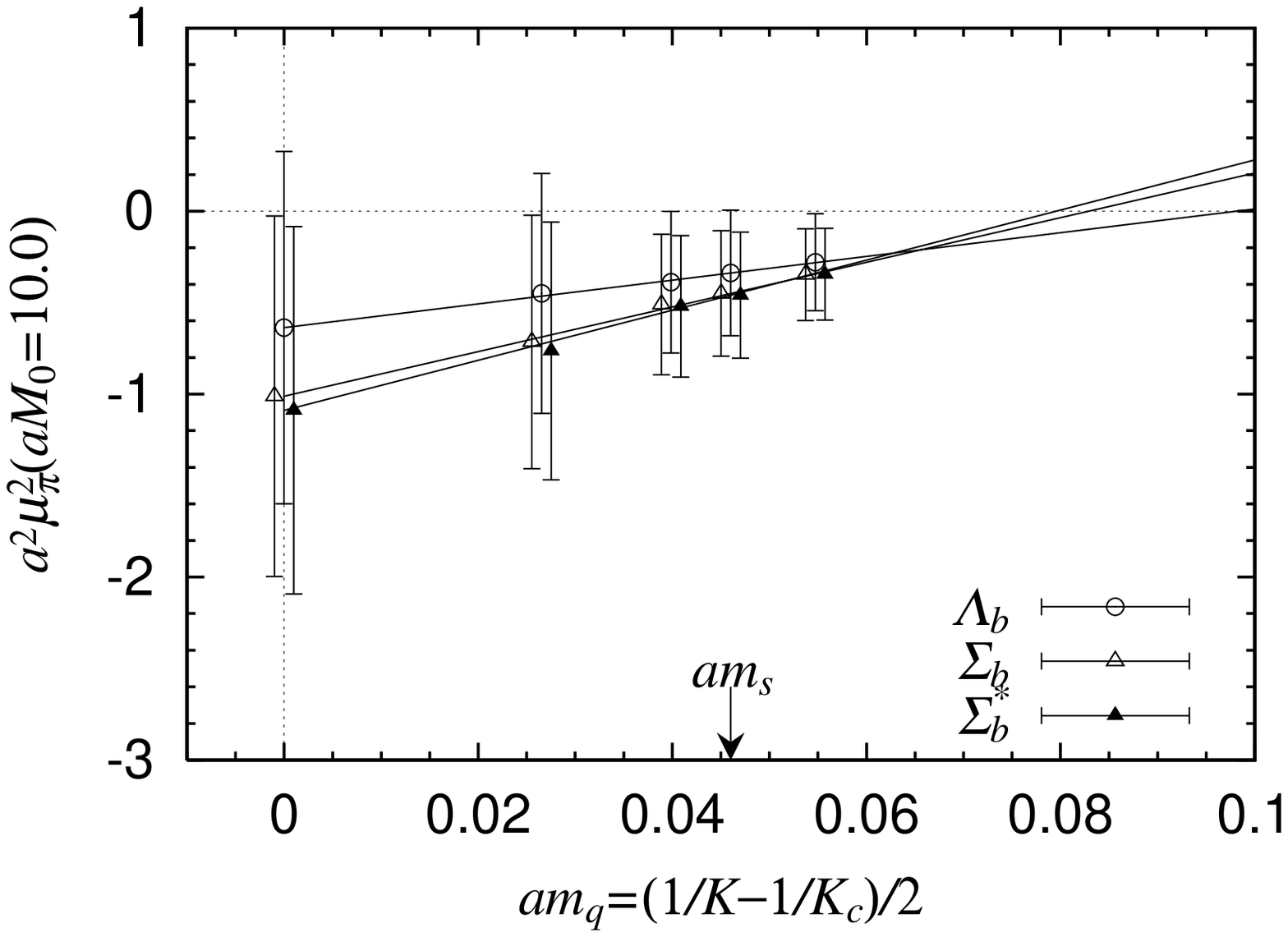}\\
  \caption{
    Matrix element $\mu_\pi^2$ for the $\Lambda_b$, $\Sigma_b$ and $\Sigma_b^*$ baryons
    as a function of light quark mass
    at $aM_0$=1.3 (top panel) and 10.0 (bottom panel).
  }
  \label{fig:lqdep3b.all.pi2}
\end{figure}

\begin{figure}
  \includegraphics[width=\figwidth,angle=-90]{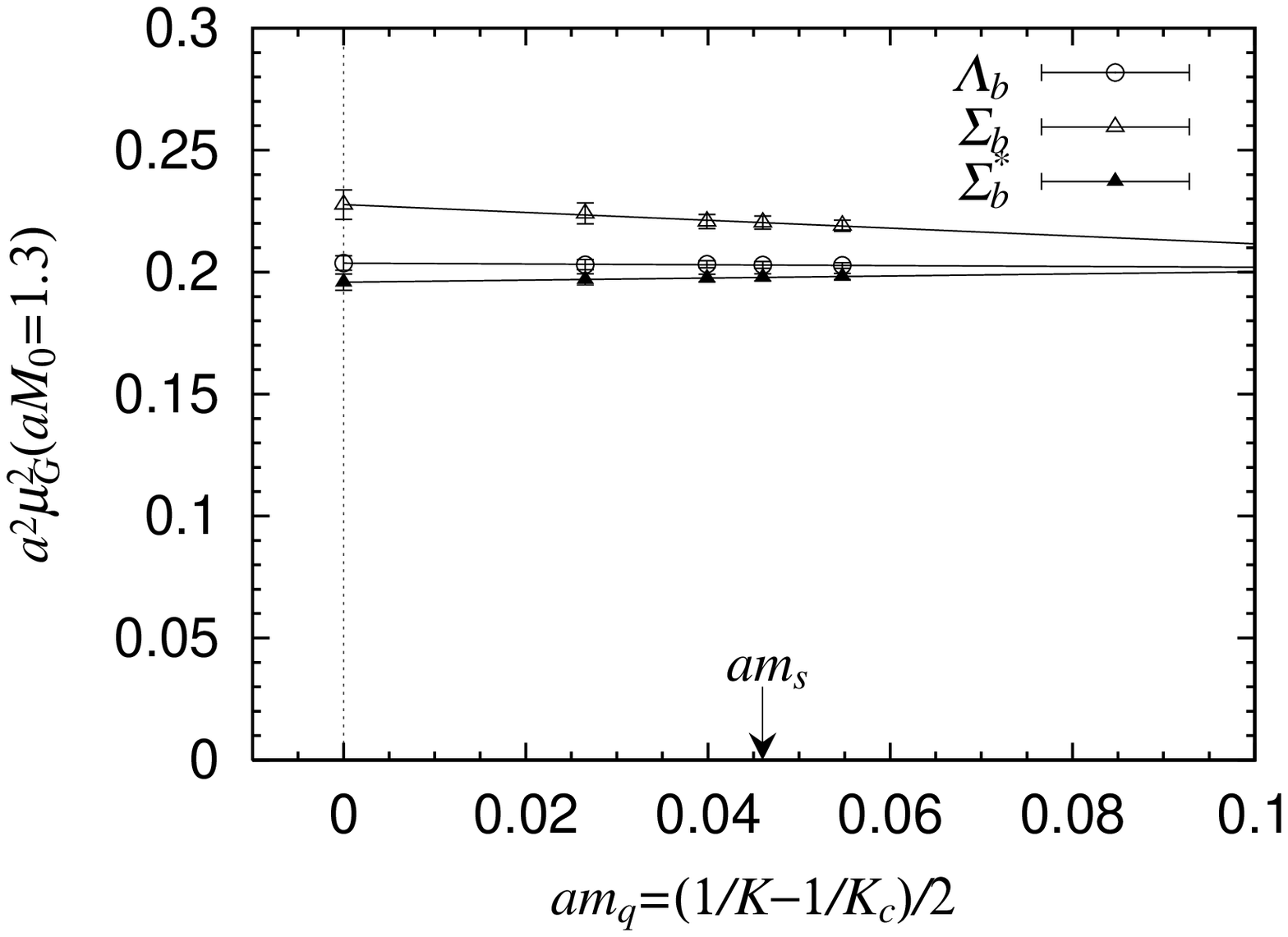}\\
  \includegraphics[width=\figwidth,angle=-90]{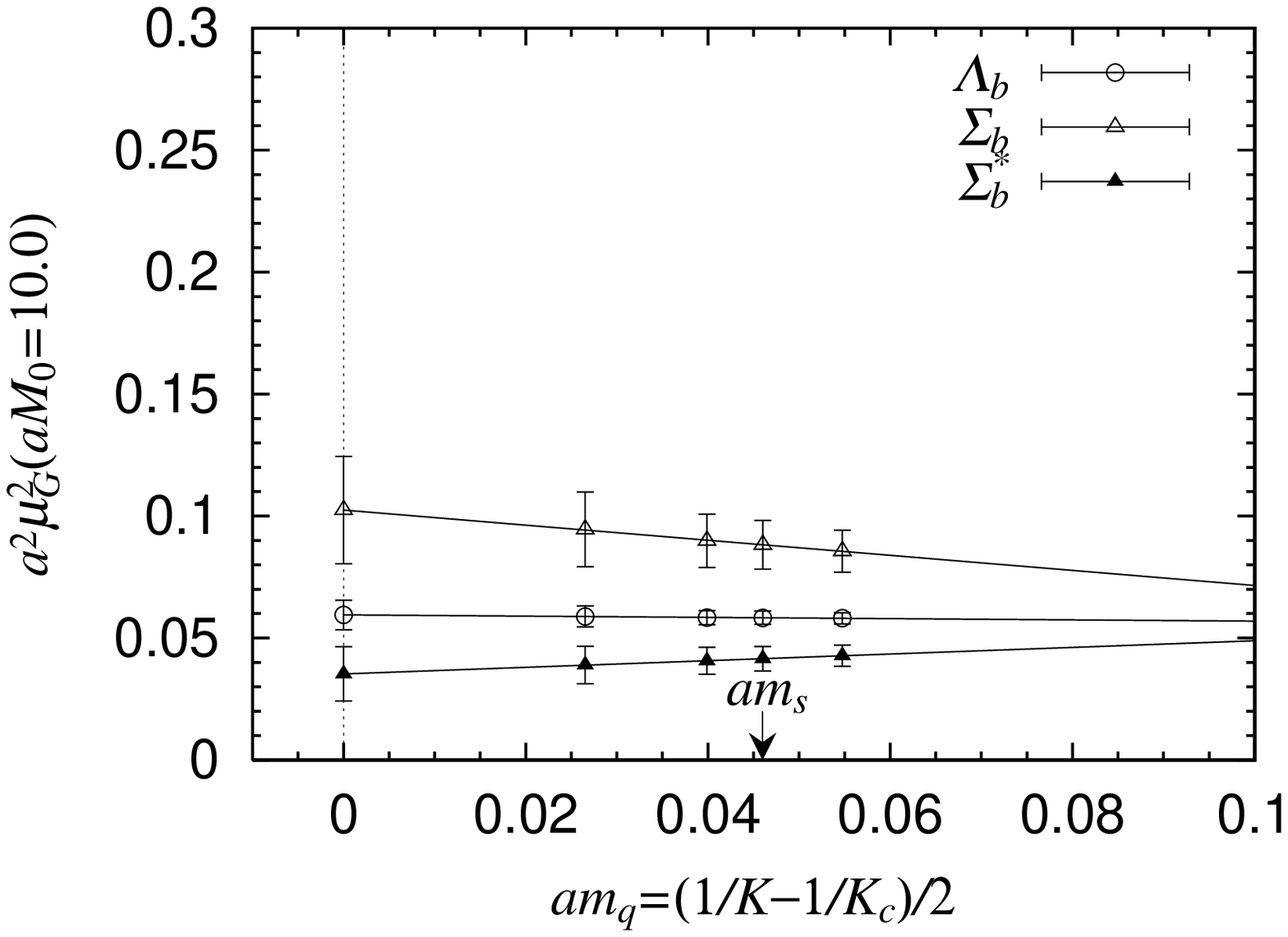}\\
  \caption{
    Matrix element $\mu_G^2$ for the $\Lambda_b$, $\Sigma_b$ and $\Sigma_b^*$ baryons
    as a function of light quark mass
    at $aM_0$=1.3 (top panel) and 10.0 (bottom panel).
  }
  \label{fig:lqdep3b.all.G2}
\end{figure}

From Figures~\ref{fig:lqdep3m.all.pi2}--\ref{fig:lqdep3b.all.G2}
we see that the light quark mass dependence of the matrix
elements is mild though the statistical error grows as light
quark mass decreases.
We therefore take a simple linear fit in the light quark
mass to obtain the results in the physical light quark
mass. 

\begin{figure}
  \includegraphics[width=\figwidth,angle=-90]{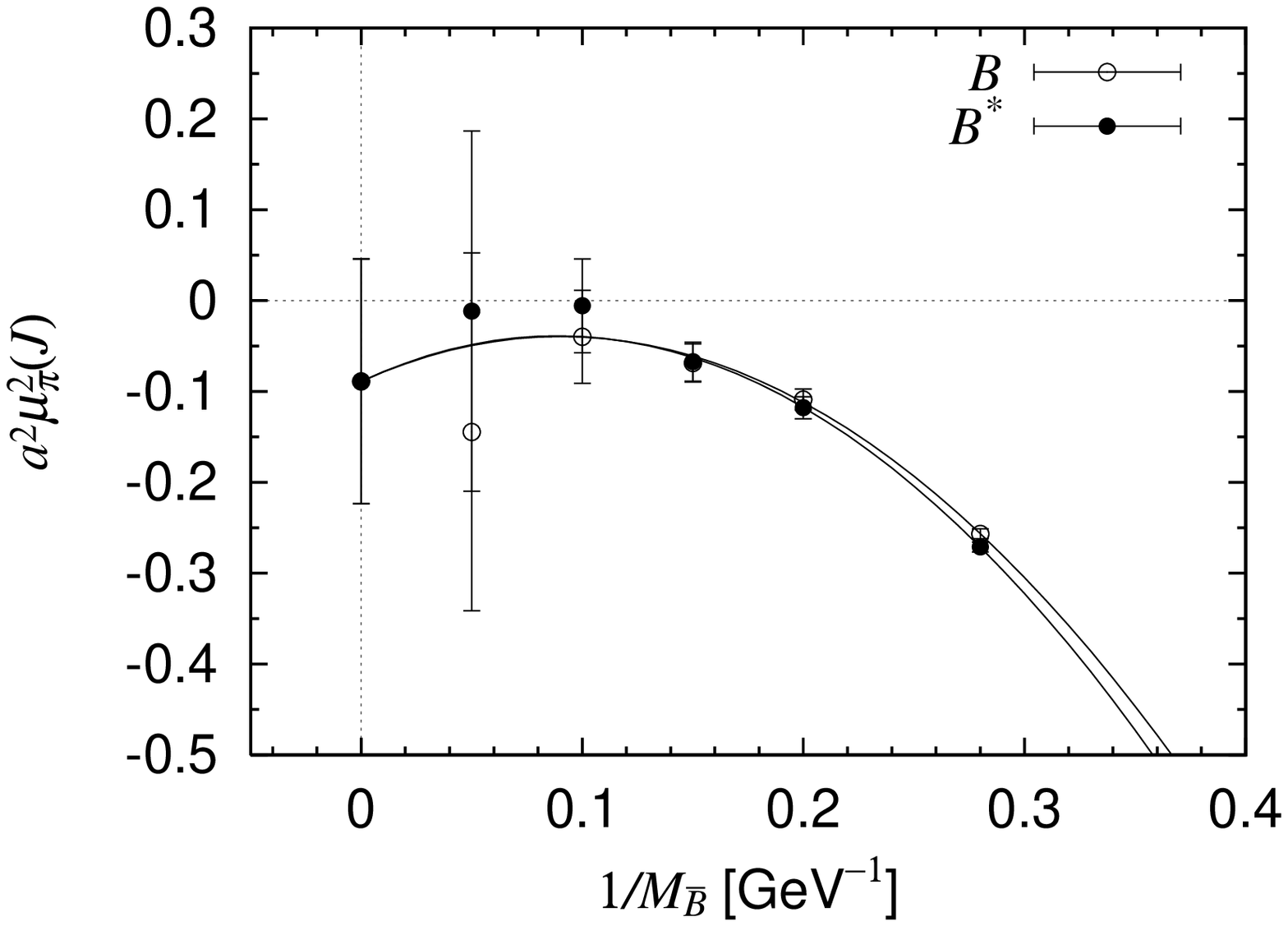}
  \caption{
    Matrix element $\mu_\pi^2$ for the $B$ and $B^*$ mesons
    as a function of $1/M_{\bar{B}}$.
    The value in the static limit is obtained from
    a fit in terms of a quadratic function in $1/M_{\bar{B}}$
    with the constarait (\ref{eq:symmetry_relation_pi}).
  }
  \label{fig:hqdep.mall.pi2.k1}
\end{figure}

\begin{figure}
  \includegraphics[width=\figwidth,angle=-90]{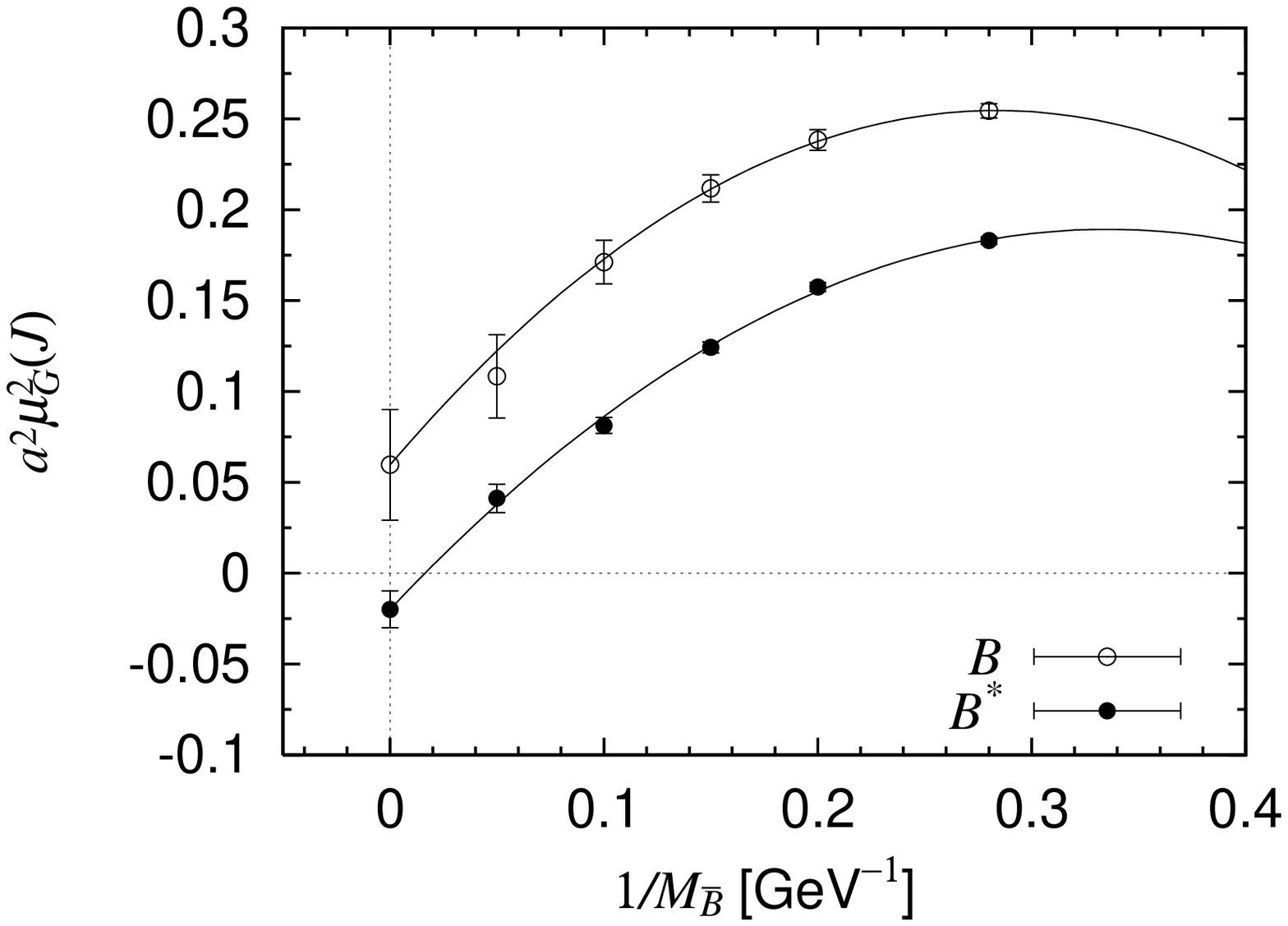}
  \caption{
    Matrix element $\mu_G^2$ for the $B$ and $B^*$ mesons
    as a function of $1/M_{\bar{B}}$.
    The values in the static limit are obtained from
    a fit in terms of a quadratic function in $1/M_{\bar{B}}$
    with the constarait (\ref{eq:symmetry_relation_G}).
  }
  \label{fig:hqdep.mall.G2.k1}
\end{figure}

\begin{figure}
  \includegraphics[width=\figwidth,angle=-90]{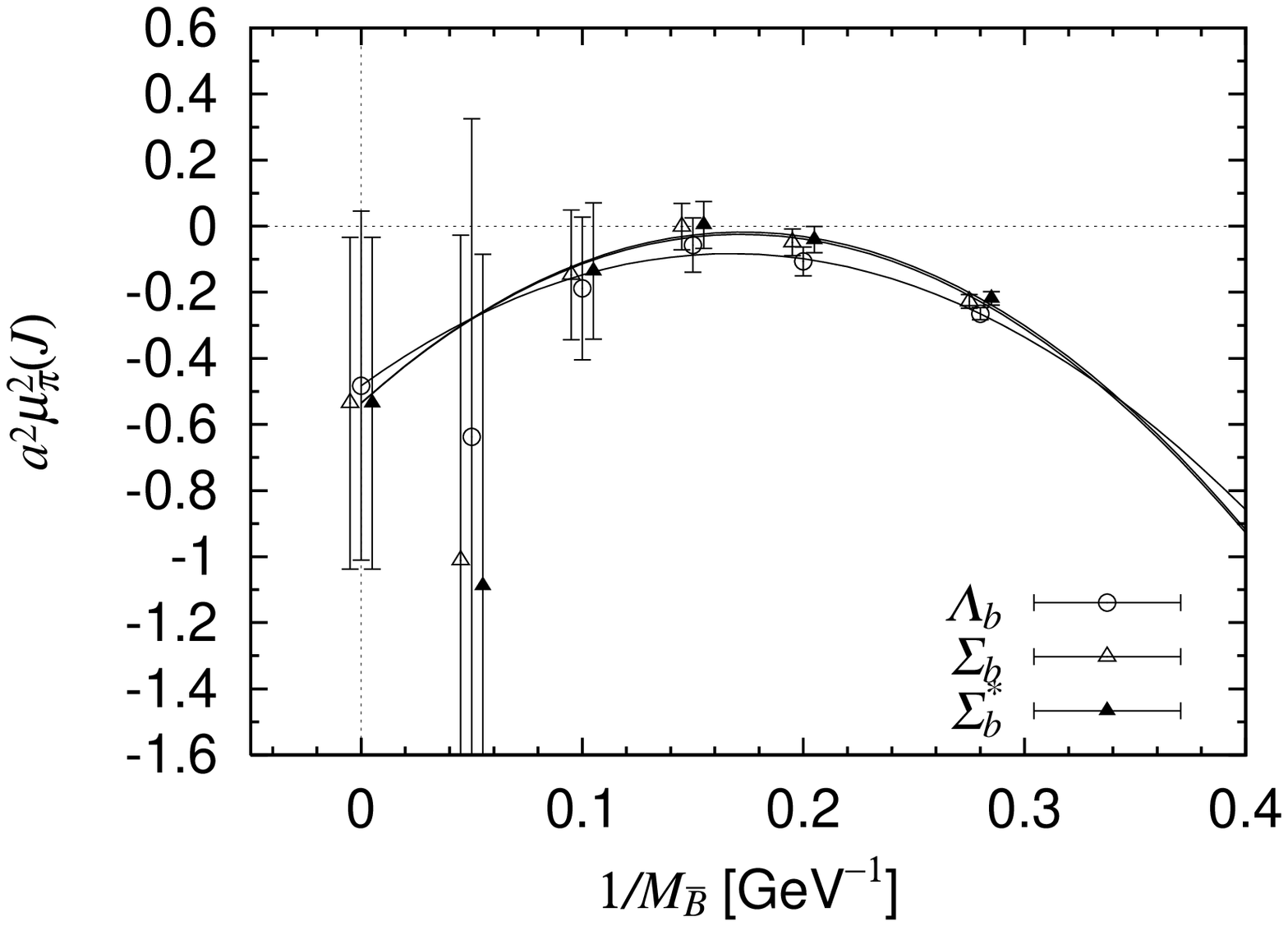}
  \caption{
    Matrix element $\mu_\pi^2$
    for the $\Lambda_b$, $\Sigma_b$ and $\Sigma_b^*$ baryons
    as a function of $1/M_{\bar{B}}$.
  }
  \label{fig:hqdep.ball.pi2.k1}
\end{figure}

\begin{figure}
  \includegraphics[width=\figwidth,angle=-90]{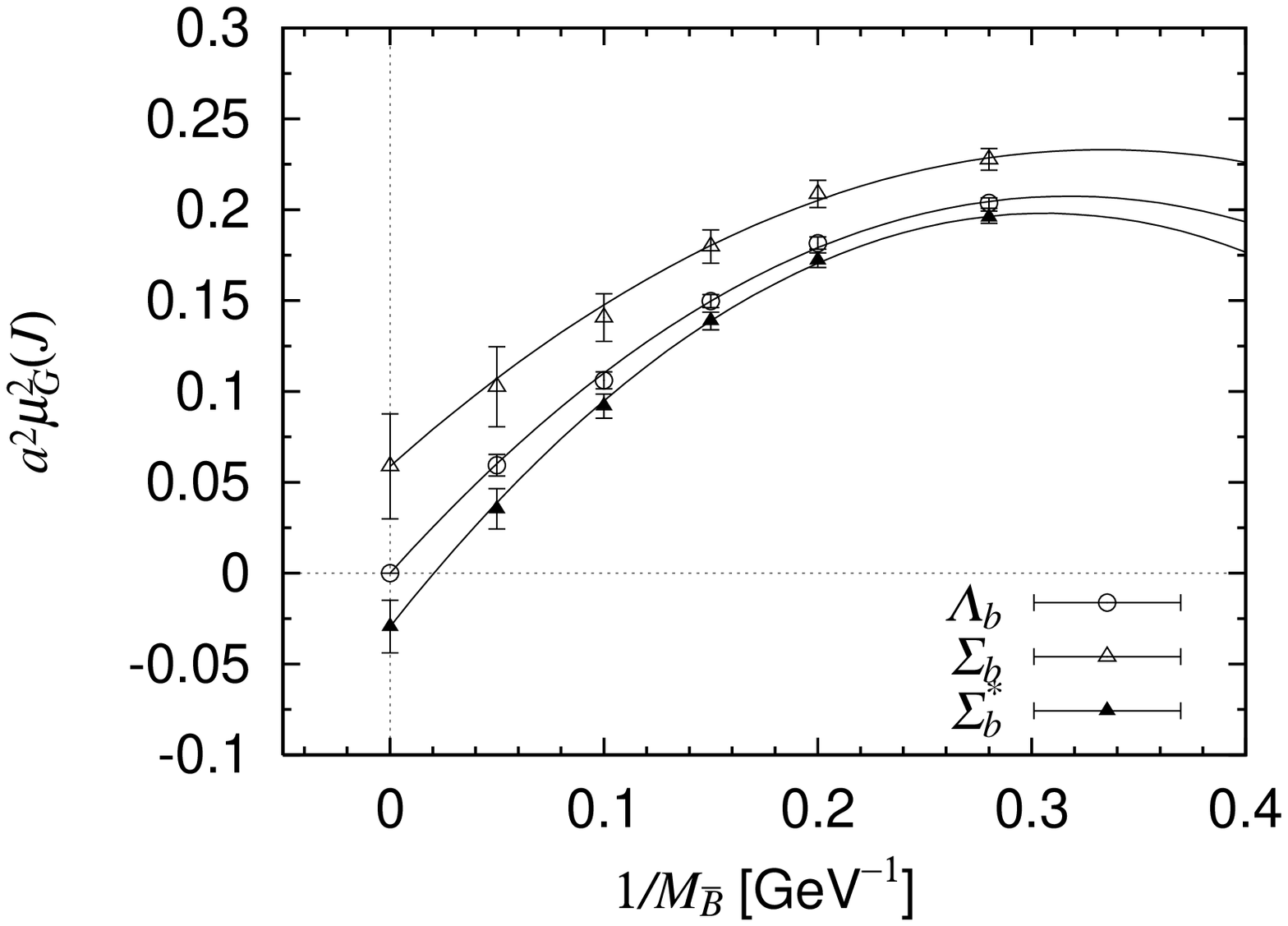}
  \caption{
    Matrix element $\mu_G^2$
    for the $\Lambda_b$, $\Sigma_b$ and $\Sigma_b^*$ baryons
    as a function of $1/M_{\bar{B}}$.
  }
  \label{fig:hqdep.ball.G2.k1}
\end{figure}

On the other hand, the heavy quark mass dependence of the
matrix elements is significant as shown in
Figures~\ref{fig:hqdep.mall.pi2.k1}--\ref{fig:hqdep.ball.G2.k1}.
In particular, the matrix elements $\mu_G^2(B)$ and
$\mu_G^2(B^*)$ in Figure~\ref{fig:hqdep.mall.G2.k1} 
are both positive at finite heavy quark masses, and hence
do not respect the symmetry relation 
$\frac{1}{3}\mu_G^2(B)=-\mu_G^2(B^*)$ given in 
(\ref{eq:symmetry_relation_G}).
This is due to the effects of operator mixing of
$\bar{Q}\vec{\sigma}\cdot\vec{B}Q$ with spin singlet
operators as mentioned in the previous section. 
The similar violation of the relation
(\ref{eq:symmetry_relation_G_for_baryon}) is found in
Figure~\ref{fig:hqdep.ball.G2.k1} for the matrix elements of 
$\Sigma_b^{(*)}$ baryons $\mu_G^2(\Sigma_b)$ and
$\mu_G^2(\Sigma_b^*)$.

In order to extract the prediction in the static limit, 
where the symmetry relations have to be satisfied, we
perform a fit of data in terms of a quadratic function in
$1/M_{\bar{B}}$ with a constraint known in the static limit.
For mesons the constraint is (\ref{eq:symmetry_relation_pi})
or (\ref{eq:symmetry_relation_G}), while for baryons we may
impose (\ref{eq:symmetry_relation_pi_for_baryon}) or
(\ref{eq:symmetry_relation_G_for_baryon}).
The fitting curves describe the data well while satisfying
the constraints as shown in 
Figures~\ref{fig:hqdep.mall.pi2.k1}--\ref{fig:hqdep.ball.G2.k1}.
The bare matrix elements extrapolated to the static limit
are listed in Tables~\ref{tab:matrix_elements_PI_static} and 
\ref{tab:matrix_elements_G_static}.
Since the chromomagnetic operator
$\bar{Q}\vec{\sigma}\cdot\vec{B}Q$ does not receive the
additive renormalization in the static limit, we may extract
the physical result from these numbers.
We obtain
\begin{eqnarray}
  \lambda_2(B) 
  \left(\equiv \frac{1}{3}\mu_G^2(B)=-\mu_G^2(B^*)\right)
  & = & 0.076(39) \;\mathrm{GeV}^2,
  \label{eq:lambda_2_method2}
  \\
  \mu_G^2(\Sigma_b) = -2\mu_G^2(\Sigma_b^*)
  & = & 0.23(11) \;\mathrm{GeV}^2,
  \label{eq:mu_G^2_Sigmab_method2}
\end{eqnarray}
after multiplying the renormalization factor $\tilde{Z}_G$ =
1.12 defined in (\ref{eq:Z_G_tilde}).

For the other matrix element $\mu_\pi^2$, the difference of
the matrix elements between different heavy hadrons has to
be considered in order to avoid the additive renormalization
due to the mixing with lower dimensional operators. 
It also helps to reduce the statistical error as it
correlates among different hadrons.
The results are
\begin{eqnarray}
  \mu_\pi^2(\Lambda_b)-\mu_\pi^2(B) 
  & = &
  - 1.3(1.8) \;\mathrm{GeV}^2,
  \label{eq:mu_pi_Lambda-B_method2}
  \\
  \mu_\pi^2(\Sigma_b)-\mu_\pi^2(\Lambda_b) 
  & = &
  - 0.2(2.5) \;\mathrm{GeV}^2,
\end{eqnarray}
which include the multiplicative renormalization factor
$\tilde{Z}_\pi$ = 0.975 as calculated in
(\ref{eq:Z_pi_tilde}).
The SU(3) breaking $\mu_\pi^2(B_s)-\mu_\pi^2(B_d)$ has also
a phenomenological importance, as it appears in the
evaluation of the lifetime ratio $\tau(B_s)/\tau(B_d)$.
Our result is
\begin{equation}
  \mu_\pi^2(B_s)-\mu_\pi^2(B_d) =
  0.09(26) \;\mathrm{GeV}^2.
\end{equation}

\begin{figure}
  \includegraphics[width=\figwidth,angle=-90]{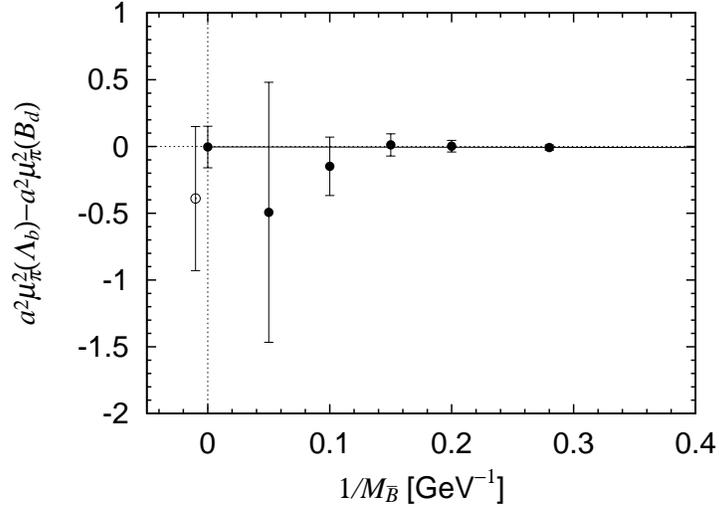}
  \caption{
    Difference of the matrix element $\mu_\pi^2(\Lambda_b)-\mu_\pi^2(B_d)$
    as a function of $1/M_{\bar{B}}$.
    Open circle denotes the result from method 2.
  }
  \label{fig:hqdep.deltamu.Lambda-B.k1}
\end{figure}

Another way to extract these physical quantities is to take
the differences before extrapolating the data to the static
limit.
As an example, we plot the difference of the matrix element
$\mu_\pi^2$ between $\Lambda_b$ baryon and $B$ meson in
Figure~\ref{fig:hqdep.deltamu.Lambda-B.k1}.
Since each matrix element $\mu_\pi^2(\Lambda_b)$ or
$\mu_\pi^2(B)$ has a quite similar heavy quark mass
dependence as seen in Figures~\ref{fig:hqdep.mall.pi2.k1}
and \ref{fig:hqdep.ball.pi2.k1}, the heavy quark mass
dependence almost cancels in the difference
(Figure~\ref{fig:hqdep.deltamu.Lambda-B.k1}).
We fit the data with a linear function in $1/M_{\bar{B}}$ and
obtain 
\begin{equation}
  \mu_\pi^2(\Lambda_b)-\mu_\pi^2(B) 
  =
  -0.01(52) \;\mathrm{GeV}^2,
  \label{eq:mu_pi_Lambda-B_method1}
\end{equation}
in the static limit.
This result is consistent with the previous analysis
(\ref{eq:mu_pi_Lambda-B_method2}) within one standard
deviation.
Since the heavy quark mass dependence is numerically better
controlled in this method, we quote
(\ref{eq:mu_pi_Lambda-B_method1}) as our final result, while
taking the other to estimate systematic uncertainty arising
from the heavy quark extrapolation.
The results for other differences of $\mu_\pi^2$ are
\begin{eqnarray}
  \mu_\pi^2(\Sigma_b)-\mu_\pi^2(\Lambda_b) 
  & = &
  0.28(68) \;\mathrm{GeV}^2,
  \\
  \mu_\pi^2(B_s)-\mu_\pi^2(B_d)
  & = &
  0.066(80) \;\mathrm{GeV}^2.
\end{eqnarray}

\begin{figure}
  \includegraphics[width=\figwidth,angle=-90]{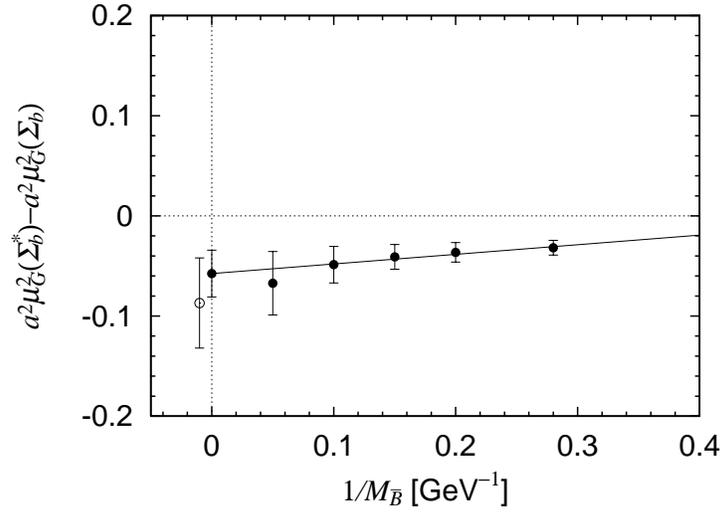}
  \caption{
    Difference of the matrix element $\mu_G^2(\Sigma_b^*)-\mu_G^2(\Sigma_b)$
    as a function of $1/M_{\bar{B}}$.
    Open circle denotes the result from method 2.
  }
  \label{fig:hqdep.SigmaStar-Sigma.G.kc}
\end{figure}

The same strategy --- differentiate then extrapolate ---
works even for $\mu_G^2$, since the additive renormalization
at finite heavy quark masses mostly cancel in the
differences like $\mu_G^2(B^*)-\mu_G^2(B)$ or
$\mu_G^2(\Sigma_b^*)-\mu_G^2(\Sigma_b)$.
Figure \ref{fig:hqdep.SigmaStar-Sigma.G.kc} shows the
difference $\mu_G^2(\Sigma_b^*)-\mu_G^2(\Sigma_b)$ as a
function of $1/M_{\bar{B}}$.
We find that the heavy quark mass dependence is much milder
than the individual matrix elements as shown in
Figure~\ref{fig:hqdep.ball.G2.k1}.
This cancellation of the $1/M_{\bar{B}}$ dependence is easily
understood from Figure~\ref{fig:hqdep.mall.G2.k1} or
\ref{fig:hqdep.ball.G2.k1}, because the mass dependence is
similar for all heavy hadrons.
The results are
\begin{eqnarray}
  \lambda_2(B) = -\frac{1}{4}(\mu_G^2(B^*)-\mu_G^2(B))
  & = & 
  0.094(19) \;\mathrm{GeV}^2,
  \label{eq:lambda_2_method1}
  \\
  \mu_G^2(\Sigma_b) =
  -\frac{2}{3}(\mu_G^2(\Sigma_b^*)-\mu_G^2(\Sigma_b))
  & = &
  0.147(60) \;\mathrm{GeV}^2,
  \label{eq:mu_G^2_Sigmab_method1}
\end{eqnarray}
which are consistent with the results obtained by taking 
the difference after the extrapolation,
(\ref{eq:lambda_2_method2}) and
(\ref{eq:mu_G^2_Sigmab_method2}) respectively.

All these results are summarized in Table~\ref{tab:summary}, 
where 
``method 1'' means our preferred method
(differentiate-then-extrapolate) while
``method 2'' denotes the other
(extrapolate-then-differentiate).

\subsection{Heavy quark expansion parameters from mass
  differences} 
\label{subsec:Heavy_quark_expansion_parameters_from_mass_differences}
The parameters $\overline{\Lambda}$, $\mu_{\pi}^2$ and
$\mu_G^2$ can also be indirectly obtained from hadron masses
using the mass formula (\ref{eq:mass_formula}).
We use the hadron masses measured on the lattice
to obtain the HQE parameters.

\begin{figure}
  \includegraphics[width=\figwidth,angle=-90]{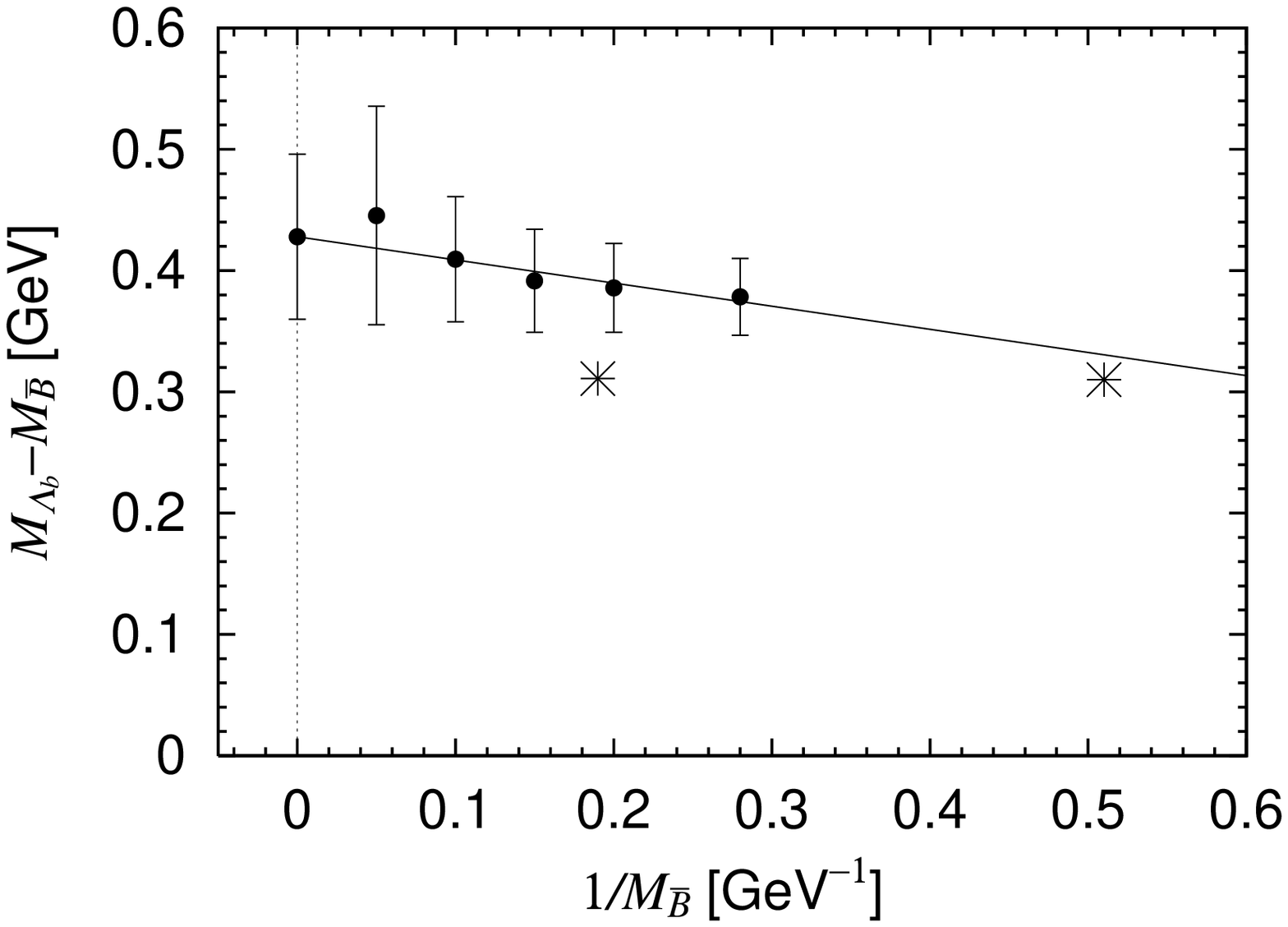}
  \caption{
    $M_{\Lambda_b}-M_{\bar{B}}$
    as a function of $1/M_{\bar{B}}$.
    The light quark mass is exptapolated to the chiral limit.
  }
  \label{fig:hqdep.L-BBar.k0}
\end{figure}

We plot the mass difference $M_{\Lambda_b}-M_{\bar{B}}$ 
as a function of the spin-averaged meson mass inverse
$1/M_{\bar{B}}$ in Figure~\ref{fig:hqdep.L-BBar.k0}.
The change from $m_Q$ to $M_{\bar{B}}$ is benign at this order
because the difference between $1/m_Q$ and $1/M_{\bar{B}}$
is of order $1/m_Q^2$ which we neglect in this analysis.
Fitting the data with a linear function of $1/M_{\bar{B}}$ 
we obtain
\begin{equation}
  \overline{\Lambda}(\Lambda_b)-\overline{\Lambda}(B) =
  428(68) \;\mathrm{MeV}, 
\end{equation}
from the intercept.
This result is in good agreement with a previous lattice
calculation by 
Ali~Khan \textit{et al.},
$\overline{\Lambda}(\Lambda_b)-\overline{\Lambda}(B)$
= 415(156)~MeV \cite{AliKhan:1999yb}.
Our result is slightly larger than the experimental value,
which is about 310~MeV for bottom and charmed hadrons as
plotted in Figure~\ref{fig:hqdep.L-BBar.k0} by bursts.
To draw a definite conclusion, however, we have to take
account of several systematic errors.
The finite volume effect is probably the most important one, 
because the physical extent of our lattice $\sim$ 2~fm may
not be large enough for baryons.

The slope of the mass difference $M_{\Lambda_b}-M_{\bar{B}}$
yields
\begin{equation}
  \mu_{\pi}^2(\Lambda_b)-\mu_{\pi}^2(B) = 
  -0.38(47) \;\mathrm{GeV}^2,
\end{equation}
which is compatible with the direct measurement of the
matrix elements (\ref{eq:mu_pi_Lambda-B_method1}) and also
with the phenomenological estimate
$-0.01(3)$~GeV$^2$ \cite{Neubert:1997we} obtained from a
combination 
$(M_{\Lambda_b}-M_{\bar{B}})-(M_{\Lambda_c}-M_{\bar{D}})$.

\begin{figure}
  \includegraphics[width=\figwidth,angle=-90]{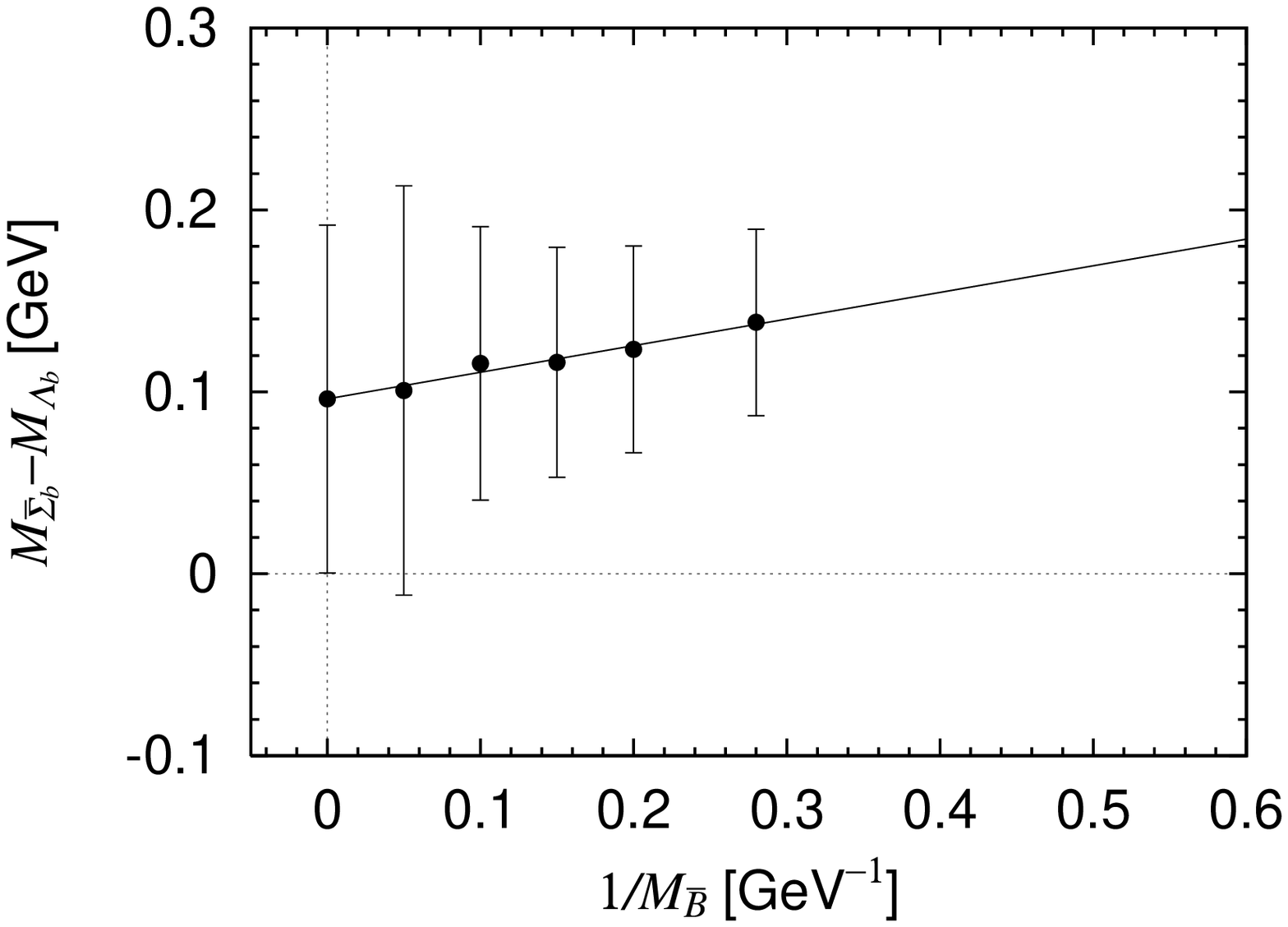}
  \caption{
    $M_{\bar{\Sigma}_b}-M_{\Lambda_b}$
    as a function of $1/M_{\bar{B}}$.
    The light quark mass is exptapolated to the chiral limit.
  }
  \label{fig:hqdep.SigmaBar-Lambda.k0}
\end{figure}

Similar analysis can be performed for
$M_{\bar{\Sigma}_b}-M_{\Lambda_b}$, which is plotted in
Figure~\ref{fig:hqdep.SigmaBar-Lambda.k0}. 
We obtain
\begin{eqnarray}
  \overline{\Lambda}(\Sigma_b)-\overline{\Lambda}(\Lambda_b)
  & = & 
  96(96) \;\mathrm{MeV},
  \\
  \mu_{\pi}^2(\Sigma_b)-\mu_{\pi}^2(\Lambda_b) 
  & = & 
  0.29(66) \;\mathrm{GeV}^2,
\end{eqnarray}
which are also consistent with the previous work
$\overline{\Lambda}(\Sigma_b)-\overline{\Lambda}(\Lambda_b)$
= 176(152)~MeV and
$\mu_{\pi}^2(\Sigma_b)-\mu_{\pi}^2(\Lambda_b)\sim$~0
\cite{AliKhan:1999yb}.

\begin{figure}
  \includegraphics[width=\figwidth,angle=-90]{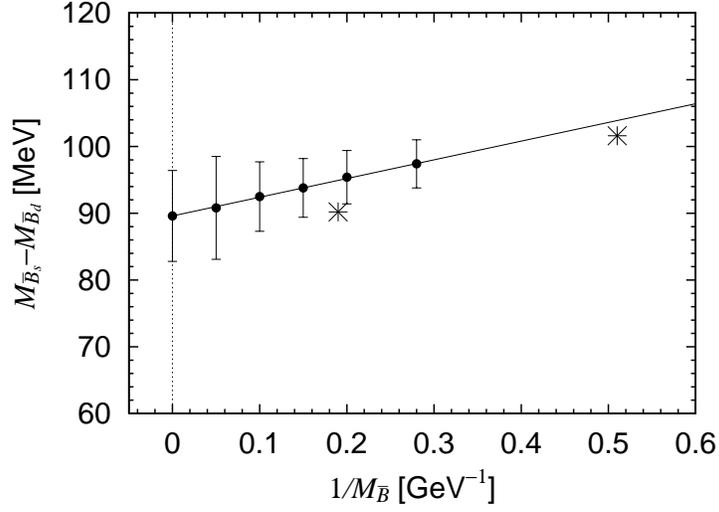}
  \caption{
    $M_{\bar{B}_s}-M_{\bar{B}_d}$
    as a function of $1/M_{\bar{B}}$.
  }
  \label{fig:hqdep.Bs-Bd.k0}
\end{figure}

The strange-nonstrange mass difference
$M_{\bar{B}_s}-M_{\bar{B}_d}$ is plotted in
Figure~\ref{fig:hqdep.Bs-Bd.k0}.
It is interesting to see that the data agree well with the
experimental value for $B_{(s)}$ and $D_{(s)}$ mesons
including the slope in $1/M_{\bar{B}}$.
A linear fit gives 
\begin{eqnarray}
  \overline{\Lambda}(B_s)-\overline{\Lambda}(B_d) 
  & = & 
  90(7) \;\mathrm{MeV},
  \\
  \mu_{\pi}^2(B_s)-\mu_{\pi}^2(B_d) 
  & = & 
  0.056(42) \;\mathrm{GeV}^2,
\end{eqnarray}
which may be compared with
$\overline{\Lambda}(B_s)-\overline{\Lambda}(B_d)$ = 81(31)~MeV and
$\mu_{\pi}^2(B_s)-\mu_{\pi}^2(B_d)$ = 0.10(28)~GeV$^2$
obtained in \cite{AliKhan:1999yb}.

\begin{figure}
  \includegraphics[width=\figwidth,angle=-90]{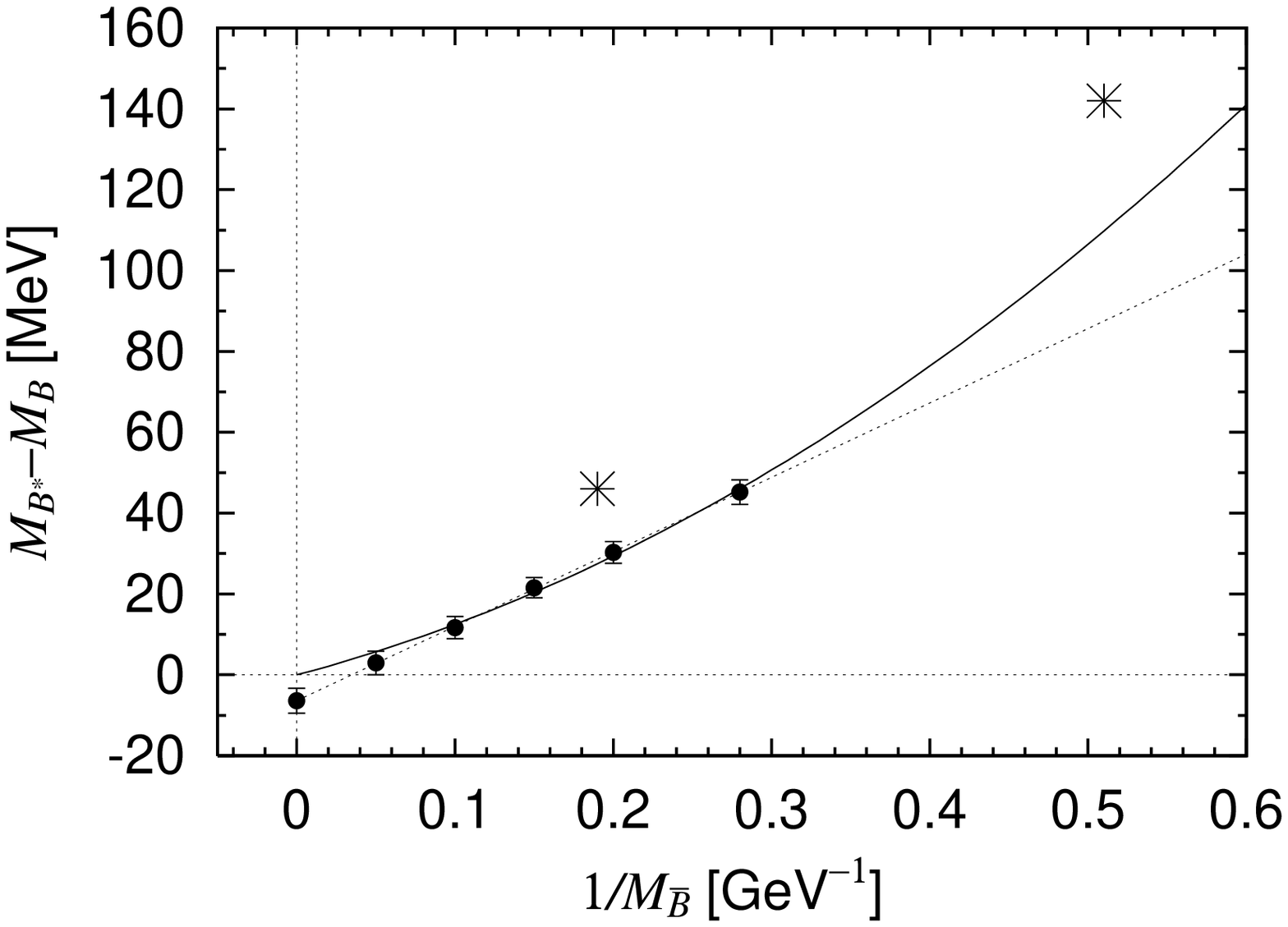}
  \caption{
    Hyperfine splitting $M_{B^*}-M_B$ as a function of
    $1/M_{\bar{B}}$. 
    The light quark mass is exptapolated to the chiral limit.
  }
  \label{fig:hqdep.hfsm.k0}
\end{figure}

\begin{figure}
  \includegraphics[width=\figwidth,angle=-90]{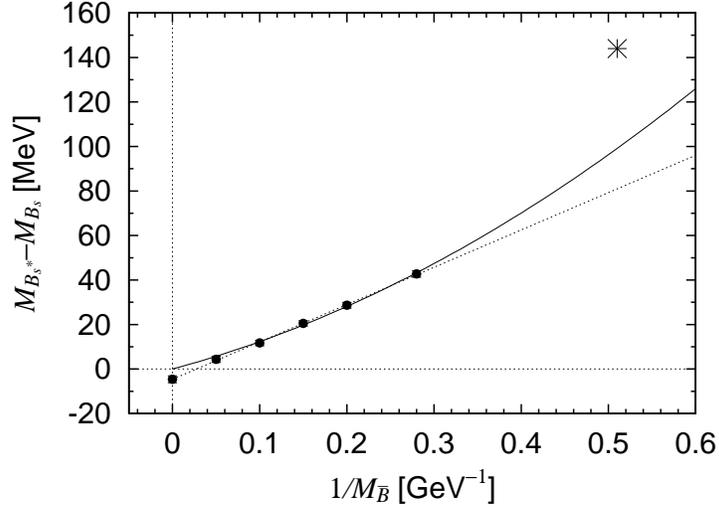}
  \caption{
    Hyperfine splitting $M_{B_s^*}-M_{B_s}$ as a function of
    $1/M_{\bar{B}}$. 
    The light quark mass is interpolated into the strange quark mass.
  }
  \label{fig:hqdep.hfsm.ks}
\end{figure}

\begin{figure}
  \includegraphics[width=\figwidth,angle=-90]{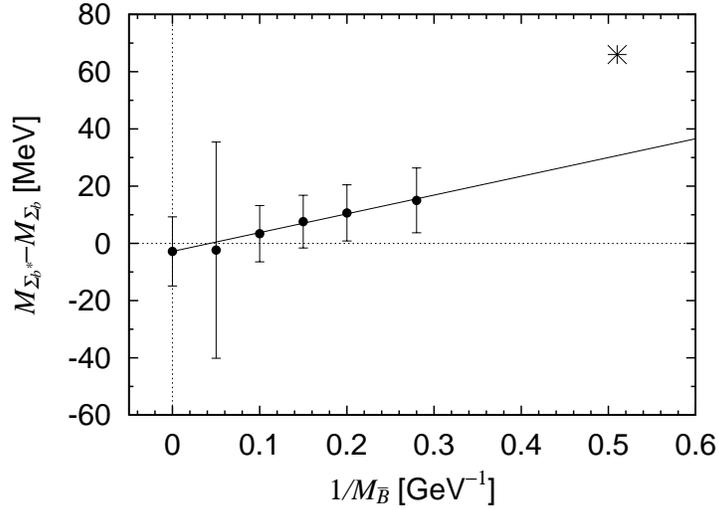}
  \caption{
    Hyperfine splitting $M_{\Sigma_b^*}-M_{\Sigma_b}$
    as a function of $1/M_{\bar{B}}$.
    The light quark mass is exptapolated to the chiral limit.
  }
  \label{fig:hqdep.hfsb.k0}
\end{figure}

The hyperfine splitting in the mesons $M_{B_d^*}-M_{B_d}$
and $M_{B_s^*}-M_{B_s}$ and in the
baryons $M_{\Sigma_b}^*-M_{\Sigma_b}$ is plotted in
Figures~\ref{fig:hqdep.hfsm.k0},
\ref{fig:hqdep.hfsm.ks} and \ref{fig:hqdep.hfsb.k0},
respectively, as a function of $1/M_{\bar{B}}$.
The numerical values at each quark masses are given in
Table~\ref{tab:mass_difference},
where the statistical error in the hyperfine splittings
is greatly reduced because it is highly correlated
within the spin multiplets.

For the $B-B^*$ splitting (Figure~\ref{fig:hqdep.hfsm.k0})
we observe a linear behavior which is consistent with the
expectation that the hyperfine splitting is proportional to
$1/m_Q$.
The intercept at $1/M_{\bar{B}}$ is, however, slightly
negative.
Since the hyperfine splitting is exactly zero in the static
limit, we attempt a constrained fit with a linear and
quadratic terms in $1/M_{\bar{B}}$, which is also shown in
Figure~\ref{fig:hqdep.hfsm.k0}.
It indicates that the quadratic term is not negligible and
amounts about 5\% at the $B$ meson mass.
From the coefficient of the linear term we obtain
\begin{equation}
  \lambda_2(B_d) = 0.051(16) \;\mathrm{GeV}^2.
\end{equation}
The similar analysis for $B_s$ gives
\begin{equation}
  \lambda_2(B_s) = 0.053(8) \;\mathrm{GeV}^2.
\end{equation}
The data and fit curves are shown in
Figure~\ref{fig:hqdep.hfsm.ks}. 
For the baryon hyperfine splitting
$M_{\Sigma_b^*}-M_{\Sigma_b}$ shown in
Figure~\ref{fig:hqdep.hfsb.k0}, the statistical error is so
large that the intercept of the linear fit is statistically
consistent with zero. 
The slope yields
\begin{equation}
  \mu_G^2(\Sigma_b^*)-\mu_G^2(\Sigma_b) 
  =
  -0.13(11) \;\mathrm{GeV}^2.
\end{equation}

The experimental values of $M_{B^*}-M_{B}$, $M_{D^*}-M_D$,
$M_{B_s^*}-M_{B_s}$
and $M_{\Sigma_c^*}-M_{\Sigma_c}$ are also shown in
Figures~\ref{fig:hqdep.hfsm.k0}, \ref{fig:hqdep.hfsm.ks} and
\ref{fig:hqdep.hfsb.k0}. 
($M_{\Sigma_b^*}-M_{\Sigma_b}$ has not yet been measured.)
The lattice data are significantly lower than these
experimental results as in many other quenched lattice
calculations.
It is partly due to the fact that the spin-chromomagnetic
 interaction term in the lattice NRQCD action
(\ref{eq:delta_H}) is matched to the continuum full theory
only at the tree level, although the mean field improvement
is applied. 
Another important uncertainty is in the quenching
approximation, whose effect is not yet entirely uncovered.

The numerical results given in this subsection are
also summarized in Table~\ref{tab:summary} together with the 
results from other groups and the experimental values.

\subsection{Consistency among matrix elements and mass differences}
Results presented so far indicate that the HQE parameters
are determined consistently with the direct measurement
of the matrix elements and with the indirect measurement
through the mass differences.
However, more stringent test is possible using the data at
fixed light quark mass, whose statistical error is smaller
than in the chiral limit. 
Although the numerical values are unphysical, there is
nothing wrong in the consistency check.
For this purpose we use the data at $K$ = 0.13331, which
corresponds to the heaviest light quark mass.

\begin{figure}
  \includegraphics[width=\figwidth,angle=-90]{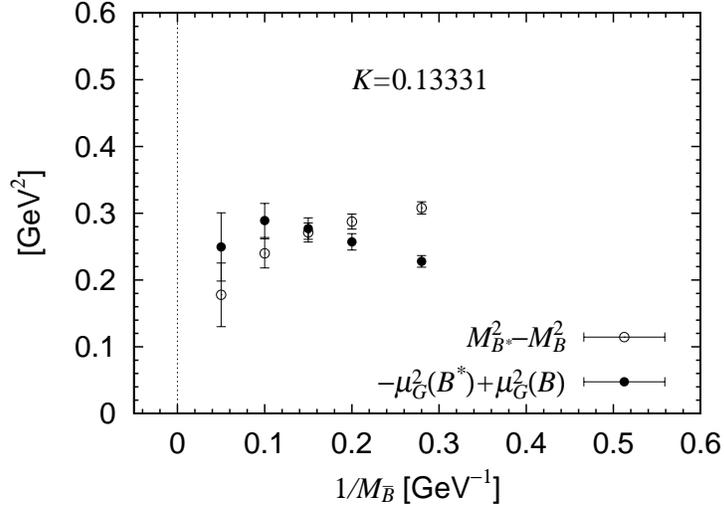}
  \caption{
    Hyperfine splitting of the heavy-light ground state mesons
    as a function of $1/M_{\bar{B}}$.
  }
  \label{fig:hqdep.hfsm2.G.k0}
\end{figure}

\begin{figure}
  \includegraphics[width=\figwidth,angle=-90]{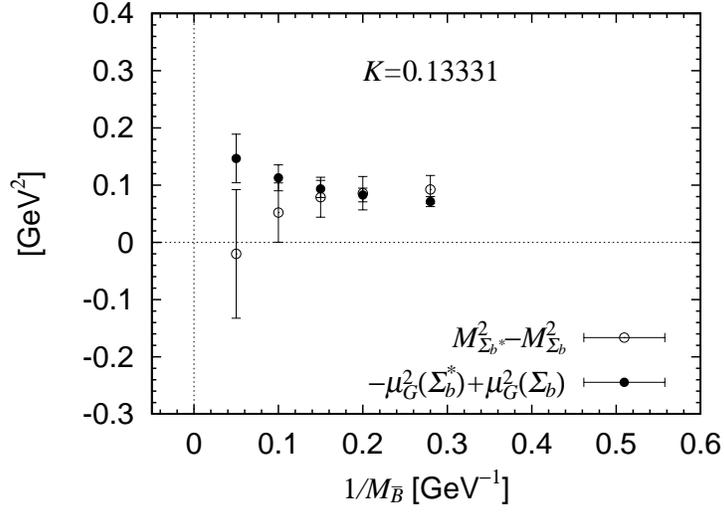}
  \caption{
    Hyperfine splitting of the heavy-light-light baryons
    as a function of $1/M_{\bar{B}}$.
  }
  \label{fig:hqdep.hfsb2.G.k0}
\end{figure}

From (\ref{eq:M_B}) and (\ref{eq:M_Bstar}) the hyperfine
splitting $M_{B^*}-M_B$ is given by $4\lambda_2/2m_b$,
or up to higher order $1/m_b$ corrections,
\begin{equation}
  M_{B^*}^2-M_B^2 
  =
  -\mu_G^2(B^*)+\mu_G^2(B).
  \label{eq:hfsm2}
\end{equation}
In Figure~\ref{fig:hqdep.hfsm2.G.k0}, we plot the results
for $-\mu_G^2(B^*)+\mu_G^2(B)$ as a function of $1/M_{\bar{B}}$
together with the lattice measurement of $M_{B^*}^2-M_B^2$.
We observe that the relation (\ref{eq:hfsm2}) is satisfied
well in the heavy quark mass region $1/M_{\bar{B}} <$
0.2~GeV$^{-1}$.
Towards lighter heavy quark mass the data deviate from the
relation (\ref{eq:hfsm2}), which is an indication of higher
order effect.
Similar analysis can be done for the hyperfine splitting of
heavy-light-light baryon, \textit{i.e.} the
$\Sigma_b^*-\Sigma_b$ splitting. 
Figure~\ref{fig:hqdep.hfsb2.G.k0} shows the mass difference
and the matrix element $-\Delta\mu_G^2$.
Both are in good agreement within the large statistical
error in the hadron mass measurement.

\begin{figure}
  \includegraphics[width=\figwidth,angle=-90]{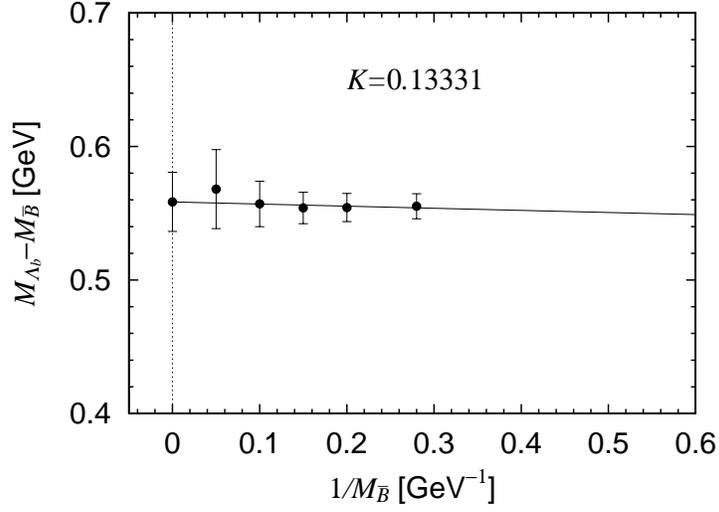}
  \caption{
    $M_{\Lambda_b}-M_{\bar{B}}$ as a function of
    $1/M_{\bar{B}}$. 
  }
  \label{fig:hqdep.L-BBar.mass}
\end{figure}

\begin{figure}
  \includegraphics[width=\figwidth,angle=-90]{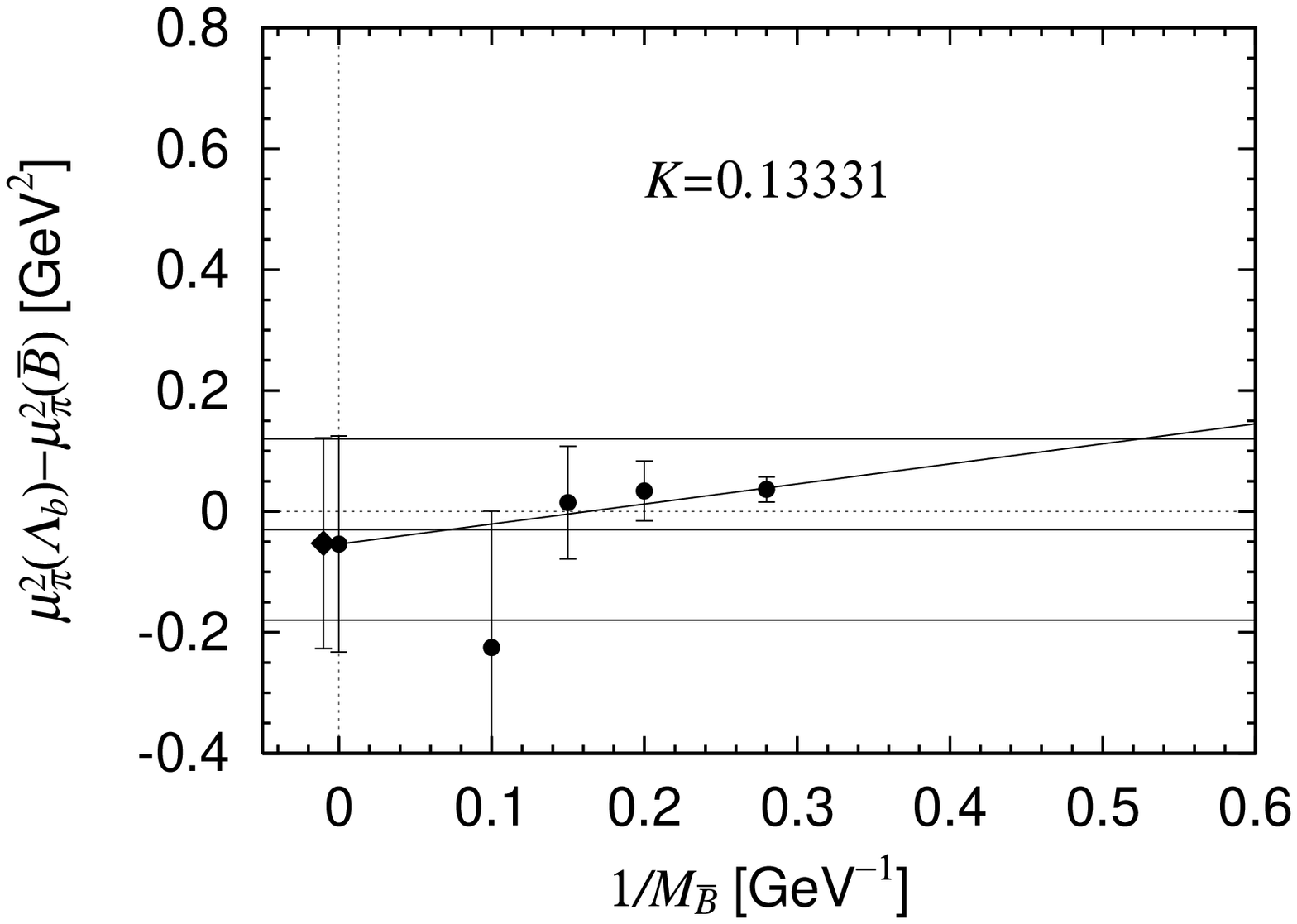}
  \caption{
    $-\mu_\pi^2(\Lambda_b)+\mu_\pi^2(\bar{B})$ measured from
    the matrix elements is compared with the indirect
    measurement from the slope of mass difference 
    $M_{\Lambda_b}-M_{\bar{B}}$, which gives
    $-0.03\pm 0.15$~GeV$^2$.
  }
  \label{fig:hqdep.L-BBar.pi.k1}
\end{figure}

The heavy-light meson-baryon mass difference 
$M_{\Lambda_b}-M_{\bar{B}}$ is given as
\begin{equation}
  M_{\Lambda_b}-M_{\bar{B}} =
  \overline{\Lambda}(\Lambda_b)-\overline{\Lambda}(B)
  +\frac{1}{2m_b}
  \left[\mu_{\pi}^2(\Lambda_b)-\mu_{\pi}^2(B)\right].
  \label{eq:meson-baryon}
\end{equation}
In Figure~\ref{fig:hqdep.L-BBar.mass} we plot 
$M_{\Lambda_b}-M_{\bar{B}}$ as a function of $1/M_{\bar{B}}$.
The slope obtained from the fit of the mass difference yields an 
indirect estimate of $\mu_{\pi}^2(\Lambda_b)-\mu_{\pi}^2(B)$ as 
$-0.03\pm 0.15$~GeV$^2$. 
Our results for the direct measurement of
$\mu_{\pi}^2(\Lambda_b)-\mu_{\pi}^2(B)$ are
plotted in Figure~\ref{fig:hqdep.L-BBar.pi.k1}, 
where the indirect measurement is shown by a band.
Both measurements are completely consistent with each
other.

\section{Lifetime ratio: a phenomenological application}
\label{sec:Phenomenological_application}

In the ratio of lifetimes of different $b$ hadrons
$H_b^{(1)}$ and $H_b^{(2)}$ the hadronic matrix elements
$\mu_\pi^2$ and $\mu_G^2$ appear as  
\begin{equation}
  \label{eq:lifetime_ratio}
  \frac{\tau(H_b^{(1)})}{\tau(H_b^{(2)})}
  = 1 + \frac{\mu_\pi^2(H_b^{(1)})-\mu_\pi^2(H_b^{(2)})}{2m_b^2}
  + c_G \frac{\mu_G^2(H_b^{(1)})-\mu_G^2(H_b^{(2)})}{m_b^2}
  + O\left(\frac{1}{m_b^3}\right),
\end{equation}
with a perturbative coefficient $c_G\simeq$ 1.2
\cite{Neubert:1997we}.
Our calculation of the differences of the matrix elements 
$\mu_\pi^2(H_b^{(1)})-\mu_\pi^2(H_b^{(2)})$ and
$\mu_G^2(H_b^{(1)})-\mu_G^2(H_b^{(2)})$ may be directly
used to evaluate the lifetime ratios at the order $1/m_b^2$.

Using our results 
$\mu_\pi^2(\Lambda_b)-\mu_\pi^2(B_d)$ = 
$-$0.01(52)~GeV$^2$
and 
$\mu_G^2(\Lambda_b)-\mu_G^2(B_d)$ ($\equiv -3\lambda_2(B_d)$) = 
$-$0.282(59)~GeV$^2$,
which are from the direct calculation (method 1),
the lifetime ratio of $\Lambda_b$ and $B_d$ is evaluated as 
\begin{equation}
    \frac{\tau(\Lambda_b)}{\tau(B_d)}
  = 0.984
  \pm 0.012
  \pm 0.003
  + O\left(\frac{1}{m_b^3}\right),
\end{equation}
with $m_b$ = 4.6 GeV,
where the first and second error comes from the statistical error of
$\mu_\pi^2(\Lambda_b)-\mu_\pi^2(B_d)$ and
$\mu_G^2(\Lambda_b)-\mu_G^2(B_d)$, respectively.
As discussed in the previous works \cite{Neubert:1997we} it
may not explain the experimental value 0.76(5) unless the
higher order effect in the $1/m_b$ expansion has a
substantially large effect.
Our calculation does not imply such a large correction to the matrix
element $\mu_\pi^2$ as shown in Figure~\ref{fig:hqdep.L-BBar.pi.k1}.
At the order $1/m_b^3$ the spectator effect arises, for
which the hadronic matrix elements of higher dimensional
operators are necessary \cite{Neubert:1997we}.
A lattice calculation \cite{DiPierro:1999tb} of those matrix
elements suggests that the spectator effects are
indeed significant but do not appear to be sufficiently
large to account for the full discrepancy.

The lifetime ratio of $B_s$ and $B_d$ is obtained as
\begin{equation}
  \frac{\tau(B_s)}{\tau(B_d)}
  = 1.001
  \pm 0.002
  \pm 0.002
  + O\left(\frac{1}{m_b^3}\right),
\end{equation}
using our results for
$\mu_\pi^2(B_s)-\mu_\pi^2(B_d)$=0.066(80)~GeV$^2$ and
$\mu_G^2(B_s)-\mu_G^2(B_d)$=$-$0.012(32)~GeV$^2$.
This result may be compared with the experimental
value $\tau(B_s)/\tau(B_d)$ = 0.949$\pm$0.038 \cite{B_lifetime}.

\section{Conclusions}
\label{sec:Conclusions}

In this article we present a lattice QCD calculation
of the heavy quark expansion parameters
$\overline{\Lambda}$, $\mu_{\pi}^2$ and $\mu_G^2$
for the heavy-light mesons and heavy-light-light baryons.
The lattice NRQCD action is used for heavy quark and the
results in the static limit are obtained by an
extrapolation. 

For $\mu_\pi^2$ and $\mu_G^2$, we performed a direct
calculation of the matrix elements through the three-point
functions. 
While the light quark mass dependence of the matrix elements
is small, the heavy quark mass dependence is significant
due to the effect of the additive renormalization.
The large heavy quark mass dependence mostly cancels by
considering the difference of the matrix elements between
different heavy hadron states, in which the additive
renormalization cancels.
We also estimate the differences of the HQE parameters
by studying the mass differences between several heavy
hadrons.

We find that the lattice measurements of the matrix elements 
$\mu_\pi^2$ and $\mu_G^2$ are consistent with the mass
relations predicted by the heavy quark expansion.
Our numerical results for the defferences of $\mu_\pi^2$ in
the heavy quark mass limit are compatible with the
previous determinations from the meson mass spectrum.
The deficit of the hyperfine splitting --- the well-known
problem of the quenched lattice calculation --- is also
reproduced in the direct calculation of the matrix element
$\mu_G^2$.

A direct phenomenological application of our results is the
evaluation of the lifetime ratios at the order $1/m_b^2$.
Previously such analysis implicitly assumed that the heavy 
quark expansion truncated at $1/m_b^2$ is valid down to the
charm quark mass, as the parameter $\mu_\pi^2$ was
determined using the combined mass difference including
charmed mesons and baryons.
Through the direct lattice calculation we have confirmed
that such analysis is justified.
The problem of the small lifetime ratio
$\tau(\Lambda_b)/\tau(B_d)$ still remains.

\begin{acknowledgments}
  This work was supported by the Supercomputer Project
  No.~66 (FY2001) and No.~79 (FY2002) of 
  High Energy Accelerator Research Organization (KEK), 
  and also in part by the Grants-in-Aid of the
  Ministry of Education (Nos. 10640246, 11640294, 12014202,
  12640253, 12640279, 12740133, 13135204, 13640259, 13640260,
  14046202, 14740173).
  N.Y. is supported by the JSPS Research Fellowship.
\end{acknowledgments}


\clearpage
\begin{table}
  \begin{tabular}{|c|ccccc|}
    \hline
    $aM_0$ & 1.3    & 2.1    & 3.0    & 5.0    & 10.0\\
    \hline
    $n$    & 3      & 3      & 2      & 2      & 2\\
    $a$    & 0.2248 & 0.2530 & 0.2711 & 0.3074 & 0.3425\\
    $b$    & 1.2484 & 1.1840 & 1.1465 & 1.0794 & 1.0294\\
    \hline
  \end{tabular}
  \caption{
    Simulation parameters. The parameter $a$ and $b$ is for
    the smeared source $e^{-a\cdot r^b}$.
  }
  \label{tab:NRQCD_parameters}
\end{table}

%
%
%
%
%
\begin{table}
\begin{tabular}{|cc|ccccc|}
    \hline
     $aM_0$ & $n$ & $A$ & $B$ & $aE_0$ &
     $Z_m$ & $a\Delta$\\
    \hline
     1.3 & 3 & 0.547 & 0.914 & 0.140 & 1.234 & 1.464\\
     2.1 & 3 & 0.754 & 0.578 & 0.193 & 1.148 & 2.218\\
     3.0 & 2 & 0.855 & 0.381 & 0.219 & 1.097 & 3.072\\
     5.0 & 2 & 0.946 & 0.176 & 0.242 & 1.045 & 4.983\\
    10.0 & 2 & 1.011 & 0.040 & 0.259 & 1.010 & 9.841\\
    \hline
  \end{tabular}
  \caption{Perturbative factors to obtain the hadron mass from
    $E_{sim}$ using (\ref{eq:binding_energy_and_mass}).
    The perturbative expansions are given as
    $aE_0=\alpha_s A$ and $Z_m=1+\alpha_s B$ where the
    coefficients $A$ and $B$ are given in 
    \cite{Ishikawa:1999xu}.
    For the numerical analysis we use a renormalized
    coupling $\alpha_V(1/a)=0.256$ for $\alpha_s$ at $\beta$
    = 6.0.
    $a\Delta$ in the last column is defined as
    $a\Delta=Z_m aM_0 - aE_0$.
  }
  \label{tab:mass_constant}
\end{table}

\begin{table}
  \begin{tabular}{|c|c|lllll|}
    \hline
     $aM_0$ & $K$ & $aE_{sim}(B)$ & $aE_{sim}(B^*)$ &
     $aE_{sim}(\Lambda_b)$ & $aE_{sim}(\Sigma_b)$ & $aE_{sim}(\Sigma_b^*)$\\
    \hline
     1.3 & 0.13331 & 0.4928(15) & 0.5156(17) & 0.8101(54) & 0.8573(60) & 0.8631(63)\\
     2.1 &         & 0.5145(17) & 0.5298(19) & 0.8256(60) & 0.8725(65) & 0.8766(67)\\
     3.0 &         & 0.5247(19) & 0.5357(21) & 0.8324(66) & 0.8787(70) & 0.8816(72)\\
     5.0 &         & 0.5327(24) & 0.5391(25) & 0.8386(92) & 0.8830(84) & 0.8843(86)\\
    10.0 &         & 0.5376(37) & 0.5401(38) & 0.847(16)  & 0.891(14)  & 0.891(14)\\
    \hline
     1.3 & 0.13384 & 0.4754(18) & 0.4987(20) & 0.7680(73) & 0.8210(85) & 0.8275(91)\\
     2.1 &         & 0.4976(21) & 0.5132(22) & 0.7849(83) & 0.8366(92) & 0.8411(97)\\
     3.0 &         & 0.5083(23) & 0.5194(25) & 0.7927(95) & 0.843(10)  & 0.847(10)\\
     5.0 &         & 0.5166(29) & 0.5229(30) & 0.801(12)  & 0.850(12)  & 0.852(12)\\
    10.0 &         & 0.5218(43) & 0.5241(44) & 0.813(21)  & 0.862(19)  & 0.862(19)\\
    \hline
     1.3 & 0.13432 & 0.4599(24) & 0.4836(26) & 0.728(12)  & 0.786(15)  & 0.793(15)\\
     2.1 &         & 0.4825(27) & 0.4983(29) & 0.746(14)  & 0.801(16)  & 0.806(16)\\
     3.0 &         & 0.4934(30) & 0.5047(32) & 0.756(15)  & 0.808(17)  & 0.812(17)\\
     5.0 &         & 0.5021(36) & 0.5085(38) & 0.767(19)  & 0.820(20)  & 0.821(20)\\
    10.0 &         & 0.5079(57) & 0.5099(55) & 0.782(32)  & 0.831(24)  & 0.830(23)\\
    \hline
    \hline
     1.3 & $K_s$   & 0.4826(17) & 0.5057(19) & 0.7850(66) & 0.8357(78) & 0.8419(82)\\
     2.1 &         & 0.5046(20) & 0.5201(21) & 0.8015(75) & 0.8510(83) & 0.8553(87)\\
     3.0 &         & 0.5151(22) & 0.5262(23) & 0.8089(84) & 0.8575(90) & 0.8606(93)\\
     5.0 &         & 0.5233(27) & 0.5296(28) & 0.816(11)  & 0.864(11)  & 0.865(11)\\
    10.0 &         & 0.5284(41) & 0.5307(42) & 0.827(18)  & 0.873(16)  & 0.873(16)\\
    \hline
     1.3 & $K_c$   & 0.4290(33) & 0.4534(36) & 0.652(17)  & 0.721(21)  & 0.729(22)\\
     2.1 &         & 0.4524(37) & 0.4688(39) & 0.673(20)  & 0.736(23)  & 0.742(24)\\
     3.0 &         & 0.4639(42) & 0.4756(43) & 0.684(23)  & 0.744(24)  & 0.749(25)\\
     5.0 &         & 0.4733(48) & 0.4796(52) & 0.699(28)  & 0.761(29)  & 0.762(29)\\
    10.0 &         & 0.4799(77) & 0.4814(74) & 0.722(49)  & 0.777(36)  & 0.776(34)\\
    \hline
  \end{tabular}
  \caption{
    Binding energy of heavy-light mesons and heavy-light-light baryons.
  }
  \label{tab:binding_energy}
\end{table}

\begin{table}
  \begin{tabular}{|c|c|lllll|}
    \hline
    $aM_0$ & $K$ & $a^2\mu_{\pi}^2(B)$ & $a^2\mu_{\pi}^2(B^*)$ &
    $a^2\mu_{\pi}^2(\Lambda_b)$ & $a^2\mu_{\pi}^2(\Sigma_b)$ & $a^2\mu_{\pi}^2(\Sigma_b^*)$ \\
    \hline
     1.3 & 0.13331 & $-$0.2507(24) & $-$0.2643(25) & $-$0.2503(62) & $-$0.2416(56) & $-$0.2449(55)\\
     2.1 &         & $-$0.0998(51) & $-$0.1075(53) & $-$0.096(15)  & $-$0.083(11)  & $-$0.085(11)\\
     3.0 &         & $-$0.0546(90) & $-$0.0558(94) & $-$0.051(27)  & $-$0.036(19)  & $-$0.037(19)\\
     5.0 &         & $-$0.016(23)  & $-$0.003(24)  & $-$0.072(63)  & $-$0.053(52)  & $-$0.048(52)\\
    10.0 &         & $-$0.030(88)  & $+$0.018(88)  & $-$0.28(27)   & $-$0.35(25)   & $-$0.34(25)\\
    \hline
     1.3 & 0.13384 & $-$0.2525(30) & $-$0.2663(30) & $-$0.2545(81) & $-$0.2416(82) & $-$0.2427(82)\\
     2.1 &         & $-$0.1027(63) & $-$0.1104(66) & $-$0.101(19)  & $-$0.081(17)  & $-$0.081(16)\\
     3.0 &         & $-$0.059(11)  & $-$0.059(12)  & $-$0.058(36)  & $-$0.035(28)  & $-$0.036(28)\\
     5.0 &         & $-$0.022(28)  & $-$0.003(29)  & $-$0.103(89)  & $-$0.082(77)  & $-$0.076(79)\\
    10.0 &         & $-$0.05(11)   & $+$0.01(11)   & $-$0.39(39)   & $-$0.51(38)   & $-$0.52(39)\\
    \hline
     1.3 & 0.13432 & $-$0.2537(40) & $-$0.2678(40) & $-$0.257(13)  & $-$0.229(14)  & $-$0.225(14)\\
     2.1 &         & $-$0.1042(84) & $-$0.1128(88) & $-$0.099(31)  & $-$0.057(27)  & $-$0.054(26)\\
     3.0 &         & $-$0.062(15)  & $-$0.062(16)  & $-$0.049(58)  & $-$0.008(47)  & $-$0.005(47)\\
     5.0 &         & $-$0.029(37)  & $-$0.004(38)  & $-$0.13(15)   & $-$0.10(13)   & $-$0.09(14)\\
    10.0 &         & $-$0.09(14)   & $+$0.00(14)   & $-$0.45(66)   & $-$0.71(69)   & $-$0.76(70)\\
    \hline
    \hline
     1.3 & $K_s$   & $-$0.2517(28) & $-$0.2654(28) & $-$0.2526(74) & $-$0.2399(73) & $-$0.2415(72)\\
     2.1 &         & $-$0.1014(58) & $-$0.1092(61) & $-$0.098(18)  & $-$0.078(14)  & $-$0.079(14)\\
     3.0 &         & $-$0.057(10)  & $-$0.058(11)  & $-$0.053(33)  & $-$0.032(25)  & $-$0.032(25)\\
     5.0 &         & $-$0.020(26)  & $-$0.003(27)  & $-$0.090(80)  & $-$0.069(68)  & $-$0.063(70)\\
    10.0 &         & $-$0.05(10)   & $+$0.01(10)   & $-$0.34(34)   & $-$0.45(34)   & $-$0.46(34)\\
    \hline
     1.3 & $K_c$   & $-$0.2567(55) & $-$0.2712(55) & $-$0.265(19)  & $-$0.227(20)  & $-$0.218(20)\\
     2.1 &         & $-$0.109(12)  & $-$0.118(12)  & $-$0.107(44)  & $-$0.049(40)  & $-$0.041(40)\\
     3.0 &         & $-$0.069(21)  & $-$0.067(21)  & $-$0.057(82)  & $-$0.001(70)  & $+$0.004(71)\\
     5.0 &         & $-$0.040(51)  & $-$0.006(52)  & $-$0.19(22)   & $-$0.15(20)   & $-$0.14(21)\\
    10.0 &         & $-$0.14(20)   & $-$0.01(20)   & $-$0.64(96)   & $-$1.01(98)   & $-$1.1(1.0)\\
    \hline
  \end{tabular}
  \caption{
    Matrix elements $\mu_{\pi}^2$ for
    heavy-light mesons and heavy-light-light baryons.
  }
  \label{tab:matrix_elements_mu_pi}
\end{table}

\begin{table}
  \begin{tabular}{|c|c|lllll|}
    \hline 
    $aM_0$ & $K$ & $a^2\mu_G^2(B)$ & $a^2\mu_G^2(B^*)$ &
    $a^2\mu_G^2(\Lambda_b)$ & $a^2\mu_G^2(\Sigma_b)$ & $a^2\mu_G^2(\Sigma_b^*)$ \\
    \hline
     1.3 & 0.13331 & 0.2507(20) & 0.1840(10) & 0.2027(12) & 0.2190(22) & 0.1982(13)\\
     2.1 &         & 0.2340(28) & 0.1588(12) & 0.1798(12) & 0.1979(28) & 0.1736(15)\\
     3.0 &         & 0.2067(35) & 0.1258(15) & 0.1482(13) & 0.1674(34) & 0.1402(18)\\
     5.0 &         & 0.1671(56) & 0.0827(21) & 0.1056(17) & 0.1273(48) & 0.0944(25)\\
    10.0 &         & 0.112(11)  & 0.0395(39) & 0.0581(23) & 0.0856(86) & 0.0427(44)\\
    \hline
     1.3 & 0.13384 & 0.2515(24) & 0.1837(11) & 0.2033(15) & 0.2208(29) & 0.1976(16)\\
     2.1 &         & 0.2350(33) & 0.1585(15) & 0.1804(15) & 0.2000(36) & 0.1732(19)\\
     3.0 &         & 0.2079(43) & 0.1253(18) & 0.1488(17) & 0.1700(43) & 0.1397(23)\\
     5.0 &         & 0.1680(68) & 0.0823(26) & 0.1060(22) & 0.1303(62) & 0.0936(32)\\
    10.0 &         & 0.111(13)  & 0.0399(46) & 0.0584(29) & 0.090(11)  & 0.0407(55)\\
    \hline
     1.3 & 0.13432 & 0.2526(29) & 0.1835(15) & 0.2031(20) & 0.2240(43) & 0.1970(23)\\
     2.1 &         & 0.2363(42) & 0.1581(19) & 0.1806(23) & 0.2040(53) & 0.1729(28)\\
     3.0 &         & 0.2094(55) & 0.1249(23) & 0.1488(25) & 0.1744(63) & 0.1395(34)\\
     5.0 &         & 0.1694(88) & 0.0820(33) & 0.1057(32) & 0.1346(91) & 0.0931(46)\\
    10.0 &         & 0.110(17)  & 0.0404(59) & 0.0589(43) & 0.095(15)  & 0.0389(77)\\
    \hline
    \hline
     1.3 & $K_s$   & 0.2512(22) & 0.1839(11) & 0.2030(14) & 0.2203(27) & 0.1978(15)\\
     2.1 &         & 0.2346(31) & 0.1586(14) & 0.1801(14) & 0.1994(33) & 0.1734(18)\\
     3.0 &         & 0.2075(40) & 0.1255(17) & 0.1484(15) & 0.1692(40) & 0.1399(21)\\
     5.0 &         & 0.1677(64) & 0.0825(24) & 0.1058(21) & 0.1293(57) & 0.0939(29)\\
    10.0 &         & 0.112(13)  & 0.0398(44) & 0.0583(27) & 0.088(10)  & 0.0415(50)\\
    \hline
     1.3 & $K_c$   & 0.2544(40) & 0.1830(20) & 0.2037(29) & 0.2277(60) & 0.1959(33)\\
     2.1 &         & 0.2384(56) & 0.1575(25) & 0.1816(33) & 0.2087(75) & 0.1722(40)\\
     3.0 &         & 0.2117(75) & 0.1242(31) & 0.1497(36) & 0.1797(91) & 0.1388(48)\\
     5.0 &         & 0.171(12)  & 0.0813(44) & 0.1061(46) & 0.141(13)  & 0.0919(66)\\
    10.0 &         & 0.108(23)  & 0.0412(78) & 0.0594(60) & 0.102(22)  & 0.035(11)\\
    \hline
  \end{tabular}
  \caption{
    Matrix elements $\mu_G^2$ for
    heavy-light mesons and heavy-light-light baryons.
  }
  \label{tab:matrix_elements_mu_G}
\end{table}

\begin{table}
  \begin{tabular}{|c|c|ccc|}
    \hline
    $aM_0$ & $K$ & $a^2\mu_{\pi}^2(B)=a^2\mu_{\pi}^2(B^*)$ &
    $a^2\mu_{\pi}^2(\Lambda_b)$ & $a^2\mu_{\pi}^2(\Sigma_b)=a^2\mu_{\pi}^2(\Sigma_b^*)$\\
    \hline
     static & 0.13331 & $-$0.057(61) & $-$0.25(17) & $-$0.22(13)\\
    \hline
     static & $K_s$   & $-$0.061(69) & $-$0.29(21) & $-$0.26(17)\\
    \hline
     static & $K_c$   & $-$0.09(13)  & $-$0.48(53) & $-$0.54(50)\\
    \hline
  \end{tabular}
  \caption{
    Matrix elements $\mu_{\pi}^2$ for
    heavy-light mesons and heavy-light-light baryons
    in the static heavy quark limit.
  }
  \label{tab:matrix_elements_PI_static}
\end{table}

\begin{table}
  \begin{tabular}{|c|c|ccccc|}
    \hline 
    $aM_0$ & $K$ & $a^2\mu_G^2(B)$ & $a^2\mu_G^2(B^*)$ &
    $a^2\mu_G^2(\Lambda_b)$ & $a^2\mu_G^2(\Sigma_b)$ & $a^2\mu_G^2(\Sigma_b^*)$ \\
    \hline
     static & 0.13331 & 0.065(15) & $-$0.022(05) & 0 & 0.039(11) & $-$0.020(06)\\
    \hline
     static & $K_s$   & 0.064(17) & $-$0.021(06) & 0 & 0.043(13) & $-$0.021(07)\\
    \hline
     static & $K_c$   & 0.060(30) & $-$0.020(10) & 0 & 0.059(29) & $-$0.029(15)\\
    \hline
  \end{tabular}
  \caption{
    Matrix elements $\mu_G^2$ for
    heavy-light mesons and heavy-light-light baryons
    in the static heavy quark limit.
  }
  \label{tab:matrix_elements_G_static}
\end{table}

\begin{table}
  \begin{tabular}{|c|c|c|c|c|}
    \hline 
     & Direct. calc. & Mass difference & Other works & Exp.\\
     & (method 1,2) & & & \\
    \hline
     $\overline{\Lambda}(\Lambda_b)-\overline{\Lambda}(B)$ [MeV]&
       & 428(68) & 415(156)~\cite{AliKhan:1999yb} &\\
     $\overline{\Lambda}(\Sigma_b)-\overline{\Lambda}(\Lambda_b)$ [MeV]&
       & 96(96)  & 176(152)~\cite{AliKhan:1999yb} &\\
     $\overline{\Lambda}(B_s)-\overline{\Lambda}(B_d)$ [MeV]&
       & 90(7)   & 81(31)~\cite{AliKhan:1999yb} &\\
    \hline
     $\mu_\pi^2(\Lambda_b)-\mu_\pi^2(B)$ [GeV$^2$] &
      $-$0.01(52), $-$1.3(1.8) & $-$0.38(47)  & 0~\cite{AliKhan:1999yb} & $-$0.01(3)\\
     $\mu_\pi^2(\Sigma_b)-\mu_\pi^2(\Lambda_b)$ [GeV$^2$] &
      0.28(68), $-$0.2(2.5) & 0.29(66)  & 0~\cite{AliKhan:1999yb} &\\
     $\mu_\pi^2(B_s)-\mu_\pi^2(B_d)$ [GeV$^2$] &
      0.066(80), 0.09(26) & 0.056(42) & 0.09(4)~\cite{Gimenez:1997av}, 0.10(28)~\cite{AliKhan:1999yb} &
      0.06(2)\\
    \hline
     $\lambda_2(B_d)$ [GeV$^2$] &
      0.094(19), 0.076(39) & 0.051(16) & 0.070(15)~\cite{Gimenez:1997av}, 0.069(19)~\cite{AliKhan:1999yb} &
      0.12(1)\\ 
     $\lambda_2(B_s)$ [GeV$^2$] &
      0.090(10), 0.082(22) & 0.053(8) & 0.078(12)~\cite{AliKhan:1999yb} &\\
     $\mu_G^2(\Sigma_b)$ [GeV$^2$] &
      0.147(60), 0.23(11) & 0.09(7) & &\\
    \hline
  \end{tabular}
  \caption{
    Results for the HQE paremeters.
  }
  \label{tab:summary}
\end{table}

\begin{table}
  \begin{tabular}{|c|c|llll|}
    \hline
     $aM_0$ & $K$ & $a(M_{B^*}-M_B)$ & $a(M_{\Sigma_b^*}-M_{\Sigma_b})$ &
     $a(M_{\Lambda_b}-M_{\bar{B}})$ & $a(M_{\bar{\Sigma}_b}-M_{\Lambda_b})$\\
    \hline
     1.3 & $K_s$   & 0.0231(08) &    0.0062(21) & 0.2851(63) & 0.055(09)\\
     2.1 &         & 0.0155(07) &    0.0043(18) & 0.2852(72) & 0.052(11)\\
     3.0 &         & 0.0111(07) &    0.0031(17) & 0.2855(81) & 0.051(12)\\
     5.0 &         & 0.0063(07) &    0.0014(18) & 0.288(11)  & 0.048(15)\\
    10.0 &         & 0.0024(08) & $-$0.0003(27) & 0.297(18)  & 0.046(24)\\
    \hline
     1.3 & $K_c$   & 0.0244(16) &    0.0081(61) & 0.205(17)  & 0.075(28)\\
     2.1 &         & 0.0164(14) &    0.0058(53) & 0.209(20)  & 0.067(31)\\
     3.0 &         & 0.0116(14) &    0.0041(50) & 0.212(23)  & 0.063(34)\\
     5.0 &         & 0.0063(15) &    0.0018(53) & 0.221(28)  & 0.063(41)\\
    10.0 &         & 0.0016(16) & $-$0.001(20)  & 0.241(49)  & 0.055(61)\\
    \hline
  \end{tabular}
  \caption{
    Mass difference between heavy hadrons.
  }
  \label{tab:mass_difference}
\end{table}

\clearpage

%



%
%

%
%

%
%

%
%

%
%
\clearpage

%
%

%
%

%
%
\clearpage

\clearpage
\begin{thebibliography}{99}

\bibitem{Neubert:1997gu}
M.~Neubert,
Adv.\ Ser.\ Direct.\ High Energy Phys.\  {\bf 15}, 239 (1998)
[arXiv:hep-ph/9702375].

\bibitem{Bigi:1997fj}
I.~I.~Bigi, M.~A.~Shifman and N.~Uraltsev,
Ann.\ Rev.\ Nucl.\ Part.\ Sci.\  {\bf 47}, 591 (1997)
[arXiv:hep-ph/9703290].

\bibitem{Chay:1990da}
J.~Chay, H.~Georgi and B.~Grinstein,
Phys.\ Lett.\ B {\bf 247}, 399 (1990).

\bibitem{Bigi:1993fe}
I.~I.~Bigi, M.~A.~Shifman, N.~G.~Uraltsev and A.~I.~Vainshtein,
Phys.\ Rev.\ Lett.\  {\bf 71}, 496 (1993)
[arXiv:hep-ph/9304225].

\bibitem{Manohar:1993qn}
A.~V.~Manohar and M.~B.~Wise,
Phys.\ Rev.\ D {\bf 49}, 1310 (1994)
[arXiv:hep-ph/9308246].

\bibitem{Blok:1993va}
B.~Blok, L.~Koyrakh, M.~A.~Shifman and A.~I.~Vainshtein,
Phys.\ Rev.\ D {\bf 49}, 3356 (1994)
[Erratum-ibid.\ D {\bf 50}, 3572 (1994)]
[arXiv:hep-ph/9307247].

\bibitem{Martinelli:1995vj}
G.~Martinelli and C.~T.~Sachrajda,
Phys.\ Lett.\ B {\bf 354}, 423 (1995)
[arXiv:hep-ph/9502352].

\bibitem{Hashimoto:1994nd}
S.~Hashimoto,
Phys.\ Rev.\ D {\bf 50}, 4639 (1994)
[arXiv:hep-lat/9403028].

\bibitem{Dikeman:1995ad}
R.~D.~Dikeman, M.~A.~Shifman and N.~G.~Uraltsev,
Int.\ J.\ Mod.\ Phys.\ A {\bf 11}, 571 (1996)
[arXiv:hep-ph/9505397].

\bibitem{Kapustin:1995nr}
A.~Kapustin and Z.~Ligeti,
Phys.\ Lett.\ B {\bf 355}, 318 (1995)
[arXiv:hep-ph/9506201].

\bibitem{Falk:1995me}
A.~F.~Falk, M.~E.~Luke and M.~J.~Savage,
Phys.\ Rev.\ D {\bf 53}, 2491 (1996)
[arXiv:hep-ph/9507284].

\bibitem{Falk:1995kn}
A.~F.~Falk, M.~E.~Luke and M.~J.~Savage,
Phys.\ Rev.\ D {\bf 53}, 6316 (1996)
[arXiv:hep-ph/9511454].

\bibitem{Gremm:1996yn}
M.~Gremm, A.~Kapustin, Z.~Ligeti and M.~B.~Wise,
Phys.\ Rev.\ Lett.\  {\bf 77}, 20 (1996)
[arXiv:hep-ph/9603314].

\bibitem{Neubert:1997we}
M.~Neubert and C.~T.~Sachrajda,
Nucl.\ Phys.\ B {\bf 483}, 339 (1997).

\bibitem{Ball:1993xv}
P.~Ball and V.~M.~Braun,
Phys.\ Rev.\ D {\bf 49}, 2472 (1994)
[arXiv:hep-ph/9307291].

\bibitem{Neubert:1996wm}
M.~Neubert,
Phys.\ Lett.\ B {\bf 389}, 727 (1996)
[arXiv:hep-ph/9608211].

\bibitem{Crisafulli:1995}
M.~Crisafulli, V.~Gimenez, G.~Martinelli and C.~T.~Sachrajda,
Nucl.\ Phys.\ B {\bf 457}, 594 (1995)
[arXiv:hep-lat/9506210].

\bibitem{Gimenez:1997av}
V.~Gimenez, G.~Martinelli and C.~T.~Sachrajda,
Nucl.\ Phys.\ B {\bf 486}, 227 (1997)
[arXiv:hep-lat/9607055].

\bibitem{AliKhan:1999yb}
A.~Ali Khan {\it et al.},
Phys.\ Rev.\ D {\bf 62}, 054505 (2000)
[arXiv:hep-lat/9912034].

\bibitem{Kronfeld:2000gk}
A.~S.~Kronfeld and J.~N.~Simone,
Phys.\ Lett.\ B {\bf 490}, 228 (2000)
[Erratum-ibid.\ B {\bf 495}, 441 (2000)]
[arXiv:hep-ph/0006345].

\bibitem{Thacker:1990bm}
B.~A.~Thacker and G.~P.~Lepage,
Phys.\ Rev.\ D {\bf 43}, 196 (1991).

\bibitem{Lepage:1992tx}
G.~P.~Lepage, L.~Magnea, C.~Nakhleh, U.~Magnea and K.~Hornbostel,
Phys.\ Rev.\ D {\bf 46}, 4052 (1992)
[arXiv:hep-lat/9205007].

\bibitem{Ishikawa:1999xu}
K.~I.~Ishikawa {\it et al.}  [JLQCD Collaboration],
Phys.\ Rev.\ D {\bf 61}, 074501 (2000)
[arXiv:hep-lat/9905036].

\bibitem{Aoki:2002bh}
S.~Aoki {\it et al.}  [JLQCD Collaboration],
arXiv:hep-lat/0208038.

\bibitem{Lepage:1992xa}
G.~P.~Lepage and P.~B.~Mackenzie,
Phys.\ Rev.\ D {\bf 48}, 2250 (1993)
[arXiv:hep-lat/9209022].

\bibitem{Sheikholeslami:1985ij}
B.~Sheikholeslami and R.~Wohlert,
Nucl.\ Phys.\ B {\bf 259}, 572 (1985).

\bibitem{Luscher:1996ug}
M.~Luscher, S.~Sint, R.~Sommer, P.~Weisz and U.~Wolff,
Nucl.\ Phys.\ B {\bf 491}, 323 (1997)
[arXiv:hep-lat/9609035].

\bibitem{Maiani:az}
L.~Maiani, G.~Martinelli and C.~T.~Sachrajda,
Nucl.\ Phys.\ B {\bf 368}, 281 (1992).

\bibitem{Flynn:1991kw}
J.~M.~Flynn and B.~R.~Hill,
Phys.\ Lett.\ B {\bf 264}, 173 (1991).

\bibitem{DiPierro:1999tb}
M.~Di Pierro, C.~T.~Sachrajda and C.~Michael  [UKQCD collaboration],
Phys.\ Lett.\ B {\bf 468}, 143 (1999)
[arXiv:hep-lat/9906031].

\bibitem{B_lifetime}
B Lifetime Working group,
http://lepbosc.web.cern.ch/LEPBOSC/lifetimes/lepblife.html

\end{thebibliography}
\end{document}